\documentclass[12pt,letterpaper]{report}
\usepackage{amsmath,amssymb,array,calc,graphicx,multirow,mathrsfs,subfigure}
\usepackage{uidiss10}
\usepackage{tensind}  
\tensordelimiter{?}
\usepackage{hyperref}
\begin{document}


\abtitlepgfalse 
\abstractpgfalse 
\titlepgtrue 
\copyrighttrue 
\signaturepagefalse 
\dedicationtrue 
\acktrue 
\abswithesistrue 
\figurespagetrue 

\title{A WALK THROUGH SUPERSTRING THEORY WITH AN APPLICATION TO YANG-MILLS THEORY: K-STRINGS AND D-BRANES AS GAUGE/GRAVITY DUAL OBJECTS} 

\author{Kory M. Stiffler} 
\advisor{Professor Vincent G. J. Rodgers} 
\dept{Physics} 
\submitdate{July 2010} 

\newcommand{\abstextwithesis}
{Superstring theory is one current, promising attempt at unifying gravity with the other three known forces: the electromagnetic force, and the weak and strong nuclear forces.   Though this is still a work in progress, much effort has been put toward this goal.  A set of specific tools which are used in this effort are gauge/gravity dualities.  This thesis consists of a specific implementation of gauge/gravity dualities to describe $k$-strings of strongly coupled gauge theories as objects dual to D$p$-branes embedded in confining supergravity backgrounds from low energy superstring field theory.

Along with superstring theory, $k$-strings are also commonly investigated with lattice gauge theory and Hamiltonian methods.  A $k$-string is a colorless combination of quark-antiquark source pairs, between which a color flux tube develops.  The two most notable terms of the $k$-string energy are, for large quark anti-quark separation $L$,  the tension term, proportional to $L$, and the Coulombic $1/L$ correction, known as the L\"uscher term.

This thesis provides an overview of superstring theories and how gauge/gravity dualities emerge from them.  It shows in detail how these dualities can be used for the specific problem of calculating the $k$-string energy in $2+1$ and $3+1$ space-time dimensions as the energy of D$p$-branes in the dual gravitational theory.  A detailed review of $k$-string tension calculations is given where good agreement is found with lattice gauge theory and Hamiltonian methods.  In reviewing the $k$-string tension, we also touch on how different representations of $k$-strings can be described with D$p$-branes through gauge/gravity dualities.  The main result of this thesis is how the L\"uscher term is found to emerge as the one loop quantum corrections to the D$p$-brane energy.  In $2+1$ space-time dimensions, we have L\"uscher term data to compare with from lattice gauge theory, where we find good agreement.}

\newcommand{\abstracttext}
{Superstring theory is one current, promising attempt at unifying gravity with the other three known forces: the electromagnetic force, and the weak and strong nuclear forces.   Though this is still a work in progress, much effort has been put toward this goal.  A set of specific tools which are used in this effort are gauge/gravity dualities.  This thesis consists of a specific implementation of gauge/gravity dualities to describe $k$-strings of strongly coupled gauge theories as objects dual to D$p$-branes embedded in confining supergravity backgrounds from low energy superstring field theory.

Along with superstring theory, $k$-strings are also commonly investigated with lattice gauge theory and Hamiltonian methods.  A $k$-string is a colorless combination of quark-antiquark source pairs, between which a color flux tube develops.  The two most notable terms of the $k$-string energy are, for large quark anti-quark separation $L$,  the tension term, proportional to $L$, and the Coulombic $1/L$ correction, known as the L\"uscher term.

This thesis provides an overview of superstring theories and how gauge/gravity dualities emerge from them.  It shows in detail how these dualities can be used for the specific problem of calculating the $k$-string energy in $2+1$ and $3+1$ space-time dimensions as the energy of D$p$-branes in the dual gravitational theory.  A detailed review of $k$-string tension calculations is given where good agreement is found with lattice gauge theory and Hamiltonian methods.  In reviewing the $k$-string tension, we also touch on how different representations of $k$-strings can be described with D$p$-branes through gauge/gravity dualities.  The main result of this thesis is how the L\"uscher term is found to emerge as the one loop quantum corrections to the D$p$-brane energy.  In $2+1$ space-time dimensions, we have L\"uscher term data to compare with from lattice gauge theory, where we find good agreement.}

\newcommand{\dedication}
{ To my parents }

\newcommand{\acknowledgement}
{I would first like to thank my advisor, Prof. Vincent G. J. Rodgers.  It has been a lot of fun working with him and I would never have been able to accomplish what I have in graduate school without him.  I would also like to thank our collaborator, Prof. Leopoldo A. Pando Zayas, the originator of the ideas of our various $k$-string projects.  Next, I thank the other members of my thesis committee, Profs.~Yannick Meurice, Wayne N. Polyzou, Craig E. Pryor, and Charles Frohman.

Looking back now at my path through undergraduate and graduate school, both at the University of Iowa, it seems an unlikely path were it not for the seemingly critical timing of the advice and help of several people.  Most notably, these people who I would now like to thank include: Prof.~M. L. Raghavan, Mart Sieren, Prof.~Michael A. Mackey, Prof.~Joon B. Park, Dr.~Ralph Adolphs, Yota Kimura, Dr.~Tony Buchanon, Dr.~Hiroto Kawasaki~M.D., Debbie Foreman, Prof.~Usha Mallik, and Prof.~William Klink.  

I must also thank my friends and colleagues with whom I've spent many hours toiling over aspects of physics from Newton's Law's, to Polyakov's writings, to formatting a University of Iowa Ph.~D. thesis with LaTex: Dr.~Alexander P. Bulmahn, Chris Doran, Dr.~Theodore R. Jaeger, Xiaolong Liu, Dr.~Tho Duc Nguyen, Leo Rodriguez, Jim Rybicki, and Tuna Yildirim.  Also, a special thanks goes to Sara Langworthy for her timely advice on how to finish a large project.  
 
Most of all, I would like to thank my parents.  

}
\beforepreface
\afterpreface




\Introduction\label{intro}
\vskip -\li
The current, most trusted theory used to describe the strong nuclear force is QCD.  In the high-energy regime, the strong force coupling is very small.  As a result, perturbative expansions of the Feynman path integral work well here, as cross-section calculations for scattering processes depend appreciably only on the first few, low order Feynman diagrams.  Because of this behavior of the strong force coupling, quarks are said to be \emph{asymptotically free} at high energies.  

The story dramatically changes in the low energy regime.  Here, the strong force coupling is very strong, and perturbative methods fail as calculations depend more and more on higher order Feynman diagrams, to the point where calculations diverge.  To solve this dilemma, a different method from perturbative QCD is implemented in this low energy regime: lattice QCD.  In a nutshell, lattice QCD solves the strong coupling problem at low energies by going back to the original definition of the path integral, fields which exist on a discretized space-time lattice, and evaluating the full path integral, before taking the continuum limit.  Because of the discrete nature of this lattice, the divergences found in perturbative QCD calculations go away, and promising results are found. Yet another method used to study QCD at low energies is to abandon the Feynman path integral approach altogether and work directly with the QCD Hamiltonian.

Finally, one can use string theory to study strongly coupled gauge fields.  What is advantageous about string theory, is that it is a candidate for unifying gravity with the gauge forces of the standard model, the strong and electroweak.  It is not known exactly how to use string theory to describe QCD and the rest of the standard model, but a lot of progress has been made toward this, most notably with the emergence of the AdS/CFT correspondence and other gauge/gravity dualities.  These gauge/gravity dualities relate supergravity theories to gauge theories through low energy effective descriptions of superstring theory, and gauge/gravity dualities exist which can be used to describe strongly coupled gauge theories.  What is particularly useful about these dualities is that they are weak/strong dualities:  where the gravitational theory is weakly coupled, the gauge theory is strongly coupled and vice-versa so one always has access to a perturbative regime to describe either.

As we don't have experimental access to the energies required to test superstring theory directly, we can instead test its theoretical predictions through gauge/gravity dualities against other theoretical methods.  One such theoretical avenue, which is the main topic for this thesis, is $k$-strings: colorless representations of QCD.   Their properties are commonly investigated with lattice gauge theories, Hamiltonian methods, and string theory.  With the AdS/CFT correspondence, one can do  calculations in AdS $\times S^5$ which are known to be dual to $\mathcal{N}=4$ Super-Yang-Mills theory in $3+1$~\cite{Maldacena:1997re}.  This thesis will expand upon this and show detailed calculations in supergravity backgrounds, which emerge from string theory, that are known to be dual to calculations in gauge theories with less supersymmetry than $\mathcal{N}=4$, and with running coupling constants~\cite{Klebanov:2000hb,Cvetic:2001ma}.  Specifically, the calculations will be dual to \emph{k}-string calculations which we can compare with Hamiltonian and lattice gauge theory methods.  It has already been shown in the literature~\cite{Herzog:2001fq,Herzog:2002ss,Ridgway:2007vh,PandoZayas:2008hw,Doran:2009pp,Stiffler:2009ma} how to do this at the classical level on the string theory side.  The main results of this thesis are the quantum corrections~\cite{PandoZayas:2008hw,Doran:2009pp,Stiffler:2009ma}.


\Chapter{K-STRINGS, A BRIEF OVERVIEW}\label{chap1}
We now give a quick overview of $k$-strings, for a more complete review, see Shifman's work in~\cite{Shifman:2005eb}.  We start with L\"uscher's fundamental string, a simple model of quark-antiquark pairs, which we also refer to as a L\"uscher string.  Next, we describe the $k$-string, an assemblage of multiple L\"uscher strings.  We describe the energy associated with $k$-strings, which consists most notably of the tension term and the L\"uscher term.  Finally, we briefly explain how $k$-string descriptions can be found through gauge/gravity dualities in superstring theory, which is the main topic of this thesis.

\section{L\"uscher's Fundamental String and the \texorpdfstring{$k$}{k}-string}\label{kstrings}
\begin{figure}[htbp]
    \centering   
    \includegraphics[width= 0.6\columnwidth]{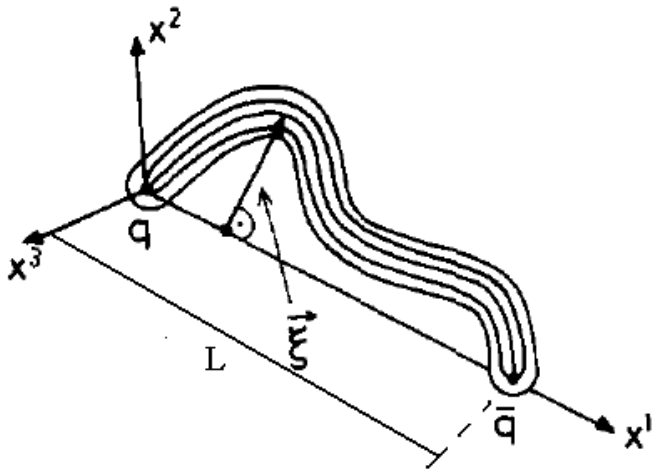}
    \caption{Luscher's fundamental string is a simple model of a quark anti-quark pair separated a large distance $L$, between which a color flux tube develops~\cite{Luscher:1980ac,Luscher:1980fr}.}
    \label{fig:LuschersFundamentalString}
\end{figure}

In~\cite{Luscher:1980ac,Luscher:1980fr}, L\"uscher created a simple model of a quark-antiquark pair tied together by a gauge flux tube as in Fig.~\ref{fig:LuschersFundamentalString}. The energy of this object, known as L\"uscher's fundamental string or simply a L\"uscher string, is given by~\cite{Luscher:1980fr,Luscher:1980ac}
\begin{equation}\label{eq:FundamentalStringTension}
   E_f = T_f L + \frac{\alpha}{L} +\beta + \mathcal{O}(1/L^2)
\end{equation}
\noindent where $\alpha = -\frac{\pi(d-2)}{24}$ with $d$ the dimension of space-time.  We see in Eq.~({\ref{eq:FundamentalStringTension}) that for large $L$, the leading term in the energy is the tension term, $T_f L$, with a Coulombic $\alpha/L$ correction.  The term $\beta$ is constant of $L$.

Now consider many L\"uscher strings parallel to each other, and spaced a distance $d << L$ apart, as depicted in Fig.~\ref{fig:kstrings}.  This configuration is known as a $k$-string.  When one side of a representation such as this has $l$ quarks and $m$ antiquarks, as in Figure~\ref{fig:kstrings}, $k$ is defined as:
\begin{equation}\label{eq:k}
 k = |l - m|
\end{equation}

\begin{figure}
\addtocontents{lof}{\protect\vspace{\li}}
 \centering
 \subfigure[ L\"uscher's fundamental string.]{\label{fig:LuscherCartoonString}\includegraphics[width=0.133\columnwidth]{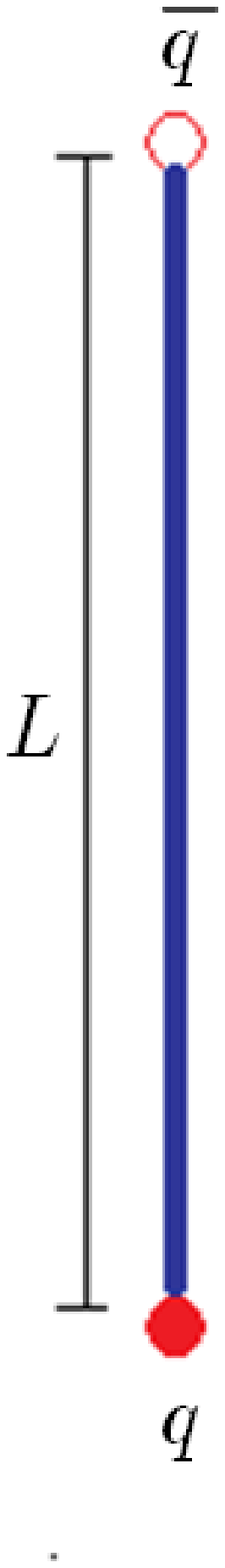}}
\quad 
\subfigure[A $k$-string.]{\label{fig:kstrings}\includegraphics[width=0.467\columnwidth]{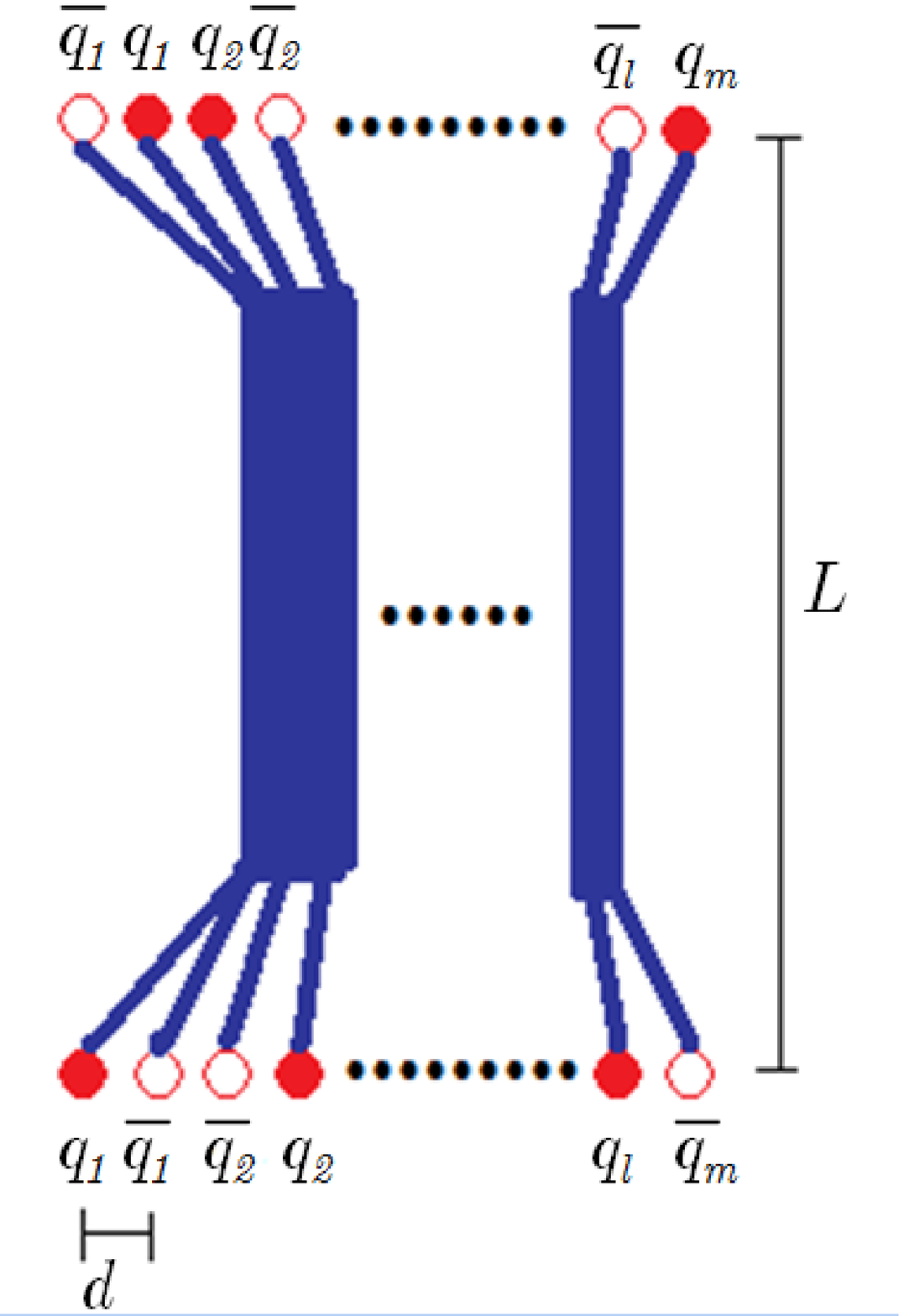}}
\caption{Multiple L\"uscher strings spaced a distance $d<<L$ apart form a $k$-string~\cite{Shifman:2005eb}.}
\end{figure}
The $k$-string energy follows a law similar to L\"uscher's fundamental string
\begin{align}\label{eq:kstringenergy}
    E_k &= T_k L + \frac{\alpha_k}{L}+\beta_k + \mathcal{O}(1/L^2)
\end{align}

\noindent where two commonly found laws for the $k$-string tension are the sine law and the Casimir law:
\begin{align}\label{eq:sinelawandCasimirlaw}
    T_k &\propto \left\{ 
           \begin{array}{ll} 
                  N\sin{\frac{k\pi}{N}} & \mbox{sine law} \\
                  k\frac{N-k}{N}         & \mbox{Casimir law}        
           \end{array}
           \right.
\end{align}   

\noindent The most important feature shared by these two laws is that the $k$-string tension vanishes when $k = 0,N$,  the sine law having an even more powerful feature, that $k$-string tension vanishes when the $N$-ality, defined as $k$ Mod $N$, vanishes.  The $k$-string tension is found to vanish when $k=0,N$ in models of lattice gauge theory~\cite{Teper:1998te,Bringoltz:2006zg,Bringoltz:2008nd}, direct Hamiltonian analysis~\cite{Karabali:1997wk,Karabali:1998yq,Karabali:2000gy,Karabali:2007mr,Karabali:2009rg}, and string theory dual models~\cite{Herzog:2001fq,Herzog:2002ss,Firouzjahi:2006vp,Ridgway:2007vh,PandoZayas:2008hw,Doran:2009pp,Stiffler:2009ma}.

From our simple picture in Fig.~\ref{fig:kstrings}, we would expect that any theory of the strong nuclear force should predict that the k-string tension vanishes when $k=N=3$ or $k=0$.  This is a statement of meson, anti-meson, baryon, and anti-baryon formation; because $d << L$, the two sides of the k-string decouple as quarks near  anti-quarks on the same side will form mesons, and $N=3$ quarks (anti-quarks) near each other on each side will form baryons (anti-baryons).

\section{\texorpdfstring{$k$}{k}-strings from String Theory}
Although $k$-strings can be studied from either Hamiltonian methods studying Yang-Mills theory or lattice gauge theory techniques, there is good reason to study them using string theory.  On a grander scale, any reason to study string theory is that it appears to be a promising theory to one day unify gravity with the other three forces.  On a lighter side, it is merely another theoretical technique to study the same thing ($k$-strings), giving more support for results from the Hamiltonian or lattice gauge theory perspective.  After all, Richard Feynman once said: ``every theoretical physicist who is any good knows six or seven different theoretical representations for exactly the same physics."~\cite{Feynman:1965,ps}

These two reasons, at either end of the emotional spectrum, can be summarized with one phrase: \emph{gauge/gravity dualities}.  Put simply, gauge/gravity dualities are tools from string theory which relate gravitational theories to gauge theories.  Inspired by the relationship predicted by the AdS/CFT correspondence~\cite{Maldacena:1997re,Aharony:1999ti}
\begin{align}\label{eq:Zgaugegravity}
    Z_{CFT} = Z_{string} \sim e^{i W_{gravity}}
\end{align}

\noindent where $W_{gravity}$ is a low energy effective action for a string theory which is known to be dual to 4-$d$ $\mathcal{N}=4$ super Yang-Mills, we will investigate the simple proposed relationship
\begin{align}\label{eq:gaugegravity}
   E_{gauge} \sim E_{gravity}.
\end{align}

\noindent In the main result of this thesis, we will test Eq.~(\ref{eq:gaugegravity}) by calculating the energy of a gravitational theory dual to $k$-strings, and compare it to $k$-string energy calculations from lattice gauge theory and Hamiltonian methods.  In this light, we will specifically find that D-branes embedded in supergravity backgrounds dual to confining gauge theories are dual descriptions of $k$-strings.

As a final note before beginning, a gravitational theory dual to the standard model has yet to be found.   The fact that one possibly exists is the driving force for much research in string theory today.  It is certainly the driving force behind the research presented in this thesis. We will develop gauge/gravity dualities slowly, starting from scratch by first showing how to construct a supersymmetric string theory by generalizing point particle mechanics to string mechanics.


\Chapter{CLASSICAL STRING THEORY} \label{stringtheory}
We will first discuss the classical point particle and then quickly move on to the classical bosonic string.  From there we will introduce the supersymmetric classical string known as the Ramond-Neveu-Schwarz (RNS) superstring.  We see in the analysis of the equations of motion a new feature of the string not present in the point particle: boundary conditions.  These boundary conditions give rise to a new object, known as a D-brane (D for ``Dirichlet", brane for ``membrane"), to which open strings can be attached. An application of Noether's Theorem will lead us to a Mass formula for the classical string.  At the end of the chapter, we introduce Green-Schwarz (GS) superstrings, an alternative but equivalent formulation of the superstring.  We also discuss Poisson brackets as a bridge to the next chapter where we discuss quantum aspects of string theory.

\section{Ramond-Neveu-Schwarz Superstrings}
As a primer to string theory, let us first discuss the point particle.  In $D$ dimensional Minkowski space-time with signature $(-,+,+,\cdots,+)$, the action for a point particle of mass $m$ is:
\begin{align}\label{eq:ppAction}
S_{pp} &= -m \int d\tau \sqrt{|\partial_{\tau} X \cdot \partial_\tau X|},
\end{align}

\noindent where
\begin{align}
  \partial_\tau X\cdot\partial_\tau X &\equiv \eta_{\mu\nu}\frac{\partial X^{\mu}}{\partial \tau}\frac{ \partial X^{\nu}}{\partial \tau},~~~~\mu,\nu = 0 \dots D-1
\end{align}

\noindent The action is proportional to the length of the world line, which is mapped out by $X^\mu(\tau)$ as shown in  Figure~\ref{fig:worldvolumeanatomy-a}.  The action for the point particle is reparameterization invariant; no matter how slowly or quickly we choose $\tau$ to flow along the world line, the physics is the same.

Because a string is one-dimensional, it will map out a two dimensional \emph{world sheet} as it moves through space and time.  As figure~\ref{fig:worldvolumeanatomy-b} illustrates, $X^\mu(\tau, \sigma)$ maps out the world sheet of the string in D dimensional Minkowski space-time.  The string's world sheet is reparameterization invariant with respect to its two parameters $\tau$ and $\sigma$.  In Figure~\ref{fig:worldvolumeanatomy-b} we have parametrized the world sheet such that at every snapshot in time, $\sigma$ flows along the string, and $\tau$ flows perpendicular to the string.   We have also chosen that the string's endpoints always coincide with $\sigma=0,\pi$.  We could have picked a different parametrization, but this choice gives a nice physical interpretation to the parameters, and we shall use it throughout this thesis.  

\begin{figure}
\addtocontents{lof}{\protect\vspace{\li}}
\centering
  \subfigure[The point particle's world line.]{\label{fig:worldvolumeanatomy-a}\includegraphics[width = 0.4\columnwidth]{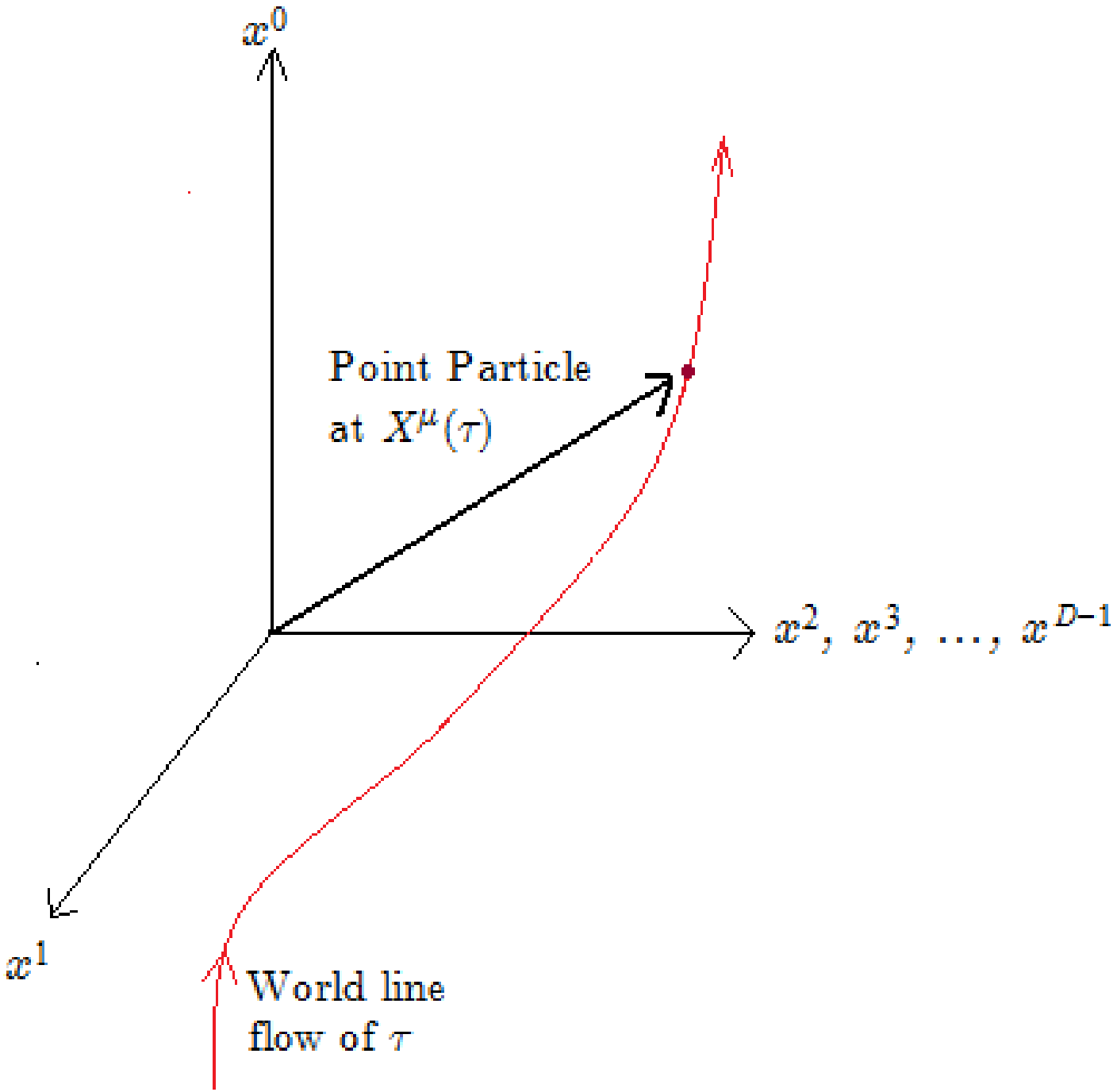}}
  \quad
  \subfigure[The string's world sheet.]{\label{fig:worldvolumeanatomy-b}\includegraphics[width = 0.4\columnwidth]{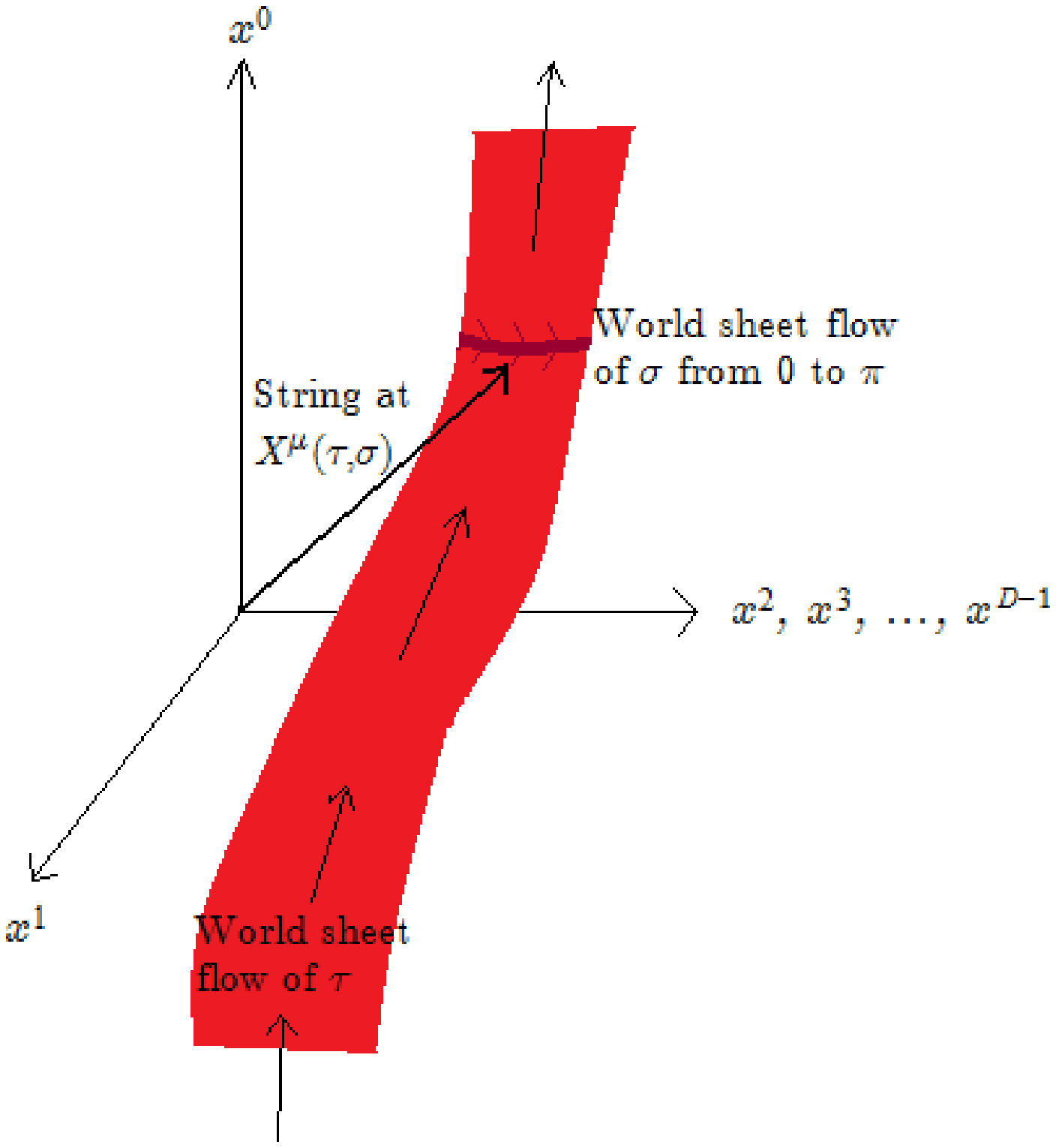}}
\caption{The point particle's world line is reparameterization invariant with respect to its one parameter $\tau$, and the string's world sheet is reparameterization invariant with respect to its two parameters $\tau$, and $\sigma$.  Here and throughout, we choose $\sigma$ to run along the string, and $\tau$ to run perpendicular to it for all times.}
\label{fig:worldvolumeanatomy}
\end{figure} 

A natural generalization from the point particle action is the Nambu-Goto (NG) string action~\cite{bbs1,Szabo:2002ca}:
\begin{align}\label{eq:NG0}
    S_{NG} &= -T_0 \int d^2\zeta \sqrt{| \det(\partial_a X \cdot\partial_b X)|} 
\end{align}

\noindent which is proportional to the \emph{area} of the world sheet.   Here the parameters are labeled $\zeta^0 = \tau$, $\zeta^1 = \sigma$, and $\partial_a \equiv \frac{\partial}{\partial \zeta^a}$.  The constant $T_0$ has units of tension, and it is a natural generalization from the constant mass, $m$, which appeared in the point particle action.  
Since the $X^\mu$ are commuting variables, they describe bosons; more specifically, they are world sheet scalars.  To have fermions in our theory, as any theory which accurately depicts our world must have, we must augment the NG string action with a fermionic piece which includes anticommuting variables.  Before we do this, however, it is useful to cast the NG string action into a different form.  The form we desire is the Polyakov string action~\cite{bbs1,Szabo:2002ca,Polchinski:1998v1,GreenSchwarzWitten:1987v1}:
\begin{align}\label{eq:Polyakov}
S_P = -\frac{T_0}{2} \int d^2\zeta \sqrt{h}h^{ab} \partial_{a}X \cdot \partial_{b}X
\end{align}

\noindent where 
\begin{align}
   h &\equiv |\det(h_{ab})|.
\end{align}

\noindent Here, $h_{ab}$ is an auxiliary metric, used to connect the Polyakov string action to the NG string action.  Varying the Polyakov string action with respect to $h_{ab}$ gives its equations of motion~\cite{bbs1,Szabo:2002ca,GreenSchwarzWitten:1987v1}:

\begin{align}\label{eq:EQMh}
    \partial_{a}X \cdot \partial_{b}X = \frac{1}{2} h_{ab} h^{cd}\partial_{c}X \cdot \partial_{d}X
\end{align}

\noindent Using this equation to eliminate $h_{ab}$ from the Polyakov string action, yields the NG string action, Eq.~\ref{eq:NG0}.

Now that we have shown the equivalence of the Polyakov and NG actions, we concentrate on simplifying the Polyakov string action.  Reparameterization invariance of $h_{ab}$ means that we are free to specify any two of its components.  Making the following choice
\begin{equation}\label{eq:reparameterization}
   h_{01} = h_{10} = 0,~h_{11} = -h_{00} = e^{\phi}
\end{equation}

\noindent allows us to write $h_{ab}$ as:
\begin{align}\label{eq:auxilliary}
h_{ab} &= e^{\phi} \eta_{ab} \nonumber\\
       &= e^{\phi} \left( \begin{array}{c c} 
                          -1 & 0 \\
                          0 & 1 
                          \end{array} \right)
\end{align}

\noindent Plugging this into Eq.~\ref{eq:Polyakov} leaves us with a simplified version of the Polyakov string action~\cite{bbs1,Szabo:2002ca}:
\begin{align}\label{eq:Polyakovsimple}
S_{P} &= -\frac{T_0}{2} \int d^2\zeta \sqrt{|\det(e^{\phi}\eta_{ab})|}e^{-\phi}\eta^{cd}\partial_{c}X \cdot \partial_{d}X \nonumber\\
      &= -\frac{T_0}{2} \int d^2\zeta \sqrt{|\det(\eta_{ab})|}\eta^{cd} \partial_{c}X \cdot \partial_{d}X \nonumber\\
      &= -\frac{T_0}{2} \int d^2\zeta \eta^{ab} \partial_{a} X \cdot \partial_{b} X
\end{align}

As previously stated, we must supersymmetrize the Polyakov action so that our theory incorporates fermions.   We include, along with the commuting bosonic fields $X^\mu$, anticommuting fermionic fields $\psi^\mu$, whose two components are Grassmann variables:
\begin{align}\label{eq:psi}
&\psi^\mu = \left( \begin{array}{l}    
                     \psi^\mu_{-} \\
                     \psi^\mu_{+}
                     \end{array}
                     \right) \\
&\left\{ \psi^\mu_{\pm}, \psi^\nu_{\pm} \right\} = 0,~~\mu,\nu = 0 \dots d
\end{align}

\noindent This method of supersymmetry is called the \emph{Ramond-Neveu-Schwarz} (RNS) \emph{formalism}.  The RNS supersymmetric action is formed by merely tacking on a Dirac-type action to the Polyakov string action:
\begin{align}\label{eq:RNSaction}
S_{RNS} &= -\frac{T_0}{2} \int d^2\zeta (\partial_{a} X \cdot \partial^{a} X + i~\overline{\psi} \cdot \rho^{a} \partial_{a} \psi) 
\end{align}

\noindent  where $\rho^{a}$ are two by two matrices satisfying a Clifford algebra,~\cite{bbs1,Szabo:2002ca}:
\begin{align}\label{eq:Clifford}
\left\{ \rho^{a}, \rho^{b} \right\} = 2\eta^{ab}
\end{align}

\noindent and $\overline{\psi}^\mu$ is defined as:
\begin{align}\label{eq:psibar}
\overline{\psi}^\mu &\equiv (\psi^{\mu*}_{+},-\psi^{\mu*}_{-} ) 
            = (\psi^\mu_{+},-\psi^\mu_{-} )
\end{align}

\noindent The last equality in Eq.~(\ref{eq:psibar}) is because the $\psi^\mu$ are all two component Majorana spinors.

One final note on the RNS action, it is invariant under the infinitesimal supersymmetric transformation:
\begin{equation}\label{eq:supertrans}
\delta X^\mu = \overline{\varepsilon} \psi^\mu,~\delta \psi^\mu = \rho^{\mu}\partial_{\mu} X^\mu \varepsilon
\end{equation}

\noindent where $\varepsilon$ is a two component infinitesimal Grassmann variable~\cite{bbs1,Szabo:2002ca}.

\subsection{Superstring Equations of Motion}
The RNS action is easiest to work with if we pick a representation for the two by two Dirac matrices $\rho^{a}$~\cite{bbs1}:
\begin{align}\label{eq:Diracrep}
\rho^0 = \left( \begin{array}{c c}
                0 & -1 \\
                1 & 0
                \end{array} \right)~~~~
\rho^1 = \left( \begin{array}{c c}
                 0 & 1 \\
                 1 & 0 
                 \end{array} \right)
\end{align}

\noindent With this choice, it is easy to show that the RNS action becomes:
\begin{align}\label{eq:RNSaction2}
S_{RNS} &= -\frac{T_0}{2} \int d^2\zeta (\partial_{a} X \cdot \partial^{a} X - 2i\psi_{-} \cdot \partial_{+} \psi_{-} - 2i\psi_{+} \cdot \partial_{-} \psi_{+}) \nonumber\\
\partial_{\pm} &\equiv \frac{1}{2} (\partial_\tau \pm \partial_\sigma)
\end{align}

This action is that of type I and II superstrings~\cite{Polchinski:1998v2}.  For completeness, we write the heterotic superstring action{\linespread{1.0}\footnote{ This is known as the fermionic construction of the heterotic string.  There is another, equivalent construction known as the bosonic construction of the heterotic string~\cite{bbs1}}}:
\begin{align}
   S_{het} &= -\frac{T_0}{2} \int d^2\zeta (\partial_a X \cdot \partial^a X - 2 i \psi_- \cdot \partial_+ \psi_- - 2 i \sum_{A=1}^{32} \psi_+^A \partial_- \psi_+^A)
\end{align}

\noindent where the main difference from the RNS action is that there are now 32 left moving fermions, $\psi_+^A$, instead of 10.  Furthermore, the world sheet supersymmetry of the heterotic string is of the right moving modes only~\cite{bbs1}:
\begin{align}
   \delta X^\mu &= i \varepsilon \psi^\mu_-,~~~\delta\psi^{\mu}_- = -2 \varepsilon \partial_- X^\mu
\end{align}

We will concentrate mostly on the type II superstring, so we will concern ourselves now with the RNS action, whose variation leads to the field equations for $X^\mu$ and $\psi^\mu$,
\begin{align}\label{eq:BosonicEQM}
\partial_{a} \partial^{a} X^\mu &= 0 \\
\label{eq:FermionicMinusEQM}
\partial_+ \psi^\mu_- &= 0 \\
\label{eq:FermionicPlusEQM}
\partial_- \psi^\mu_+ &= 0
\end{align}

\noindent and the boundary conditions~\cite{bbs1,Szabo:2002ca}:
\begin{align}\label{eq:BosonicBC}
\delta X \cdot \partial_\sigma X |_{\sigma = 0}^{\pi} &= 0 \\
\label{eq:FermionicBC}
 \psi_+ \cdot \delta \psi_+ - \psi_- \cdot \delta \psi_-|_{\sigma = 0}^{\pi} &= 0
\end{align}

\noindent We now discuss satisfying these boundary conditions separably for open and closed strings.

\subsubsection{Open Strings}
\noindent We can satisfy the open bosonic boundary conditions in two ways:
\begin{align}\label{eq:DirichletBC}
\delta X^\mu|_{\sigma=0,\pi} &= 0 ~~~~\mbox{Dirichlet}  \\
\label{eq:NeumannBC}
\partial_{\sigma} X^\mu|_{\sigma=0,\pi} &= 0 ~~~~\mbox{Neumann}
\end{align}

\noindent To illustrate clearly the implications of these boundary conditions, consider Fig.~\ref{fig:D1brane}, which shows a specific choice of a string's boundary conditions in $D = 2+1$ dimensions.  Here the string's end points are drawn as rings; a dramatization which illustrates that they are free to slide along the $x$-direction, but are fixed in the $y$-direction at $y=a,b$.  This string's bosonic field $X^1 = x$ satisfies Neumann boundary conditions, $\partial_\sigma x|_{\sigma=0,\pi} = 0$, and its bosonic field $X^2 = y$ satisfies Dirichlet boundary conditions, $\delta y|_{\sigma = 0,\pi} = 0$. 

\begin{figure}[htbp]
\addtocontents{lof}{\protect\vspace{\li}}
\centering
\includegraphics[width = 0.8\columnwidth]{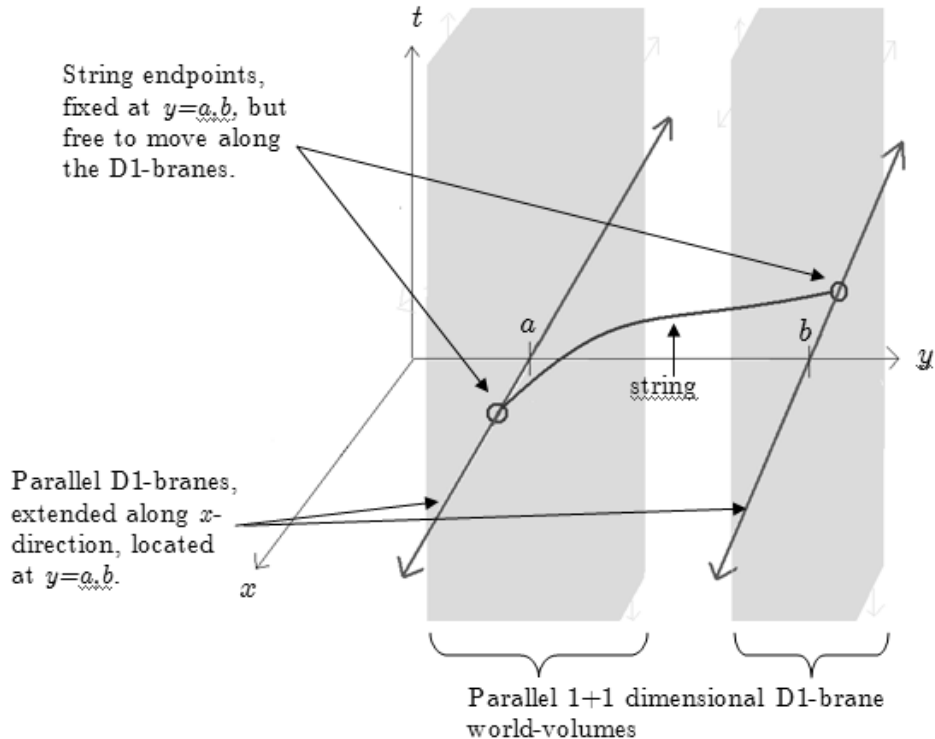}
\caption{String with end-points lying on parallel D1-branes.  The D1-branes map out 1+1 dimensional world volume planes.  The strings world sheet is not shown to avoid clutter.}
\label{fig:D1brane}
\end{figure}

In Fig.~\ref{fig:D1brane}, the two objects that the string's endpoints slide along are called D-branes (D for Dirichlet).  The most general definition of a D-brane is an object that extends along all the coordinates of any number of spatial dimensions.  If a D-brane spans all of $p$ spatial dimensions, then it is called a \emph{Dp-brane}.  Both D-branes in Fig.~\ref{fig:D1brane} are D1-branes; they both span the entire $x$-direction.  

Drawing an analogy from the world line of the point particle, a D$p$-brane augmented with the time coordinate makes up the \emph{world volume} of a D$p$-brane.  For a D$1$-brane, the world volume will be a $1+1$ dimensional plane.  The world volume for each D$1$-brane in Fig.~\ref{fig:D1brane} is an $x-t$ plane; one located at $y=a$ and the other at $y=b$.

We see in Fig.~\ref{fig:D1brane} that there are two different types of spatial coordinates that describe D-branes: normal and tangential.  Normal coordinates specify the D-branes location, and satisfy Dirichlet boundary conditions for the string, where as tangential coordinates make up the contents of the D-branes world volume, and satisfy Neumann boundary conditions for the string.  The time coordinate $t$, is \emph{always} a tangential coordinate, and so in $D$ dimensional space-time, a D$p$-brane will have $p+1$ tangential coordinates, and $D-(p+1)$ normal coordinates.  For example, D$1$-branes in $D=3$ dimensional space-time, as in Fig.~\ref{fig:D1brane}, each have $1+1 = 2$ tangential coordinates, $x$ and $t$, and $3-(1+1) = 1$ normal coordinate, $y$.

For an open bosonic string in $D$ space-time dimensions attached between two parallel D$p$ branes, one located at $x^i = d_1^i$, the other at $x^i = d_2^i$, the boundary conditions for the string would be
\begin{align}
X^i|_{\sigma = 0} = d_1^i,~~~~X^i|_{\sigma=\pi} = d_2^i,~~~&\mbox{normal coordinates},\\
\partial_\sigma X^j|_{\sigma = 0,\pi} = 0,~~~&\mbox{tangential coordinates}.
\end{align}

\noindent with solutions to the bosonic string equations of motion, Eq.~(\ref{eq:BosonicEQM}):
\begin{align}
   X^i = d_1^i + (d_2^i - d_1^i) \frac{\sigma}{\pi} + l_s\sum_{n \ne 0} \frac{1}{n}\alpha_n^i \sin n\sigma~e^{-i n\tau}~~~&i= p+1,\cdots,D-1\\
   X^j = x^j_0 + l_s^2 p_0^j \tau + i l_s \sum_{n \ne 0} \frac{1}{n}\alpha_n^i \cos n\sigma~e^{-i n\tau},~~~&j = 0,\cdots,p.
\end{align}

\noindent where $l_s = (\pi T_0)^{-1/2} = \sqrt{ 2 \alpha'}$. The string's bosonic center of mass momentum, $p_0^\mu$, is related to the zero modes, $\alpha_0^{\mu}$,
\begin{align}
    p_0^{\mu} &\equiv \int_0^{\pi} d\sigma P^{\mu}(0,\sigma) \equiv \frac{1}{\sqrt{2 \alpha'}} \alpha_0^{\mu},
\end{align}

\noindent with the string's bosonic momentum defined as (see section~\ref{Noether})
\begin{align}\label{eq:bp}
   P^{\mu}(\tau,\sigma) &\equiv  T_0 \dot{X}^{\mu}(\tau,\sigma),~~~\dot{X}^{\mu} \equiv \partial_\tau X^{\mu}.
\end{align}

\subsubsection{Closed Strings}
We will now discuss closed string solutions.  Naively, one would expect the solutions to the string equations of motion for closed strings to be periodic.  We see the boundary conditions for closed string bosonic fields, Eq.~(\ref{eq:BosonicBC}), are automatically satisfied if periodicity is demanded:
\begin{equation}\label{eq:Bosonicperiodicity}
X^\mu(\tau,\sigma) = X^\mu(\tau,\sigma + \pi)
\end{equation}

With this periodicity condition, the general solution to Eq.~(\ref{eq:BosonicEQM}) for closed string bosonic fields is a linear combination of independent right moving and left moving oscillators $\alpha_m^\mu$ and $\tilde{\alpha}_m^\mu$, respectively~\cite{bbs1,Szabo:2002ca}:
\begin{equation}\label{eq:BosonicSoln}
X^\mu(\tau,\sigma) = x_0^\mu + l_s^2 p_0^\mu \tau + \frac{i}{2}l_s \sum_{n \ne 0} \frac{1}{n}[\alpha^\mu_n e^{-2in(\tau-\sigma)} + \tilde{\alpha}^\mu_n e^{-2in(\tau+\sigma)}]
\end{equation}

\noindent now with the string's bosonic center of mass momentum $p_0^{\mu}$, shared equally between the right and left moving zero modes, $\alpha_0^\mu$ and $\tilde{\alpha}_0^\mu$:
\begin{align}\label{eq:p0}
    p_0^{\mu}&\equiv \int_0^{\pi} d\sigma P^{\mu}(0,\sigma) \equiv \frac{2}{\alpha'}\alpha_0^{\mu} \equiv \frac{2}{\alpha'}\tilde{\alpha}_0^\mu.
\end{align}

The situation is slightly more complicated for the fermionic fields.  The boundary conditions for closed string fermionic fields, Eq.~(\ref{eq:FermionicBC}), are automatically satisfied if either periodic (Ramond) or anti-periodic (Neveu-Schwarz) boundary conditions are used~\cite{bbs1,Szabo:2002ca}:
\begin{align}\label{eq:RamondBC}
\psi_{\pm}^\mu(\tau,\sigma) &= \psi_{\pm}^\mu(\tau,\sigma+\pi)~~~~\mbox{Ramond}\\
\label{eq:NeveuSchwarzBC}
\psi_{\pm}^\mu(\tau,\sigma) &= -\psi_{\pm}^\mu(\tau,\sigma+\pi)~~~~\mbox{Neveu-Schwarz}
\end{align}

\noindent Solutions to Eqs.~(\ref{eq:FermionicMinusEQM}) and~(\ref{eq:FermionicPlusEQM}) are left and right moving solutions, respectively, and can have either Ramond (R) or Neveu-Schwarz (NS) boundary conditions:
\begin{align}\label{eq:PsimR}
&\psi_-^\mu(\tau,\sigma) = \sum_{m \in \mathbb{Z}} d_m^\mu e^{-2im(\tau-\sigma)},~~~
&\psi_+^\mu(\tau,\sigma) = \sum_{m \in \mathbb{Z}} \tilde{d}_m^\mu e^{-2im(\tau+\sigma)}~~~~~~
&\mbox{R}\\
\label{eq:PsimNS}
&\psi_-^\mu(\tau,\sigma) = \sum_{r \in \mathbb{Z} + 1/2} b_r^\mu e^{-2ir(\tau-\sigma)},~~~
&\psi_+^\mu(\tau,\sigma) = \sum_{r \in \mathbb{Z} + 1/2} \tilde{b}_r^\mu e^{-2ir(\tau+\sigma)}~~~
&\mbox{NS}
\end{align}



\subsection{Noether's Theorem and Conserved Quantities of the RNS action}\label{Noether}
Noether's theorem says for every local symmetry there is a conserved quantity.  Two conserved quantities of the RNS action are the stress tensor, $T_{ab}$, and the super current, $J_A^a$.  The stress tensor is derived from local parameterization invariance of the action
\begin{align}
   \delta S_{RNS} &= 0 = \int d^2\zeta \epsilon^a \partial^b T_{ab},~~~\mbox{for}~~~ \zeta \to \zeta^a + \epsilon^a(\zeta)
\end{align}

\noindent and the supercurrent is derived from local supersymmetric invariance of the action
\begin{align}
   \delta S_{RNS} &= 0 = \int d^2 \zeta \overline{\varepsilon}\partial_a J^a
\end{align}

\noindent where $\varepsilon = \varepsilon(\zeta)$~\cite{bbs1,Szabo:2002ca}.

The solutions of these conserved currents are~\cite{bbs1,Szabo:2002ca}
\begin{align}
   T_{ab} &= \partial_a X \cdot \partial_bX + \frac{1}{4}\overline{\psi} \cdot \rho_{(a} \partial_{b)} - \frac{1}{2}\eta_{ab}\left( \partial_c X \cdot \partial^c X + \frac{1}{2}\overline{\psi}\cdot \rho^c \partial_c \psi \right) \\
   J^a_A &= -\frac{1}{2}(\rho^b\rho^a\psi)_A \cdot \partial_b X,~~~A =-,+.
\end{align}

\noindent Using Eqs.~(\ref{eq:BosonicEQM}), (\ref{eq:FermionicMinusEQM}), and (\ref{eq:FermionicPlusEQM}) for the closed string, and light cone coordinates
\begin{align}
   \zeta^+ &= \tau + \sigma,~~~\zeta^- = \tau - \sigma,
\end{align}

\noindent the non-vanishing components of these currents can be written
\begin{align}\label{eq:Tlightcone}
   T_{--}(\zeta^-) &= \sum_{m=-\infty}^{\infty} L_m e^{-2 i m \zeta^-},~~~T_{++}(\zeta^+) =~\sum_{m=-\infty}^{\infty} \tilde{L}_m e^{-2 i m \zeta^+} \\
   J_-^+(\zeta^-) &= \sum_r G_r e^{-2 i r \zeta^-},~~~~~~~J_+^-(\zeta^+) = \sum_r \tilde{G}_r e^{-2 i r \zeta^+}
\end{align}

\noindent where
\begin{align}
   L_n &= \frac{1}{2}\sum_{m=-\infty}^\infty \alpha_{n-m}\cdot\alpha_m + \frac{1}{4}\sum_r(2r - n)c_{n-r}\cdot c_r, \\
   \tilde{L}_n &= \frac{1}{2}\sum_{m=-\infty}^\infty \tilde{\alpha}_{n-m}\cdot\tilde{\alpha}_m + \frac{1}{4}\sum_r(2r - n)c_{n-r}\cdot c_r \\
   G_r &= \sum_{m=-\infty}^{\infty} \alpha_m \cdot c_{r-m},~~~\tilde{G}_r = \sum_{m=-\infty}^{\infty} \tilde{\alpha}_m \cdot \tilde{c}_{r-m}
\end{align}

\noindent and $c_{r}$ are either the fermionic oscillators $b_r$ or $d_r$ for NS or R boundary conditions, respectively~\cite{Szabo:2002ca}. 

Notice the bosonic part of the stress tensor vanishes by the equations of motion for $h_{ab}$, Eq.~(\ref{eq:EQMh}).  Similarly, it can be shown that the remaining components of the full supersymmetric stress tensor and the supercurrent vanish:~\cite{Szabo:2002ca,bbs1}
\begin{align}
   T_{++} &= T_{--} = J_-^+ = J_+^- = 0
\end{align}

\noindent From equation~(\ref{eq:Tlightcone}), we see this means that $G_r = 0$ and $L_m = 0$.  Calculating
\begin{align}\label{eq:L0classicalconstraint}
0 &= L_0 + \tilde{L}_0 \nonumber\\     
  &= \alpha_0^2 + \left(\frac{1}{2}\sum_{n\ne 0}\alpha_{-n}\cdot\alpha_n  + \frac{1}{2}\sum_{r}r c_{-r}\cdot c_{r}\right) + \left(\frac{1}{2}\sum_{n \ne 0 }\tilde{\alpha}_{-n}\cdot\tilde{\alpha}_n + \frac{1}{2}\sum_{r}\tilde{c}_{-r}\cdot \tilde{c}_{r}\right) \nonumber\\
  &= \alpha_0^2 + N + \tilde{N}
\end{align}

\noindent and using $2\alpha_0^2 = \alpha' p_0^2$, Eq.~(\ref{eq:p0}), we solve for the closed superstring mass
\begin{align}\label{eq:classicalmass}
   \alpha' M^2 &= -\alpha' p_0^2 = - 2 \alpha_0^2 = 2 (N + \tilde{N})
\end{align}

\noindent where
\begin{align}\label{eq:NumberOperators}
    N &=  \sum_{n = 1}^\infty \alpha_{-n}\cdot\alpha_n + \sum_{r>0}r c_{-r}\cdot c_r, \nonumber\\
    \tilde{N} &= \sum_{n = 1}^\infty \tilde{\alpha}_{-n}\cdot\tilde{\alpha}_n + \sum_{r>0}r \tilde{c}_{-r}\cdot \tilde{c}_r.
\end{align}

\subsection{Poisson Brackets}\label{PoissonBrackets}
As we will soon quantize the superstring, it is now prudent to discuss Poisson brackets for the bosonic, classical theory, described by the Polyakov action, Eq. (\ref{eq:Polyakovsimple}).  The momentum conjugate to $X^{\mu}$ is
\begin{align}
   P^{\mu}(\sigma,\tau) &= \frac{\delta S_P}{\delta \dot{X}_{\mu}(\sigma,\tau)} = T_0 \dot{X}^{\mu}(\sigma,\tau)
\end{align}

\noindent where 
\begin{align}
   \frac{\delta X^{\mu}(\tau,\sigma)}{\delta X^{\nu}(\tau',\sigma')} &\equiv \delta^{\mu}_{~\nu}\delta(\tau - \tau')\delta(\sigma-\sigma').
\end{align}

Defining the equal $\tau$ Poisson brackets as
\begin{align}
   [A^{\mu}(\sigma,\tau),B^{\nu}(\sigma',\tau)]_{P.B.} &\equiv \int d^2\tilde{\sigma} \left(\frac{\partial A^{\mu}(\sigma,\tau)}{\partial P^{\alpha}(\tilde{\sigma},\tilde{\tau})}\frac{\partial{B^{\nu}(\sigma',\tau)}}{\partial X_{\alpha}(\tilde{\sigma},\tilde{\tau})} + \right. \nonumber\\
   &\left.~~~~~~~~~~~~~~~~~ -\frac{\partial A^{\mu}(\sigma,\tau)}{\partial X_{\alpha}(\tilde{\sigma},\tilde{\tau})}\frac{\partial{B^{\nu}(\sigma',\tau)}}{\partial P^{\alpha}(\tilde{\sigma},\tilde{\tau})}\right),
\end{align}

\noindent a straightforward calculation shows
\begin{align}
   [ P^{\mu}(\sigma,\tau),P^{\nu}(\sigma',\tau)]_{P.B.} &= [X^{\mu}(\sigma,\tau),X^{\nu}(\sigma',\tau)]_{P.B.} = 0, \\
[ P^{\mu}(\sigma,\tau),X^{\nu}(\sigma')]_{P.B.}&=\eta^{\mu\nu}\delta(\sigma - \sigma').
\end{align}

Inserting the bosonic solutions into these equations gives the Poisson brackets for the bosonic modes:
\begin{align}
   [\alpha_{m}^{\mu},\alpha_n^{\nu}]_{P.B.} &= [\tilde{\alpha}_m^\mu,\tilde{\alpha}_n^{\nu}]_{P.B.} = i m \eta^{\nu\nu}\delta_{m+n,0} \\
   [\alpha_m^{\mu},\tilde{\alpha}_n^{\nu}]_{P.B.} &= 0
\end{align}

\noindent where in the open string case, there are only one set of modes. In deriving these relationships, it is useful to first derive the Poisson Brackets for the center of mass variables
\begin{align}
  x_0^{\mu} &=  \frac{1}{\pi} \int_0^{\pi} d\sigma X^{\mu}(0,\sigma),~~~p_0^{\mu} =  \int_0^{\pi} d\sigma P^{\mu}(0,\sigma) \\
  &[p_0^{\mu},x_0^{\mu}]_{P.B.} = \eta^{\mu\nu}
\end{align}

\noindent where $2 \alpha_0^\mu = 2 \tilde{\alpha}_0^\mu = l_s p_0^\mu$ for the closed string and $\alpha_0^\mu = l_s p_0^\mu$ for the open string.  

\section{Green Schwarz Superstrings}
In the previous section, we showed the supersymmetric version of the Polyakov string: the RNS superstring.  The supersymmetric version of the Nambu-Goto string is the Green-Schwarz (GS) superstring, with supersymmetric action:
\begin{align}
    S_{GS} &= -\frac{1}{\pi}\int d^2\sigma \sqrt{-\det M_{ab}} +\int \Omega_2, \nonumber\\ 
    M_{ab} &= (\partial_a X^{\mu} - \bar{\Theta}^A\Gamma^{\mu}\partial_a\Theta^A)(\partial_b X_{\mu} - \bar{\Theta}^A\Gamma_{\mu}\partial_b\Theta^A), \nonumber\\
   \Omega_2 &= c(\bar{\Theta^1} \Gamma_{\mu} d\Theta^1 - \bar{\Theta^2}\Gamma_{\mu}d\Theta^2)dX^{\mu} - c \bar{\Theta^1}\Gamma_{\mu} d\Theta^1 \bar{\Theta^2}\Gamma^{\mu}d\Theta^2,
\end{align} 

\noindent where c is a constant, and $A = 1,2$.  The action is supersymmetric with respect to the transformations
\begin{align}
   \delta \Theta^{A} = \varepsilon^A,~~~\delta X^{\mu} = \overline{\varepsilon}^A\Gamma^\mu \Theta^A.
\end{align}

This action has $\mathcal{N} = 2$ supersymmetries for closed strings, as in types IIA, and IIB.  For open strings, $\varepsilon^1 = \varepsilon^2$, and so type I superstring theory has $\mathcal{N}=1$ supersymmetries as it contains open strings~\cite{bbs1}.


\Chapter{QUANTUM STRING THEORY}\label{chap3}
To move to the quantum theory of strings, we will start by finding oscillator commutation relations from the quantum limit of the Poisson brackets.  Taking these oscillators to be creation and annihilation operators acting on a Fock vacuum, the string mass will now take the form of an operator.  As the operators no longer all commute with each other, normal ordering of the mass operator now slightly modifies it from the classical theory, with a constant proportional to $D-10$.  Acting with this mass operator on the ground states of the string, we will investigate the field theory for type II superstrings, which is an infinite tower of massive fields.  We find the type II ground states to have zero mass for $D=10$, the critical dimension of superstring theory. 

The ground states of the type II theory are shown to be the field theory of a supergravity, evidence that the low energy effective action for superstrings is precisely supergravity for type II superstrings and supergravity coupled to Yang-Mills theory for type I and Heterotic string theory.  As the Ramond-Ramond charges of type II superstring theory are carried by D-branes, we find that augmenting the type II theory with D-branes comes naturally, and that they can contain dynamics.  Furthermore, D-brane interactions with open strings show that they carry $U(1)$ gauge fields on their world volume, which can be extended to $U(N)$ gauge fields in the case of $N$ coincident D-branes (parallel, identical D-branes, stacked infinitesimally close together).  At this point we will have all the ingredients for the AdS/CFT correspondence: \emph{$\mathcal{N} = 4~U(N)$ super Yang-Mills theory in $3+1$ dimensions is dual to type IIB superstring theory on $AdS_5 \times S^5$}


\section{The Quantized Type II Superstring}
Here we will quantize the closed superstring, leading us to the type II superstring theory.
In moving from a classical theory to a quantum theory, we take the Poisson brackets from section~\ref{PoissonBrackets}, and let
\begin{align}
   [~~~,~~~]_{P.B.} \to i [~~~,~~~]
\end{align}

\noindent where the right hand side is the quantum theory. The commutator is of the Fourier coefficients $\alpha$ and $\tilde{\alpha}$ in Eq.~(\ref{eq:BosonicSoln}), which are now interpreted as creation and annihilation operators acting on a Fock space~\cite{bbs1}.  For the fermionic coefficients, we propose the equal $\tau$ anticommutator relation for the solutions to Eqs.~(\ref{eq:PsimR})-(\ref{eq:PsimNS})
\begin{align}
   \{ \psi_{A}^{\mu}(\tau,\sigma),\psi_B^\mu(\tau,\sigma') \} &= \pi \eta^{\mu\nu}\delta_{AB}\delta(\sigma - \sigma'),~~~A,B = +,-
\end{align}
and solve for the anticommuting relations of the modes:~\cite{bbs1,Szabo:2002ca}:
\begin{align}\label{eq:Bosoniccommutators}
[ \alpha_m^\mu, \alpha_n^\nu ] &= [ \tilde{\alpha}_m^\mu, \tilde{\alpha}_n^\nu ] = m\delta_{m+n,0} \eta^{\mu\nu} \\
\label{eq:NeveuSchwarzcommutators}
\{ b_r^\mu, b_s^\nu \} &= \{ \tilde{b}_r^{\mu}, \tilde{b}_s^{\nu} \} = \eta^{\mu\nu} \delta_{r+s,0} \\
\label{eq:Ramondcommutators}
\{ d_m^\mu, d_n^\nu \} &= \{ \tilde{d}_m^\mu, \tilde{d}_n^\nu \} = \eta^{\mu\nu} \delta_{m+n,0}. 
\end{align}  

We now interpret these as oscillators acting on Fock states.  We define the ground state for a string of center of mass momentum $p$ as $|p,0\rangle_R$ for Ramond boundary conditions and $|p,0\rangle_{NS}$ for Neveu-Schwarz boundary conditions.  The positively moded oscillators annihilate their ground state
\begin{align}
    \alpha_m^\mu |p, 0 \rangle_R &= d_m^\mu |p,0 \rangle_R = 0,~~~ m>0 \\
    b_r^\mu |p, 0 \rangle_{NS} &= 0,~~~\alpha_m^\mu | p,0\rangle_{NS} = 0,~~~r,m>0     
\end{align} 

\noindent and the negatively moded oscillators build mass states out of the ground state:
\begin{align}
    \alpha_m^\mu |p, 0 \rangle_R &= d_m^\mu |p, 0 \rangle_R =  |p,|m|\rangle_R^\mu,~~~m<0 \\
    \alpha_m^\mu |p,0 \rangle_{NS} &= |p,|m|\rangle_{NS}^\mu,~~~b_r^\mu |p, 0 \rangle_{NS} = |p,|r|\rangle_{NS}^\mu,~~~m,r<0
\end{align}

\noindent where these are states of mass $|m|$ or $|r|$ units above the ground state.  The same relations hold for the left moving, tilded oscillators.

This leads us to the quantum mass operator.  Borrowing from the classical theory the constraint $L_0 + \tilde{L}_0 = 0$, we postulate that this manifests itself quantum mechanically as the operator condition
\begin{align}
   (L_0 + \tilde{L}_0)| phys \rangle = 0
\end{align}

\noindent where $| phys \rangle$ is a physical state that is a tensor product of a left moving state, $|p,|\tilde{m}|\rangle_R^\mu$ or $|p,|\tilde{r}|\rangle_{NS}^\mu$, and a right moving state, $|p,|m|\rangle_R^\mu$ or $|p,|r|\rangle_{NS}^\mu$, of which there are four possible sectors~\cite{Szabo:2002ca,bbs1}:
\begin{align}\label{eq:physicalstates}
   |phys\rangle &= \left\{ 
      \begin{array}{ll}
        |p,|\tilde{m}|\rangle_R^\mu    \otimes |p,|n|\rangle_R^\nu    & \mbox{R-R sector} \\
        |p,|\tilde{r}|\rangle_{NS}^\mu \otimes |p,|s|\rangle_{NS}^\nu & \mbox{NS-NS sector} \\
        |p,|\tilde{r}|\rangle_{NS}^\mu \otimes |p,|m|\rangle_R^\nu    & \mbox{NS-R sector} \\
        |p,|\tilde{m}|\rangle_R^\mu \otimes |p,|r|\rangle_{NS}^\nu    & \mbox{R-NS sector}
      \end{array}
      \right.
\end{align}

\noindent where the left moving operators act on the left states and the right moving operators act on the right states.

Demanding that the stress energy tensor still vanishes quantum mechanically, we follow Eq.~(\ref{eq:L0classicalconstraint}) and calculate
\begin{align}\label{eq:L0quantumconstraint}
0 &= (L_0 + \tilde{L}_0)|phys\rangle \nonumber\\     
  &= \left(\alpha_0^2 + \frac{1}{2}\sum_{n\ne 0}\alpha_{-n}\cdot\alpha_n  + \frac{1}{2}\sum_{r}r c_{-r}\cdot c_{r} + \frac{1}{2}\sum_{n \ne 0 }\tilde{\alpha}_{-n}\cdot\tilde{\alpha}_n + \frac{1}{2}\sum_{r}\tilde{c}_{-r}\cdot \tilde{c}_{r}\right)| phys \rangle \nonumber\\
  &= (\alpha_0^2 + N + \tilde{N} + a_b + \tilde{a}_b + a_f + \tilde{a}_f) |phys\rangle
\end{align}

\noindent  and solving for the mass operator, we find:
\begin{align}
   \alpha' M^2 &= -\alpha' p_0^2 = -2 \alpha_0^2 \nonumber\\
               &= 2(N + \tilde{N} + a_b + \tilde{a}_b + a_f + \tilde{a}_f) 
\end{align}

\noindent where the number operators, $N$ and $\tilde{N}$, are as in Eq.~(\ref{eq:NumberOperators}).

Here we notice that the quantum mass operator is different from the classical mass, Eq.~(\ref{eq:classicalmass}).  This is because taking into account the commutation relations in Eqs. \ref{eq:Bosoniccommutators}, \ref{eq:NeveuSchwarzcommutators}, and \ref{eq:Ramondcommutators} when we normal order the \emph{quantum} oscillators leads to the formally infinite constants:
\begin{align}
  \label{eq:abdefinition}
  a_b &= \tilde{a}_b = \frac{D-2}{2}\sum_{n=1}^\infty n \\
  a_f,~\tilde{a}_f &= \left\{\begin{array}{l} -\frac{D-2}{2}\sum_{n=1}^\infty n,~~~\mbox{R}\\ -\frac{D-2}{2}\sum_{r=\frac{1}{2},\frac{3}{2},\cdots}^\infty r~~~\mbox{NS}\end{array}\right.
\end{align}

\noindent Using the finite, analytic continuations of the infinite sums:
\begin{align}
    \sum_{n=1}^\infty n &\to \zeta(-1) = -\frac{1}{12} \\
    \sum_{r=\frac{1}{2},\frac{3}{2},\cdots}^\infty r &\to \frac{1}{24}
\end{align}

\noindent the infinite constants are redefined as
\begin{align}
  a_b &\to -\frac{D-2}{24} \\
  a_f,~\tilde{a}_f &\to \left\{\begin{array}{l} \frac{D-2}{24},~~~\mbox{R}\\                 
         -\frac{D-2}{48},~~~\mbox{NS.}\end{array}\right.
\end{align}

\noindent See Appendix~\ref{app:AnalyticContinuation} for a discussion on finite analytic continuation of infinite sums.  A final note here is that the factor $D-2$ in these constants traces back to reparameterization invariance of the world sheet, which results in only $D-2$ independent oscillators in each set $\alpha_m^{\mu}$, $d_m^{\mu}$, $b_r^{\mu}$, etc.

\subsection{The Physical Ground States}
The Gliozzi-Scherk-Olive (GSO) projection keeps only the states with positive G-parity in the NS sector as physical states.  The G-parity operator is defined for NS states by~\cite{bbs1,GreenSchwarzWitten:1987v1,Szabo:2002ca}
\begin{align}
    G &= (-1)^{\sum_{r=1/2}^{\infty} b_{-r} \cdot b_r + 1} ~~~(NS)
\end{align}

\noindent This means that $|p,0\rangle_{NS}$ is not physical, and that the first excited state, $b_{-1/2}^{\mu} |p,0\rangle_{NS} $, is the physical ground state, with $D-2$ real propagating degrees of freedom, making it a space-time vector.  

The classical constraint $G_0 = 0$ becomes a quantum mechanical constraint. For the Ramond ground state, this constraint is~\cite{Szabo:2002ca}
\begin{align}
   G_0 |p, 0 \rangle &= \left( \alpha_0 \cdot d_0 + \sum_{n \ne 0} \alpha_n \cdot d_{-n}\right) |p, 0 \rangle = 0 \nonumber\\
          &= 0 = \alpha \cdot d_0 |p, 0 \rangle \nonumber\\
          &= 0 = p^\mu \Gamma_\mu |p, 0 \rangle
\end{align}

\noindent as $\alpha_0^{\mu} \propto p^\mu$ and $d_0^\mu \propto \Gamma^\mu$ since the $d_0's$ must have a representation as Dirac matrices as they furnish a $D$ dimensional Clifford algebra, Eq.~(\ref{eq:Ramondcommutators}).  The Ramond ground state, therefore, satisfies a massless Dirac equation, meaning it is a spinor in $D$ space-time dimensions, with $2^{\tilde{D}/2 +1}$ real degrees of freedom, with
\begin{align}
   \tilde{D} &= \left\{ \begin{array}{l l}
                            D, & D~\mbox{is even} \\
                            D - 1, & D~\mbox{is odd}
                       \end{array}
                \right.
\end{align}
Enforcing the Dirac equation as a constraint, while at the same time forcing it to be a Majorana-Weyl spinor (real with definite chirality), reduces this number of degrees of freedom by a real factor of eight.

The G-parity operator for the R sector is defined as~\cite{bbs1}
\begin{align}
 \label{eq:RGparity}
G &= \Gamma_{\tilde{D}+1}(-1)^{\sum_{n=1}^{\infty} d_{-n} \cdot d_{n}}~~~(R),\\
\label{eq:Gamma11}
\Gamma_{\tilde{D}+1} &= \Gamma_0 \Gamma_1 \cdots \Gamma_{\tilde{D}-1}.
\end{align} 
Here we keep either positive or negative G-parity states as physical states.  Looking at Eq.~(\ref{eq:Gamma11}), we see this boils down to keeping either states with positive or negative parity with respect to $\Gamma_{\tilde{D}+1}$ as physical states.  We then define the physical Ramond ground states as either
\begin{align}
   |p,+\rangle_R &\equiv \Gamma_{\tilde{D}+1} |p, 0 \rangle_{R} = + |p, 0 \rangle_R,
\end{align}

\noindent or
\begin{align}
   |p,-\rangle_R &\equiv \Gamma_{\tilde{D}+1} |p, 0 \rangle_{R} = - |p, 0 \rangle_R.
\end{align}

The type II superstring is a theory of closed superstrings, and so its ground states are the four sectors shown in Eq. (\ref{eq:physicalstates}).  The type II physical ground states are therefore tensor products of left and right moving ground states.  Choosing the left moving Ramond ground states to have the same chirality, with respect to $\Gamma_{\tilde{D}+1}$, yields type IIA superstring theory, choosing opposite chirality yields type IIB superstring theory.  For IIA, the ground states are~\cite{bbs1,Szabo:2002ca}
\begin{align}\label{IIARRsector}
|p,\tilde{-}\rangle_R &\otimes |p,+\rangle_R &\mbox{R-R sector} \\
\label{IIANSNSsector}
\tilde{b}_{-1/2}^\mu |p,\tilde{0}\rangle_{NS} &\otimes b_{-1/2}^\nu |p,0\rangle_{NS} &\mbox{NS-NS sector} \\
\label{IIANSRsector}
\tilde{b}_{-1/2}^\mu |p,\tilde{0}\rangle_{NS} &\otimes |p,+\rangle_R &\mbox{NS-R sector} \\
\label{IIARNSsector}
|p,\tilde{-}\rangle_R &\otimes b_{-1/2}^\mu |p,0\rangle_{NS} &\mbox{R-NS sector}
\end{align}

\noindent where as for IIB the ground states are~{\linespread{1.0}\footnote{The choice of positive or negative chirality is arbitrary, only the relative sign between the left and right moving modes matters. This is, again, opposite chirality for IIA, and the same chirality for IIB~\cite{bbs1}}}:

\begin{align}\label{IIBRRsector}
|p,\tilde{+}\rangle_R &\otimes |p,+\rangle_R &\mbox{R-R sector} \\
\label{IIBNSNSsector}
\tilde{b}_{-1/2}^\mu |p,\tilde{0}\rangle_{NS} &\otimes b_{-1/2}^\nu |p,0\rangle_{NS} &\mbox{NS-NS sector} \\
\label{IIBNSRsector}
\tilde{b}_{-1/2}^\mu |p,\tilde{0}\rangle_{NS} &\otimes |p,+\rangle_R &\mbox{NS-R sector} \\
\label{IIBRNSsector}
|p,\tilde{+}\rangle_R &\otimes b_{-1/2}^\mu |p,0\rangle_{NS} &\mbox{R-NS sector}
\end{align}

Acting on these states with the mass operator, where $N$ only acts on the right moving part of the ground state and $\tilde{N}$ acts on the left moving part of the ground state, we find that the masses of the ground states are
\begin{align}
   \alpha' M^2 &= 0,~~~\mbox{R-R sector} \\
   \alpha' M^2 &= -\frac{1}{4}(D-10),~~~\mbox{NS-NS sector} \\
    \alpha' M^2 &= -\frac{1}{8}(D-10),~~~\mbox{NS-R sector} \\
    \alpha' M^2 &=-\frac{1}{8}(D-10),~~~\mbox{R-NS sector}
\end{align}

\noindent which all vanish for $D=10$, the critical space-time dimension for superstring theory: for dimensions smaller than this, some of the ground states are massive, for dimensions larger than this, some of the the ground states are tachyonic (NS-NS, NS-R, and R-NS sectors).  While this is not a proof of the critical dimension of string theory, it shows evidence for it.  A more complete discussion of the critical dimension of superstring theory is given by Polyakov~\cite{Polyakov:1981re} involving conformal and diffeomorphism invariance of the path integral measure, a condition sometimes referred to as the vanishing of the trace anomaly of the stress energy tensor.

In ten space-dimensions, the four massless sectors each contain $64$ degrees of freedom, which can be reorganized into the fields shown in Table~\ref{tab:typeIImasslessfields}~\cite{bbs1,Szabo:2002ca}
\begin{table}[htbp]
\centering
\caption{Field content for the massless modes of type IIA and IIB superstrings.}
\label{tab:typeIImasslessfields}
\begin{tabular}{|c|c|c|} 
\hline Sector & Massless Fields & Respective Name of Fields\\
\hline \multirow{2}{*}{R-R} & $F_2 = dC_1$, $F_4 = dC_3$, type IIA & \multirow{2}{*}{Ramond-Ramond Fields}\\
                            & $F_1 = dC_0$, $F_3 = dC_2$, $F_5 = dC_4$,  type IIB & \\
\hline \multirow{2}{*}{NS-NS}  & \multirow{2}{*}{$G_{\mu\nu}$, $H_3 = dB_2$, $\Phi$} & graviton, Kalb-Ramond Field,\\
 & & dilaton \\
\hline NS-R & $\Psi_{\mu}$, $\lambda$ & gravitino, dilatino\\
\hline R-NS & $\Psi_{\mu}'$, $\lambda'$ & gravitino, dilatino\\
\hline
\end{tabular}
\end{table}

\noindent Notice that the massless fields of type IIA correspond to the field content of a particular supergravity, and the massless fields of type IIB correspond to the field content of a slightly different particular supergravity.  This is evidence that a low energy, effective action for string theory exists which contains supergravity~\cite{bbs1,Polchinski:1998v2}.

\section{Low Energy Effective Actions of String Theory}\label{SUGRAactions}
The field content of type II superstring theory was a finite number of massless fields and an infinite tower of massive fields which we now denote by $\phi_0$ and $\phi_H$, respectively.  The massive fields have masses at the Planck scale, and so don't need to be considered if we are to investigate string theory at currently available energy scales.  If we knew the entire string field theory, $S[\phi_0,\phi_H]$, we could, in principle, integrate out the massive fields, and study the effective action of just the massless fields~\cite{GreenSchwarzWitten:1987v2}:
\begin{align}\label{eq:LowEnergyEffectiveAction}
   e^{iS_{eff}[\phi_0]} \sim \int D\phi_H e^{iS[\phi_0,\phi_H]}
\end{align}

As the full string field theory $S[\phi_0,\phi_H]$ is not currently known, we can instead construct effective actions for low energy superstrings by considering the known massless field content, as described in the previous section for type II superstring theory, and using guiding principles such as supersymmetry, gauge invariance, and comparing scattering matrix elements between  theories~\cite{GreenSchwarzWitten:1987v2}.

The low energy effective actions for the five different superstring theories can be split up into their bosonic and fermionic components as
\begin{align}
S_{eff} &= S_b +S_f.
\end{align}

\noindent As the fermionic actions are considered to vanish classically~\cite{VanNieuwenhuizen:1981ae,bbs1}, we discuss only the bosonic part of the actions here.  The full actions, including the fermionic contributions, can be found in~\cite{VanNieuwenhuizen:1981ae,GreenSchwarzWitten:1987v2}.  The bosonic action takes the form
\begin{align}\label{eq:GeneralLowEnergyEffectiveAction}
  S_b &= \frac{1}{2 \kappa^2}\int d^{10}x \sqrt{G}R - \frac{1}{4 \kappa^2}\int \mathcal{L}_m,~~~2\kappa^2 = (2 \pi)^7\alpha'^4 
\end{align}

\noindent and the bosonic matter actions for the different superstring theories are~\cite{Polchinski:1998v2}{\linespread{1.0}\footnote{A nice discussion of the equations of motion and various solutions can be found in~\cite{Herzog:2000rz}, where the type II actions can be derived from those listed here with $G_{\mu\nu} \to g_s^{-1/2} G_{\mu\nu}$ and $\kappa \to \kappa/g_s$.}}
\begin{align}\label{eq:typeIIA}
  \mathcal{L}_m^{(IIA)} &= d\Phi \wedge *d\Phi + e^{-\Phi}H_3\wedge *H_3 + e^{3\Phi/2} F_2 \wedge *F_2 + \tilde{F}_4 \wedge *\tilde{F}_4 + B_2 \wedge F_4 \wedge F_4 \\
  \label{eq:typeIIB}
  \mathcal{L}_m^{(IIB)} &= d\Phi \wedge *d\Phi + e^{2\Phi} F_1 \wedge *F_1 + e^{-\Phi}H_3\wedge *H_3 + e^{\Phi} \tilde{F}_3 \wedge *\tilde{F}_3 + \frac{1}{2}\tilde{F}_5 \wedge *\tilde{F}_5 + \nonumber\\ 
    &~~~+ C_4 \wedge H_3 \wedge F_3 \\
   \label{eq:typeI} 
\mathcal{L}_m^{(I)} &=  d\Phi \wedge *d\Phi + e^{\Phi} \tilde{F}_3 \wedge * \tilde{F}_3 + \frac{2\kappa^2}{g_{10}^2} e^{\Phi/2} Tr_{v}\left(F_2 \wedge * F_2\right),~~~g_{10}^2 = 2(2\pi)^{7/2} \alpha' \kappa \\
   \label{eq:heterotic}
\mathcal{L}_m^{(het)} &=  d\Phi \wedge *d\Phi + e^{-\Phi} \tilde{H}_3 \wedge * \tilde{H}_3 + \frac{2\kappa^2}{g_{10}^2} e^{-\Phi/2} Tr_{v}\left(F_2 \wedge * F_2\right),~~~g_{10} = \frac{2\kappa}{\sqrt{\alpha'}}
\end{align}

\noindent For type II, we have pure supergravity, with only $U(1)$ gauge fields~{\linespread{1.0}\footnote{Alternate conventions for the self dual five form are $\tilde{F}_5 = F_5 - C_2 \wedge H_3$ and $\tilde{F}_5 = F_5 - \frac{1}{2} C_2 \wedge H_3 + \frac{1}{2} B_2 \wedge F_3$.}}
\begin{align}
  H_3 &= dB_2,~~~F_p = d C_{p-1},~~~\tilde{F}_3 = F_3 - C_0 H_3,~~~\tilde{F}_4 = F_4 - C_1 \wedge H_3,\nonumber\\
 \tilde{F}_5 &= *\tilde{F}_5 = F_5 + B_2 \wedge F_3.
\end{align}

\noindent For type I and heterotic superstring theory, we have supergravity coupled to Yang-Mills theory, with 
\begin{align}
  \tilde{F}_3 &= d C_2 - \frac{\kappa^2}{g_{10}^2} \omega_3,~~~\tilde{H}_3 = d B_2 - \frac{\kappa^2}{g_{10}^2} \omega_3, \nonumber\\
\omega_3 &= Tr_{v}\left(C_1 \wedge F_2 - \frac{2i}{3}C_1\wedge C_1 \wedge C_1 \right),
\end{align}

\noindent where the Yang-Mills field $F_2 = d C_1$ is matrix valued with gauge group $SO(32)$ in type I, and either $SO(32)$ or $E_8 \times E_8$ in the heterotic theory~\cite{GreenSchwarzWitten:1987v2,Szabo:2002ca,Polchinski:1998v2}.  A nice summary of the bosonic equations of motion derived from the IIA and IIB actions can be found in~\cite{Herzog:2000rz}.  

We have written everything in the Einstein frame, which is related to the string frame via:
\begin{align}
   (G_{\mu\nu})_{Einstein} &= e^{-\Phi/2}(G_{\mu\nu})_{string} 
\end{align}

\noindent To switch these actions between the Einstein and string frames, the following conformal identities for D dimensional space-time are helpful~\cite{bbs1}:
\begin{align}\label{eq:conformalidentities}
   &R_{\mu\nu} \to R_{\mu\nu}-\frac{as}{2}\left[((D-2)\delta^{\alpha}_{~\mu}\delta^{\beta}_{~\nu} + G^{\alpha\beta}G_{\mu\nu})\Phi_{;\alpha;\beta} + \frac{a(D-2)}{2}(G_{\mu\nu}G^{\alpha\beta} - \delta^{\alpha}_{~\mu}\delta^{\beta}_{~\nu})\Phi_{;\alpha}\Phi_{;\beta}  \right]\nonumber\\
 &\int d^Dx\sqrt{G} e^{b\Phi} R \to \int d^Dx \sqrt{G} e^{((D-2)a/2 + b)\Phi}\bigg[R + \nonumber\\
  &\left.~~~~~~~~~~~~~~~~~~~~~~~~~~~~~~~~~~~~~~~~~ +s\frac{a^2(D-1)(D-2)}{4}\left(1 + \frac{4b}{a(D-2)}\right)(\partial_{\mu}\Phi)^2\right]\nonumber\\
 &F_p \wedge * F_p \to e^{(D/2-p)a\Phi}F_p \wedge *F_p
\end{align}

\noindent for $G_{\mu\nu} \to e^{a\Phi}G_{\mu\nu}$ with arbitrary constants $a$ and $b$ and where
\begin{align}
   R_{\mu\nu} &= s(\Gamma^{\alpha}_{~\mu\nu,\alpha} - \Gamma^{\alpha}_{~\mu\alpha,\nu} + \Gamma^{\alpha}_{~\mu\nu}\Gamma^{\beta}_{~\alpha\beta} - \Gamma^{\alpha}_{~\mu\beta}\Gamma^{\beta}_{~\nu\alpha}),~~~s=\pm 1.
\end{align}

\subsection{Black \texorpdfstring{$p$}{p}-branes}\label{Blackpbranes}
We now discuss a famous solution to the supergravity equations of motion, known as the black $p$-brane~\cite{Horowitz:1991cd,Aharony:1999ti}.  The action for a type II supergravity with only one non-vanishing R-R source, in string frame{\linespread{1.0}\footnote{We have, in addition, made the redefinition $e^{-2\Phi} \to g_s^2 e^{-2\Phi}$ to be consistent with some of the literature, where $g_s$ is the string coupling constant.}}, is
\begin{align}
   S &= \frac{1}{2\kappa^2}\int \sqrt{G} d^{10}x \mathcal{L},~~~\mathcal{L} = g_s^2 e^{-2\Phi}(R + 4 (\partial\Phi)^2) - \frac{c_p}{2} |F_{p+2}|^2
\end{align}

\noindent whose equations of motion are 
\begin{align}
   &R = -4(\nabla\Phi)^2 -4 \nabla^2\Phi, \\
   &d*F_{p+2} = 0, \\
   &R_{\mu\nu} = -8\nabla_{\mu}\Phi\nabla_{\nu}\Phi + \frac{1}{2}G_{\mu\nu}g_s^{-2}e^{2\Phi}\mathcal{L} +  g_s^{-2}e^{2\Phi}\frac{c_p}{2(p+1)!}F_{\mu\mu_1\cdots\mu_{p+1}}F_{\nu}^{\mu_1\cdots\mu_{p+1}}
\end{align}

\noindent where $c_p = 1$ for all R-R forms except for the five form from IIB, where $c_p=1/2$.

We investigate a specific solution to a R-R source at the center of an $S^{8-p}$
\begin{align}
   *F_{p+2} &= \left\{\begin{array}{l} Q \omega_{8-p},~~~p \ne 3\\
                       F_5 = Q (\omega_5 + * \omega_5),~~~p=3
                      \end{array}
               \right.
\end{align}

\noindent where $Q = q g_s(\alpha')^{(7-p)/2}N$, the charge per unit volume of the $S^{8-p}$,
\begin{align}
   \int_{S^{8-p}} * F_{p+2} &= Q \int\omega_{8-p},
\end{align}

\noindent $N$ is an integer, $q$ and $g_s$ are unitless, and $\omega_n$ is the volume form for an $S^n$, with volume $\int \omega_n = 2 \pi^{(n+1)/2}/\Gamma((n+1)/2)$.

The solution to this source is
\begin{align}
   ds^2 = \sqrt{f_-(\rho)}\left(-\frac{f_+(\rho)}{f_-(\rho)} dt^2 + dx_p^2\right) &+ \frac{f_-(\rho)^{-1/2 - (5-p)/(7-p)}}{f_+(\rho)}d\rho^2 + \nonumber\\
       &+\rho^2 f_-(\rho)^{1/2 - (5-p)/(7-p)} d\Omega_{8-p}^2, \\
    e^{-2\Phi} = g_s^{-2}f_{-}(\rho)^{-(p-3)/2},~~~~~~~~~~~~~~~~~~~~&f_{\pm}(\rho) = 1 - \left(\frac{r_{\pm}}{\rho}\right)^{7-p}
\end{align}

\noindent where $dx_p^2$ is the $p$ dimensional Euclidean line element, whose volume is known as a black the $p$-brane: a black $p$-brane is the $p$ dimensional analogy of a black hole.

The constants $r_\pm$ are related to the Mass $M$, and charge $Q$ by
\begin{align}
   M \propto (8-p)r_+^{7-p} - r_-^{7-p},~~~Q = (r_+ r_-)^{(7-p)/2}
\end{align}

The singularity at $\rho = r_{+}$ is an event horizon and the singularity at $\rho = r_{-}$ is a curvature singularity.  For $p=0$, this solution describes a black hole, without a naked singularity for $r_+ > r_-$.  For general $p \le 6$, a $p$ dimensional \emph{brane} exists at the curvature singularity, and the solution is said to describe a black $p$-brane~\cite{Horowitz:1991cd}.

\subsection{Extremal black \texorpdfstring{$p$}{p}-branes}
Extremal black $p$-brane solutions are those for which $r_+ = r_-$, where the charge $Q$ becomes related to the horizon $r_+$ as
\begin{align}
   r_+^{7-p} = Q = q g_s N (\alpha')^{(7-p)/2}.
\end{align} 
Applying the coordinate transformation
\begin{align}
  r^{7-p} \equiv \rho^{7-p} - r_+^{7-p}
\end{align}

\noindent the extremal black $p$-brane solution becomes
\begin{align}
    ds^2 &= H(r)^{-1/2}(-dt^2 +dx_p^2) + H(r)^{1/2}(dr^2 + r^2 d\Omega_{8-p}^2), \\
    e^{\Phi} &= g_s H(r)^{(3-p)/4},~~~H(r) = \frac{1}{f_+(\rho)} = 1 + \left(\frac{r_+}{r}\right)^{7-p}, \\
    F_{p+2} &= \left\{\begin{array}{l}
                      \frac{Q}{H(r)^2 r^{8-p}} dx^0\wedge \cdots \wedge dx^p\wedge dr,~~~p \ne 3\\ 
                      Q\left(\omega_5 + \frac{1}{H(r)^2r^5} dx^0\wedge \cdots \wedge dr\right),~~~p=3,
                       \end{array}
               \right.
\end{align}

\noindent where now the horizon is located at $r=0$~\cite{Aharony:1999ti}.

\section{The Open Superstring and Non-Abelian Gauge Theories}\label{SUYMaction}
We have seen that superstring theory can be thought of as an effective theory of gravity, thus bearing the possibility that it can describe the quantum gravity of our world.  Superstring theory can also possibly describe the standard model, as it contains chiral fermions and non-abelian gauge theories~\cite{Polchinski:1998v1}.  The latter feature is clearly evident in three of the superstring theories presented in the previous section: the type I theory and one of the heterotic theories contain an $SO(32)$ gauge theory, and the other heterotic theory contains an $E_8 \times E_8$ gauge theory.  Though the type II theories presented thus far only contain $U(1)$ gauge theories as part of their supergravity,  we will see that they can be modified to contain non-abelian gauge theories.  For instance, if we add open strings to the type IIB theory ending on multiple, parallel D-branes, non-abelian gauge theories will appear on the world volumes of the D-branes~\cite{Polchinski:1995mt,z1}.

\subsection{D-branes and Gauge Theories}
Much work has been done relating D-branes to gauge theories, the inception being t' Hooft's work~\cite{'tHooft:1973jz}, where he showed the duality between open strings and gauge theories for a large $N$ number of colors.  As D-branes are the end points of open strings, it is not surprising that they too are found to be related to gauge theories.    

Analysis of vertex operators in the closed type II superstring reveals that superstring carries NS-NS charge~\cite{Polchinski:1995mt}.  We can therefore write an interaction between the superstring and the NS-NS two-form $B_2$ as~\cite{z1}
\begin{align}\label{eq:Bint}
   S_{B} &= -\frac{1}{2}\int dX^\mu \wedge dX^\nu B_{\mu\nu}(X(\tau,\sigma)) \nonumber\\
         &= -\frac{1}{2}\int d^{10}x B_{\mu\nu}(x)j^{\mu\nu}(x)
\end{align}

\noindent where the NS-NS charge is carried by the string current
\begin{align}\label{eq:stringcurrent}
   j^{\mu\nu}(x) &\equiv \int dX^{\mu} \wedge dX^{\nu} \delta^{10}(x - X(\tau,\sigma)).
\end{align}

\noindent The action, Eq.~(\ref{eq:Bint}), is gauge invariant with respect to the transformation
\begin{align}\label{eq:Btransformation}
   B_{\mu\nu} \to B_{\mu\nu} + \partial_{\mu}\Lambda_\nu - \partial_\nu\Lambda_\mu
\end{align}
\noindent for closed superstrings and open superstrings with \emph{Neumann} boundary conditions.  If we add some open superstrings ending on a D$p$-brane to the type IIB theory, we must add a $U(1)$ gauge field, $A_a(X)$, at the string endpoints to maintain this gauge invariance
\begin{align}
   S_{B,p} &= S_B + 2\pi\alpha'\int A_a(X) (dX^a|_{\sigma = \pi} - dX^a|_{\sigma = 0}),~~~a = 0,\cdots,p
\end{align} 
\noindent where $A_a(X)$ transforms as
\begin{align}\label{eq:Atransformation}
    A_a \to A_a - \frac{1}{2\pi\alpha'}\Lambda_a
\end{align}

This can be viewed as the NS-NS open superstring current, Eq.~(\ref{eq:stringcurrent}), turning into a vector $U(1)$ gauge current at its endpoints sourcing the gauge invariant field
\begin{align}\label{eq:curlyF}
  \mathcal{F}_{ab} \equiv B_{ab} + 2\pi\alpha' F_{ab} \\
   F_{ab}(\zeta) = \partial_a A_b(\zeta) - \partial_b A_a(\zeta) 
\end{align} 

\noindent which propagates along the D$p$-branes world volume, parametrized by $\zeta$~\cite{z1}.  This process can be depicted pictorially as in Figure~\ref{fig:DpBraneGaugeFlux-a}. 
\begin{figure}[htbp]
\addtocontents{lof}{\protect\vspace{\li}}
   \centering
   \subfigure[$U(1)$ gauge flux flowing through $N=1$ Dp-brane.]{\label{fig:DpBraneGaugeFlux-a}\includegraphics[width=0.4\columnwidth]{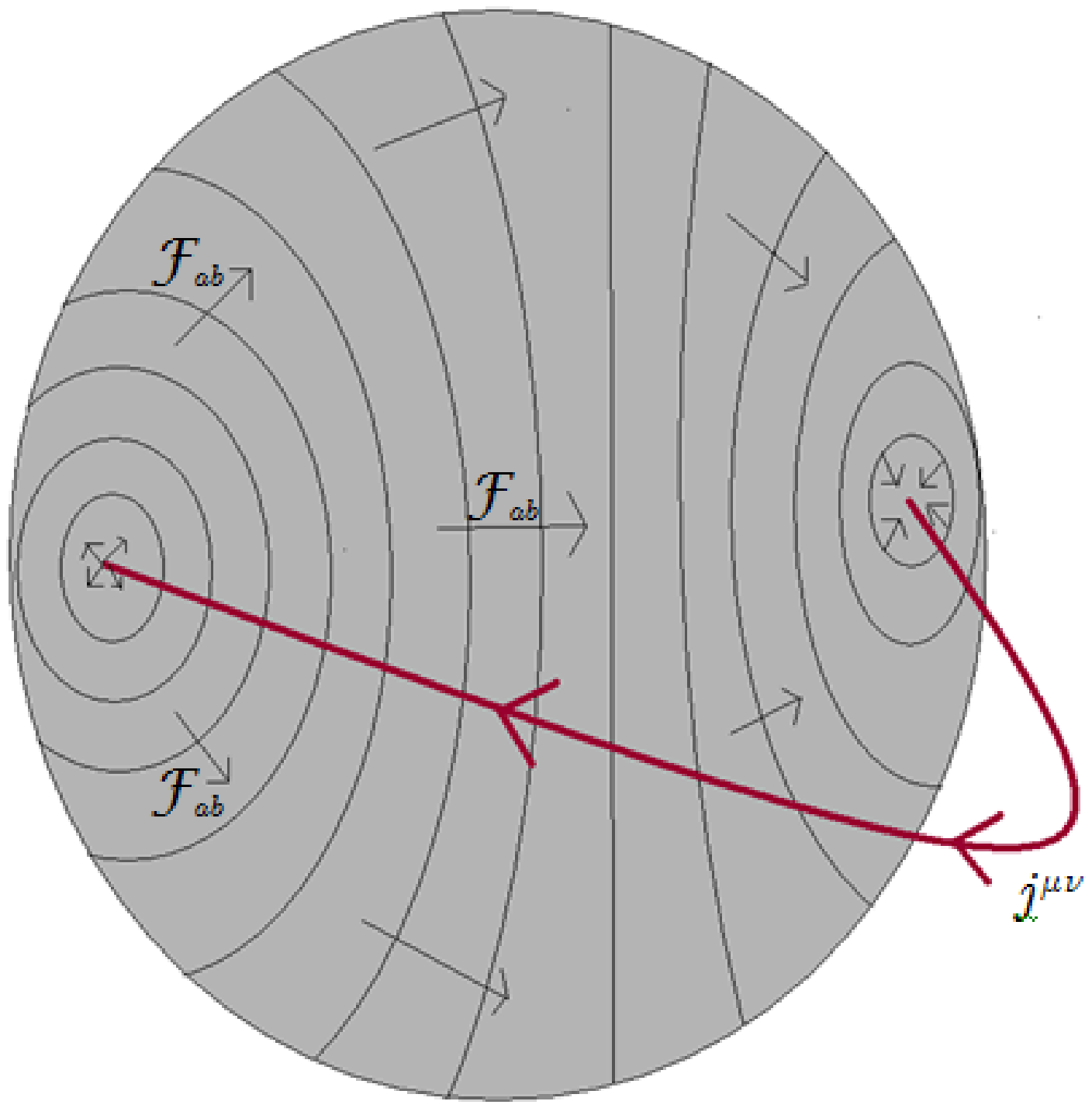}}    
\quad   
\subfigure[From the perspective of $N$ coincident Dp-branes, the $U(N)$ gauge theory looks like a $U(N)$ 1-string flux tube in the dual gauge theory picture.]{\label{fig:DpBraneGaugeFlux-b}\includegraphics[width=0.4\columnwidth]{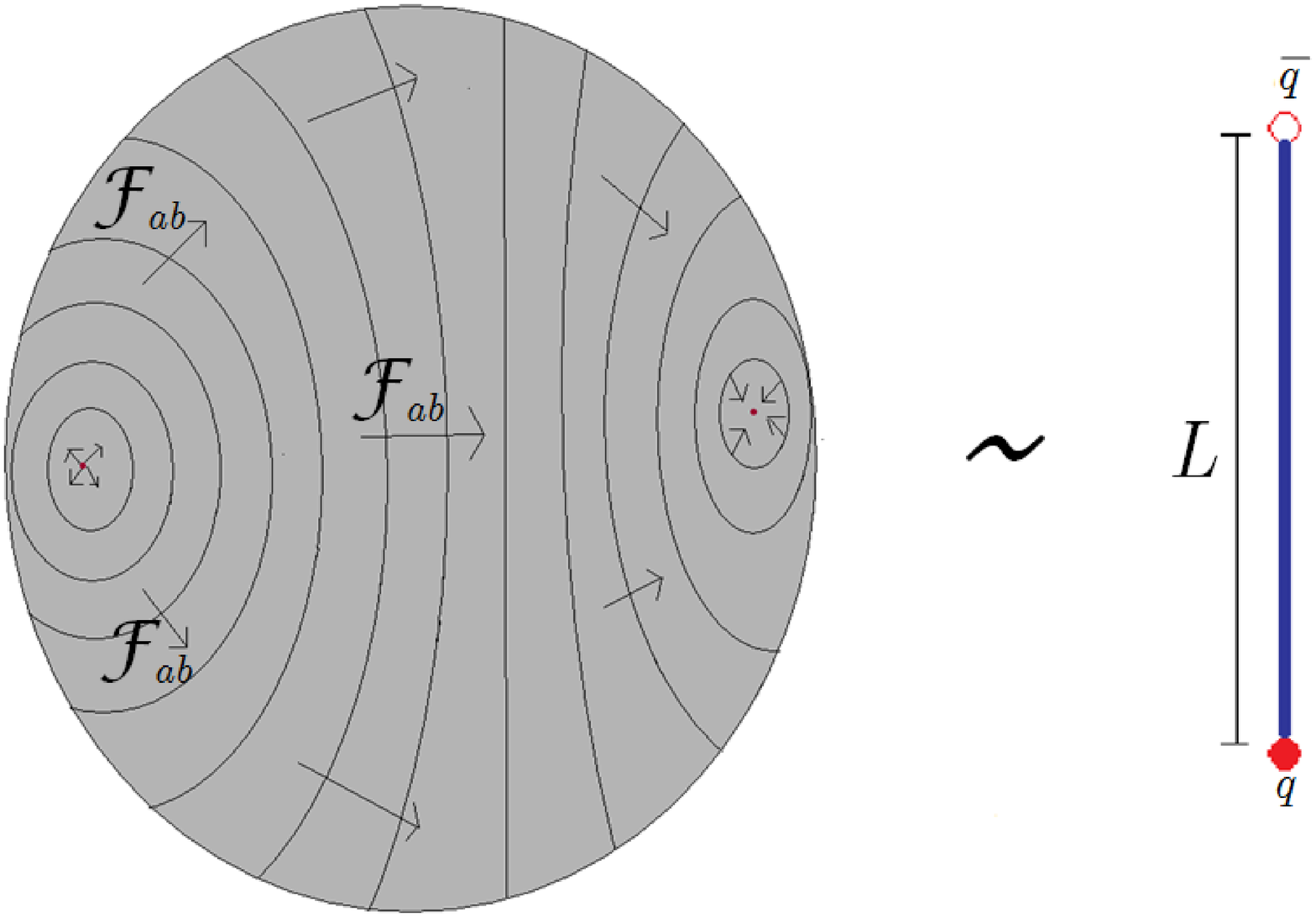}} 
   \caption{String current, $j^{\mu\nu}$, flowing into $N$ coincident Dp-branes as $U(N)$ gauge flux, (essentially $N$ copies of $\mathcal{F}_{ab}$), then flowing back out at the other string endpoint, turning once again into string current.}
   \label{fig:DpBraneGaugeFlux}
\end{figure}

\begin{figure}
\addtocontents{lof}{\protect\vspace{\li}}
   \centering
   \includegraphics[width=0.6\columnwidth]{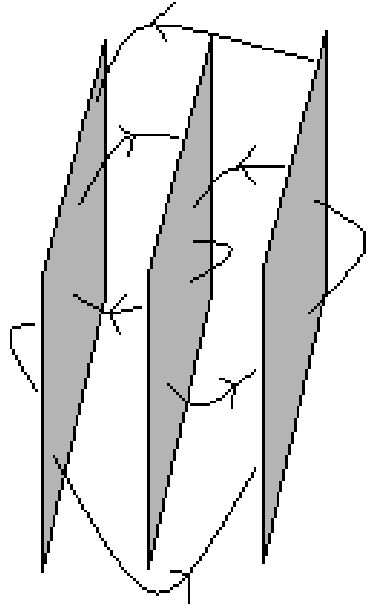}
   \caption{Nine distinct strings can stretch between three parallel D-branes.}  
  \label{fig:stacked}
\end{figure}

\newpage
Consider now three identical, parallel D$p$-branes, spaced an infinitesimal distance apart, as in Figure~\ref{fig:stacked}.  Collections of D-branes such as this are known as \emph{coincident} D-branes. As there are $3^2 = 9$ distinct types of strings attached to the D-branes in this configuration, there are $9$ distinct gauge fields, which compose as $U(3)$ gauge field.  For $N$ coincident D-branes, we have a $U(N)$ gauge field propagating on their coincident world volume. 

\subsection{D-branes as Dynamical Objects}
We will now explain how D-branes are dynamical objects as well as objects which carry $U(N)$ gauge flux.  Perhaps the quickest and most succinct argument as to why D-branes possess dynamics is due to Polchinski~\cite{Polchinski:1995mt}:

\begin{enumerate}
    \item Analysis of vertex operators in the closed type II superstring shows that while strings carry NS-NS charge and interact with the NS-NS two form field $B_2$ via Eq.~(\ref{eq:Bint}), they \emph{do not} carry NS-NS charge and thus \emph{can not} interact with the R-R forms~\cite{Polchinski:1995mt,Witten:1995im}.
    \item Since S-duality in type IIB superstring theory switches the NS-NS and R-R two forms ($B_2 \leftrightarrow C_2$), there must be something with R-R charge for the R-R form to interact with.  Since this can't be the string, it makes sense to assume that it is the D-brane~\cite{Polchinski:1995mt}.
\end{enumerate}  

\noindent Furthermore, just as there is a natural coupling between a string and $B_2$, there is a natural coupling between a D$p$-brane and $C_{p+1}$:
\begin{align}\label{eq:CpCoupling}
   S_{C_p} &\sim \int d^{p+1} \zeta C_{\mu_1\cdots\mu_{p+1}}dX^{\mu_1} \wedge \cdots \wedge d X^{\mu_{p+1}}.
\end{align}

\noindent and we now can interpret the black $p$-brane sources of R-R charge in section~\ref{Blackpbranes}, as D-branes on which open strings can end.  The D-brane's dynamics are manifest in Eq. (\ref{eq:CpCoupling}) in it's bosonic coordinates $X=X(\zeta)$ becoming scalar \emph{fields} on its world volume, parameterized by $\zeta$.  

A supersymmetric effective action encompassing all these features of a D$p$-brane in a curved type II background is given by~\cite{Martucci:2005rb}
\begin{eqnarray}\label{eq:DpbraneAction}
  S_{p}& = & -\mu_p \int d^{p+1}\zeta e^{-\Phi}\sqrt{\mathcal{M}} +\mu_p \int \sum_n C_n\wedge e^\mathcal{F} + \nonumber\\
       & &+ \frac{\mu_p}{2}\int d^{p+1}\zeta e^{-\Phi}\sqrt{\mathcal{M}}~\mathcal{L}_p^f(\Theta)
\end{eqnarray}

\noindent where
\begin{align}
   \mathcal{M} &= |\det \mathcal{M}_{ab}|,~~~\mathcal{M}_{ab} = g_{ab} + \mathcal{F}_{ab}, \nonumber\\
 \mu_p &= (2 \pi)^{-p}(\alpha')^{-(p+1)/2}, 
\end{align}

\noindent and $g_{ab}$ and the more general version of $B_{ab}$ in Eq. (\ref{eq:curlyF}), are both \emph{pullbacks} to the Dp-brane world volume:
\begin{align}
   g_{ab} = \partial_a X^\mu \partial_b X^\nu G_{\mu\nu},~~~B_{ab} = \partial_a X^\mu \partial_b X^\nu B_{\mu\nu}.
\end{align}
\noindent The first term in this action can be seen as the generalization from the Nambu-Goto string action to higher dimensional objects endowed with a $U(1)$ gauge field $\mathcal{F}_{ab}$.

As $\zeta$ is the parametrization of the D$p$-brane's world-volume, the D$p$-brane action is a \emph{field theory} of the bosonic scalar fields, $X^{\mu}(\zeta)$, the bosonic $U(1)$ vector fields $A_a(\zeta)$, and fermionic fields of the 32-component Green-Schwarz spinor, $\Theta(\zeta)$.  The fermionic fields give dynamics to the Dp-brane through the fermionic Lagrangian, $\mathcal{L}_p^f$, which is different for the type IIA and IIB supergravity theories~\cite{Martucci:2005rb}:
\begin{align}\label{eq:LfIIA}
   \mathcal{L}^{f}_p(\Theta) &=  \bar{\Theta} \left(1-\Gamma_{D_p}\right)\biggl[(\mathcal{M}^{-1})^{ab}\left(\Gamma_a D_b^{(0)} +\Gamma_b W_a\right) - \Delta^{(1)}-\Delta^{(2)}\biggr]\Theta,~~~\mbox{type IIA}\\
\label{eq:LfIIB}
\mathcal{L}^{f}_{p}(\Theta)&=\bar{\Theta}\biggl[\left(\mathcal{M}^{-1}\right)^{ab}\left(\Gamma _{a}D_{b}^{(0)}-\Gamma_{D_p}^{-1}\Gamma_{b}W_{a}\right) - \Delta^{(1)} + \Gamma_{D_p}^{-1}\Delta^{(2)}\biggr]\Theta,~~~\mbox{type IIB}
\end{align}

\noindent The rest of the definitions in these Lagrangians are found in appendix ~\ref{app:FermionicDefinitions}.

The action, Eq.~(\ref{eq:DpbraneAction}), shows how a $U(1)$ gauge theory manifests itself on a D$p$-brane embedded in a supergravity background. This is the low energy effective action from open strings ending on one D$p$-brane added to a type IIB theory.   As explained previously, $U(N)$ gauge theories manifest themselves as the low energy effective action from open strings ending on $N$ coincident D$p$-branes~\cite{Witten:1995im,Aharony:1999ti,z1}. Previewing the connection to $k$-strings, consider $N$ such coincident D-branes. Removing the attached string from the picture, as in Figure~\ref{fig:DpBraneGaugeFlux-b}, the strings end points look like quarks in the $k$-string dual picture, and the $U(N)$ flux through the D-brane look like the $U(N)$ flux tube of the $k$-string for $k$=1.  For arbitrary $k$, the analogous picture would be the same, but with multiple strings attached to the D-brane.

\subsection{The AdS/CFT correspondence}\label{AdSCFTcorrespondence}
In section~\ref{SUGRAactions}, we saw that the low energy field theory of type II closed superstrings is effectively supergravity.  So far in section~\ref{SUYMaction}, we have seen evidence that the field theory of open strings and D-brane excitations contains Yang-Mills theories.  This should not at all be a surprise, as the bosonic, low energy effective action of type I open and closed superstrings, Eqs. (\ref{eq:typeI}) and (\ref{eq:GeneralLowEnergyEffectiveAction}), is that of supergravity coupled to Yang-Mills theory.
Consider now the theory of type IIB closed strings and add some open superstrings ending on $N$ coincident D$3$-branes.
Integrating out all the massive modes, as in Eq.~(\ref{eq:LowEnergyEffectiveAction}), results in a low energy effective action~\cite{Aharony:1999ti}
\begin{align}
   S_{eff} &= S_{SUGRA} + S_{branes} +S_{int} + \mbox{higher order derivative terms},
\end{align} 
\noindent where $S_{SUGRA}$ contains supergravity, due to the closed strings, $S_{branes}$ contains four dimensional $\mathcal{N}=4$ $U(N)$ super Yang-Mills theory, due to the open strings and D-brane excitations, and $S_{int}$ contains the perturbative couplings, in powers of $\kappa \sim g_s \alpha'^2$~\cite{Maldacena:1997re,Aharony:1999ti}. 
Taking the weak coupling limit $\alpha' \to 0$ results in decoupled theories of free supergravity ($S_{gravity}$) and free super Yang-Mills theory ($S_{SUYM}$):  
\begin{align}\label{eq:SIIBeff1}
   S_{eff} &= S_{gravity} + S_{YM}[\phi_{YM}] + \mathcal{O}(\kappa)
\end{align}

\noindent where we have denoted the super Yang-Mills fields, collectively as $\phi_{SUYM}$.  At this point, we can construct a partition function for this string theory by \emph{evaluating} $S_{eff}$ on a particular supergravity background, and path integrating over the super Yang-Mills fields, we denote collectively by $\phi_{YM}$:
\begin{align}\label{eq:Zequivalences}
   Z_{string} = Z_{CFT} &= \int D\phi_{YM} e^{i S_{eff}} \sim e^{iW_{gravity}}.
\end{align} 


Now consider the effective action $W_{gravity}$ from the rightmost term in Eq. (\ref{eq:Zequivalences}).   This should be related to a classical supergravity theory with D-branes as the black p-brane sources, as in section~\ref{Blackpbranes}.  Considering decoupled low energy excitation in this new perspective, let us match which corresponds to the free gravity and which corresponds to the free SUYM theory in the path integral in Eq. (\ref{eq:Zequivalences}). Specifically, we consider $N$ parallel D$3$-branes as black $3$-branes, with R-R charge $Q \propto N$, \emph{sourcing} the supergravity fields as described in section~\ref{Blackpbranes}.  Considering extremal solutions, $r_+ = r_-$, the background is
\begin{align}
    ds^2 &= H(r)^{-1/2}(-dt^2 +dx_3^2) + H(r)^{1/2}(dr^2 + r^2 d\Omega_5^2), \\
    \phi &= \phi_0 = \mbox{constant},~~~H(r) = 1 + \frac{q g_s N\alpha'^2}{r^4}.
\end{align}

Considering low energy probes embedded into this geometry, we can separate them into two types: those that decouple from the near horizon region and those that don't.  Massless particles propagating through the space-time will decouple from the near horizon geometry and we interpret these as the free gravity.   We are left to conclude that low energy excitations that live near the horizon of the black brane must be dual descriptions of the gauge theory~\cite{Maldacena:1997re,Aharony:1999ti,z1}.  

Investigating the near horizon geometry, we pull out the perturbative $\alpha'$ dependence by switching coordinates to 
\begin{align}
    U & = \frac{r}{\alpha'}
\end{align}
\noindent and carefully take the near horizon limit
\begin{align}
   r \to 0~~~&\mbox{as}~~~\alpha' \to 0 \\
   U &= \mbox{fixed}
\end{align} where the metric becomes
\begin{align}
   ds^2 &\to \alpha' \left[ \frac{U^2}{\sqrt{qg_sN}}(-dt^2 + dx_3^2) + \frac{\sqrt{qg_sN}}{U^2}dU^2 + \sqrt{qg_sN} d\Omega_5^2 \right] \\
        &\to \alpha'\left[ \frac{\sqrt{qg_sN}}{z^2}(-dt^2 + dx_3^2 + dz^2) +\sqrt{qg_sN}d\Omega_5^2\right],~~~z = \frac{\sqrt{qg_sN}}{U},
\end{align}

\noindent which is $AdS_5 \times S^5$ (see appendix~\ref{app:AdSspace}).  As the $\mathcal{N} = 4~U(N)$ super Yang-Mills theory is dual to the excitations in this region, the AdS/CFT correspondence is stated:
\emph{$\mathcal{N} = 4~U(N)$ super Yang-Mills theory in $3+1$ dimensions is dual to type IIB superstring theory on $AdS_5 \times S^5$}~\cite{Maldacena:1997re,Aharony:1999ti}.  

The real power of this correspondence is that it is a strong-weak correspondence between the two theories.  The correspondence relates the coupling constants between the two theories akin to the famous t'Hooft coupling relationship
\begin{align}
   g_{YM}^2 N \sim g_s N.
\end{align}
The product $g_s N$ is proportional to the radius of the $AdS_5$ space and when this is large the gravitational coupling is small, and one can do perturbative gravity calculations.  These calculation will be dual to strongly coupled gauge theory calculations.  On the other hand, if the the $AdS_5$ radius is small, the gravitational coupling is large, and the gauge theory coupling is small, so one can here do calculations on the gauge theory side and relate them to the gravitational theory side.  In this way, the $AdS/CFT$ correspondence is a way to do perturbative calculations in one theory and relate them to the perturbatively intractable calculation in the other theory. 

Taking a probe to be the dilaton field evaluated at the boundary of the $AdS_5$ space, $\Phi_0(x) \equiv \Phi(x,z)|_{z=0}$, we can be more precise with Eq.~(\ref{eq:Zequivalences}) and write the $AdS/CFT$ correspondence as:~\cite{Gubser:1998bc,Witten:1998qj,Aharony:1999ti}
\begin{align}\label{eq:AdSCFT_Partition_Correspondence}
   Z_{CFT}[\Phi_0(x)] &\equiv \langle e^{\int d^4x \Phi_0(x) \mathcal{O}(x)}\rangle = Z_{String}[\Phi_0(x)] \sim e^{-S_{IIB}[\Phi]}|_{\Phi=\Phi_0(x)}
\end{align}

\noindent where $x = (t,x_3)$ is the parametrization of the black D-branes.  Here, $CFT$ is an abbreviation for Conformal Field Theory, and refers to the four dimensional $\mathcal{N} = 4$ $SUYM$ theory.  We have approximated the full string partition function with its low energy effective action partition function of type IIB supergravity.  On the gauge theory side, the dilaton manifests itself as a source of correlation functions for the conformal operator, $\mathcal{O}$:
\begin{align}
   \langle \mathcal{O}(x_1)\cdots\mathcal{O}(x_n) \rangle = \frac{1}{Z_{CFT}[0]} \frac{\delta}{\delta \Phi_0(x_1)}\cdots \frac{\delta}{\delta \Phi_0(x_n)}Z_{CFT}[\Phi_0(x)]|_{\Phi_0(x) = 0}
\end{align}

A formula such as Eq.~(\ref{eq:AdSCFT_Partition_Correspondence}) can be generalized to generating fields from the supergravity side that are spinors, vectors, and  $p$-forms of mass $m$.  For an operator of dimension $\Delta$, the corresponding generating field has dimensions $4-\Delta$, where~\cite{Gubser:1998bc,Aharony:1999ti}
\begin{align}
   \Delta_{\pm} &= 2 \pm \sqrt{(p-2)^2 +m^2},~~~p \ge 0 \\
   \Delta &= 2 + |m|,~~~\mbox{spinors}
\end{align}

\noindent and for the the $p$-forms, either $\Delta$ = $\Delta_+$ or $\Delta$ = $\Delta_-$.



\Chapter{K-STRING TENSION VIA GAUGE/GRAVITY DUALITIES}\label{oldresults}

The AdS/CFT correspondence, laid out in section~\ref{AdSCFTcorrespondence}, shows that calculations in a supergravity background can be dual to calculations in Yang-Mills theories.  This is well thrashed out for $AdS_5$ and four dimensional $U(N)$ $\mathcal{N}=4$ SUYM theory, including ways of building gauge invariant operators on the CFT side using gravity calculations dual to the functional variations of the CFT partition function.  Though this is a step toward describing the standard model with string theory, it's still very far away from this goal.  For one thing, the dual gauge theory is conformal and so the coupling constant doesn't run.
 
\begin{table*}[!hbp]
\addtocontents{lot}{\protect\vspace{\li}}
	\centering
\caption{Gauge Theory States and Their String Theory Configurations.}
\label{tab:GaugeTHeoryStatesAndTheirStringTheoryConfigurations}
			\begin{tabular}{|c|c|}\hline
\textbf{Gauge Theory State} & \textbf{String Theory Configuration}  \\ \hline \hline
Glueballs & Spinning Folded Closed String  \\ \hline
Mesons of heavy quarks & Spinning open strings ending on boundary\\ \hline
~Baryons of heavy quarks~& Strings attached to baryonic vertex \\ \hline
Dibaryons & Strings attached to wrapped branes \\ \hline
Mesons of light quarks & ~Spinning open strings ending on D7 branes~ \\ \hline
	$k$-strings & Wrapped branes with flux\\ \hline	
		\end{tabular}
\end{table*}

An interesting research topic which has been pursued is constructing supergravity solutions which have dualities with gauge theories with less supersymmetry and coupling constants that run, which are much closer to the standard model than CFT's.  Investigating brane solutions similar to the black $p$-brane solution in section~\ref{Blackpbranes}, we find that D-brane solutions which carry the smallest possible Ramond-Ramond charge break half the supersymmetry on the gauge theory side of the correspondence~\cite{Polchinski:1995mt}.  These branes are known as Bogomolny-Prasad-Sommerfield (BPS) branes, the algebra of the corresponding gauge theory being the corresponding BPS state~\cite{vgjrp,bbs1}.

We can also find supergravity solutions with $H_3$ sources, sourced by magnetically charged Neveu-Schwarz 5-branes (NS5-branes) which break supersymmetry~\cite{vgjrp,Douglas:1995nw,Klebanov:2000hb,Maldacena:2000yy}.  Furthermore, conifold theories are clever ways of wrapping these brane configurations around certain geometries ($S^n$, $S^n \times S^m$, $Y^{pq}$ with even $J$) to break supersymmetry and maintain the regulating features of the gauge/gravity duality~\cite{vgjrp,Klebanov:2000hb,Maldacena:2000yy,Canoura:2005uz}.  Sometimes, infrared divergence problems exist on conifolds and they must be deformed to regulate this behavior~\cite{vgjrp,Klebanov:2000hb}.   We will now investigate calculations in two such backgrounds:  the Klebanov-Strassler (KS)~\cite{Klebanov:2000hb} and Cvetic, Gibbons, L\"u, and Pope (CGLP)~\cite{Cvetic:2001ma} backgrounds.   The gauge theory duals of these backgrounds have running coupling constants, the energy scale of the dual gauge theory being dual to a coordinate on the supergravity side: small (large) distances from a singularity in the supergravity correspond to the IR (UV) in the gauge theory~\cite{Klebanov:2000nc,Klebanov:2000hb,Cvetic:2001ma}.

 String theory objects embedded in such backgrounds have correspondences with gauge theory states, as shown in table \ref{tab:GaugeTHeoryStatesAndTheirStringTheoryConfigurations} \cite{PandoZayas:2003yb,PandoZayas:2008hw,Doran:2009pp,Stiffler:2009ma}. The correspondence between $k$-strings and $D$-branes composes the main result of this thesis: \emph{$k$-strings are dual configurations of charged Dp-branes embedded in supergravity backgrounds}, as evidenced in~\cite{Herzog:2001fq,Herzog:2002ss,PandoZayas:2008hw,Doran:2009pp,Stiffler:2009ma}.  
We briefly saw a glimpse of this correspondence in Figure~\ref{fig:DpBraneGaugeFlux-b}, and devote the final two chapters to more concrete evidence.  Specifically, we will calculate the theoretical D-brane energy, at small distances  (IR), and directly compare it to various $k$-string energy calculations in the literature~\cite{Luscher:1980ac,Luscher:1980fr,Ambjorn:1984me,Ambjorn:1984mb,Ambjorn:1984yu,Bringoltz:2006zg,Bringoltz:2008nd,Karabali:1998yq,Karabali:2000gy,Karabali:2007mr,Karabali:2009rg}.

\section{General Formula for D-brane Energy}\label{generalEformula}

As our theory of D-branes is one of scalar fields, spinor fields, and vector fields, let us begin simply by calculating the free energy of a scalar field, with the hopes that it may shed light on a formula for D-brane energy.  From statistical mechanics, we expect the free energy for a scalar field $\varphi$ to be given by:
\begin{align}\label{eq:1dpropogator}
   Z = e^{-\beta E} = \langle \varphi | e^{-\beta H} | \varphi\rangle \nonumber\\
\end{align}

\noindent The partition function, $Z$, should be dominated by the classical, minimum Hamiltonian, $H_{min}^{(0)}$ and so we have the approximate relationship~\cite{Schulman:1981pi}
\begin{align}
   Z = e^{-\beta E} &\sim \langle \varphi | e^{-\beta H_{min}^{(0)}} | \varphi \rangle \nonumber\\
     &\sim e^{-\beta H_{min}^{(0)}} \langle \varphi | \varphi \rangle \nonumber\\
     &\sim e^{-\beta H_{min}^{(0)}}.
\end{align} 

\noindent We approximate that the free energy of this theory will be equal to $H_{min}^{(0)}$ plus quantum corrections: 
\begin{align}
   E \sim H_{min}^{(0)} + \mbox{quantum corrections}.
\end{align}

For the quantum corrections, we write the partition function in terms of the path integral
\begin{align}
    Z &= \langle \varphi | e^{- \beta H} | \varphi \rangle \nonumber\\
      &= \int D\varphi~e^{iS} \nonumber\\
      &= \int D\varphi~e^{-S_E}
\end{align}

\noindent where we have Wick rotated to a Euclidean action, $S_E$.  We now expand around the classical solution $\varphi = \varphi_{(0)} + \delta\varphi$
\begin{align}
    Z = e^{-\beta E} &\sim \int D\delta\varphi~e^{-S^{(0)}_E - \delta S_E} \nonumber\\
      &\sim e^{-\beta H_{min}^{(0)}} \int D\delta\varphi~e^{-\delta S_E}
\end{align}

\noindent Solving for the free energy, we find
\begin{align}\label{eq:Escalarfree}
    E &\sim H_{min}^{(0)} + \delta E
\end{align}

\noindent where
\begin{align}\label{eq:DeltaEscalar}
    \delta E = - \frac{1}{\beta}\log \int D\delta\varphi ~e^{-\delta S_E}
\end{align}

Guided by Eq.~(\ref{eq:Escalarfree}) and also~\cite{Herzog:2001fq,Herzog:2002ss,Firouzjahi:2006vp,Ridgway:2007vh}, we now define the energy of a probe Dp-brane as
\begin{align}\label{eq:DpbraneEnergy}
     E & \equiv H_{min}^{(0)} + \delta E
\end{align}

The probe D$p$-branes we will investigate will be parameterized by some coordinates $t,x,\theta,\phi \in \zeta^a$, and will be either electrically($Q$) or magnetically($M$) charged
\begin{equation}
   F = dA = Q dt \wedge dx + M d\theta \wedge d\phi,
\end{equation}

\noindent and also embedded in a classical SUGRA background, as in Figure~\ref{fig:embedding}, typically of the form
\begin{equation}\label{eq:typicalbackground}
ds_{10}^2 = H^q dx^{a}dx^{b}\eta_{ab} +H^p ds_{10-d}^2,~~~a,b = 0 \dots d-1
\end{equation}

\noindent sourced by
\begin{eqnarray}
   F_{n+1}(X^{\mu}) &=& d C_n(X^{\mu}), ~~~\Phi(X^{\mu}), \nonumber\\
   H_3(X^{\mu}) &=& d B_2(X^{\mu}),
\end{eqnarray}

\begin{figure}[htbp]
\addtocontents{lof}{\protect\vspace{\li}}
\centering
\includegraphics[width=0.8\columnwidth]{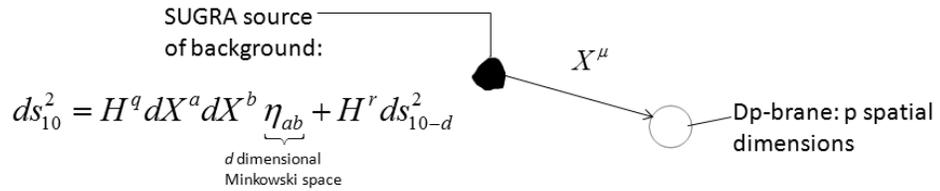}
\caption{A probe Dp-brane embedded in a SUGRA background.} \label{fig:embedding}
\end{figure}

\noindent The two different embeddings we will investigate will each have the probe brane sitting right on the supergravity source, which will be the location of a singularity, and therefore, will correspond to IR calculations in the dual gauge theory.

It is important to note here that $d$ is the space-time dimension of the Minkowski space-time portion of the metric in Eq.~(\ref{eq:typicalbackground}), which will be the space-time dimension in which the $k$-string will be embedded in the dual gauge theory. Also, $X^\mu = X^\mu(\zeta^a)$ maps the world volume of the Dp-brane, as in Fig.~\ref{fig:embedding}.   The bosonic supergravity coordinates of the D$p$-brane, $X^\mu(\zeta^a)$, are therefore \emph{scalar fields} on the D$p$-brane, the field theory dynamics governed by the D$p$-brane action
\begin{align}
  S_{p}& =  -\mu_p \int d^{p+1}\zeta e^{-\Phi}\sqrt{\mathcal{M}} +\mu_p \int \sum_n C_n\wedge e^\mathcal{F} + \nonumber\\
       &~~~+ \frac{\mu_p}{2}\int d^{p+1}\zeta e^{-\Phi}\sqrt{\mathcal{M}}~\mathcal{L}_p^f(\Theta)\nonumber\tag{\ref{eq:DpbraneAction}}
\end{align}

\noindent To find $\mathcal{H}_{(0)}$ for a probe D$p$-brane, we consider the action classically, defined by $\Theta_{(0)} = 0$~\cite{VanNieuwenhuizen:1981ae,bbs1}:
\begin{align}\label{eq:Spclassical}
      S_{p}^{(0)}& =  -\mu_p \int d^{p+1}\zeta e^{-\Phi}\sqrt{\mathcal{M}} +\mu_p \int \sum_n C_n\wedge e^\mathcal{F}
\end{align} 

\noindent and investigate the specific classical solutions{\linespread{1.0}\footnote{ We will demonstrate that these truly are classical solutions in the next chapter, when we fluctuate around the proposed classical solution, and find that the resulting part of the action linear in the fluctuations vanishes, up to total derivatives.  This is clearly seen for the fermionic solution, Eq.~(\ref{eq:fluctuate}): the action, Eq. (\ref{eq:DpbraneAction}), being trivially quadratic to lowest order in the fermionic fluctuations, $\delta \Theta$.}} $A^m = A^m_{(0)}$ and $X^\mu = X^\mu_{(0)}$, where only the gauge fields, $A^m_{(0)}$, have dynamics
\begin{equation}
    S_{p}^{(0)} = \int d^{p+1}\zeta~ \mathcal{L}^{(0)}(A^m_{(0)},\dot{A}^m_{(0)},X^\mu_{(0)}).
\end{equation} 
\noindent

 We next apply a Legendre transformation to this classical D$p$-brane action, Eq.~(\ref{eq:Spclassical}), yielding the Hamiltonian density:
\begin{equation}\label{eq:LegTransform}
   \mathcal{H}^{(0)} = D_{m}\dot{A}_{(0)}^{m} - \mathcal{L}^{(0)},~~~D_{m} = \frac{\partial \mathcal{L}^{(0)}}{\partial \dot{A}_{(0)}^m} 
\end{equation}

\noindent Minimization of this Hamiltonian density leads to ~\cite{Herzog:2001fq,Herzog:2002ss}
\begin{equation}
  H_{min}^{(0)} = \int d^p\zeta~\mathcal{H}_{min}^{(0)} = T_k L
\end{equation}

\noindent where $L$ is a large distance in one of the $p$ spatial coordinates.  In the dual gauge theory, this corresponds to the large distance, $L$, between quark-antiquark pairs, as in figure~\ref{fig:kstrings}, and we interpret $T_k$ as the $k$-string tension. 

The first quantum corrections are found by fluctuating around the classical solution
\begin{equation}\label{eq:fluctuate}
   X^\mu = X^\mu_{(0)} + \delta X^\mu,~~~A^{m} = A^{m}_{(0)} + \delta A^{m},~~~\Theta = 0 + \delta \Theta,
\end{equation}

\noindent expanding out the action to second order in these fluctuations
\begin{equation}\label{eq:expandaction}
  S_{p} = S_{p}^{(0)} + \delta S_{p}[\delta X, \partial \delta X, \partial \delta A, \Theta, \partial \Theta] + \mathcal{O}(\delta^3),
\end{equation}

\noindent and constructing a formula analogous to Eq.~(\ref{eq:DeltaEscalar})
\begin{align}\label{eq:deltaEgeneral}
     \delta E &= - \frac{1}{\beta}\int D \delta X \int D\delta A \int D \delta\overline{\Theta}D\delta\Theta e^{i\delta S_{p}}.
\end{align}

\noindent Through this procedure, we find the one loop corrections to the energy to be~\cite{PandoZayas:2008hw,Doran:2009pp,Stiffler:2009ma}
\begin{equation}\label{eq:Luscherterm}
  \delta E^{(d,p)} = -\frac{\pi(d + p - 3)}{24 L} + \beta_d,
\end{equation}

\noindent where $\beta_d$ is constant of $L$, and the first term is dual to the L\"uscher term for $k$-strings in the $d$ dimensional dual gauge theory.        

In the next few sections we will show explicit calculations of the $k$-string tension in SUGRA backgrounds dual to $3+1$ and $2+1$ $k$-strings.  In the next chapter, we will show explicit calculations of the one loop quantum corrections.  Let us first make a note on our method of demonstrating classical solutions of the D$p$-brane action, Eq.~(\ref{eq:DpbraneAction}).  In the rest of this chapter, we will simply use known, classical solutions from the literature~\cite{Herzog:2001fq,Herzog:2002ss,Firouzjahi:2006vp,Ridgway:2007vh,PandoZayas:2008hw,Doran:2009pp,Stiffler:2009ma}.  

As the rest of this chapter is with respect to classical field theories, we will suppress all classical subscripts and superscripts, $(0)$, for the remainder of this chapter.  This is not to be confused with the subscript or superscript 0 \emph{without} parenthesis, whose meaning should be obvious when it occurs (usually as a tensor index, or indicating the value of a function at a specific space-time point).
\section{Duality with the Klebanov-Strassler Background: \texorpdfstring{$k$}{k}-strings in 3+1}
In this section we will show the explicit calculation for $3+1$ $k$-string tensions using the supergravity dual theory of a D3-brane embedded in the type IIB supergravity background of Klebanov and Strassler(KS)~\cite{Klebanov:2000hb}.  We will first briefly review the KS background, then quickly move on to the tension calculation.  This section is a summary of work previously published in~\cite{PandoZayas:2008hw}.

\subsection{The Klebanov-Strassler Background}
In the Einstein frame, a type IIB supergravity source of $M$ D5-branes and $N$ D3-branes, characterized by the fields
\begin{align}\label{eq:KSsources}
 H_3 &= dB_2= \frac{g_s M\alpha'}{2} d[ f(\tau) g^1\wedge g^2 +k(\tau) g^3 \wedge g^4] \nonumber\\
 F_1 &= 0,~~~~\Phi = 0\nonumber\\
F_3 &= dC_2= \frac{M \alpha'}{2} \left\{ g^5 \wedge g^3 \wedge g^4 + d[F(\tau)(g^1 \wedge g^3 + g^2 \wedge g^4)] \right\}\nonumber\\
F_5 &= dC_4 = \frac{4 g_s M^2 \alpha'^2}{ \epsilon^{8/3}} \frac{l(\tau)}{K(\tau)^2 h(\tau)^2 \sinh^2\tau} dx^0 \wedge dx^1 \wedge dx^2 \wedge dx^3 \wedge d\tau
\end{align}

\noindent with
\begin{align}\label{eq:Bfield}
C_0 &= 0 \nonumber\\
B_2 &= \frac{g_s M \alpha'}{2} [f(\tau)g^1\wedge g^2 + k(\tau)g^3\wedge g^4]\nonumber\\
 C_2 &= \frac{M\alpha'}{4}[ 2 F(\tau)(g^1 \wedge g^3 + g^2\wedge g^4) + (\cos \psi \sin\theta_1 \sin\theta_2 -\cos\theta_1 \cos\theta_2)d\phi_1\wedge d\phi_2 \nonumber\\
&~~~~- \cos \psi d\theta_1\wedge d\theta_2
        +\psi(\sin\theta_1 d\theta_1\wedge d\phi_1- \sin\theta_2 d\theta_2\wedge d\phi_2)\nonumber\\
        &~~~~-\sin \psi \sin\theta_1 d\phi_1\wedge d\theta_2 + \sin \psi \sin\theta_2 d\phi_2\wedge d\theta_1] \nonumber\\
  C_4 &= \left(\frac{\epsilon^{8/3}}{2 g_s^3 M^2 \alpha'^2 3^{4/3}}\right)\tau^2~dx^0 \wedge dx^1 \wedge dx^2 \wedge dx^3 + \mathcal{O}(\tau^3).
\end{align}

\noindent results in a background known as the Klebanov-Strassler (KS) background~\cite{Klebanov:2000hb}:
\begin{align}\label{eq:KSmetric}
 ds_{KS}^2 &= h^{-1/2}(\tau)dx^a dx^b \eta_{ab} + h^{1/2}(\tau)ds_6^2,~~~a,b=0,1,2,3\nonumber\\
ds_6^2 &= \frac{1}{2}\epsilon^{4/3}K(\tau) \bigg[\frac{1}{3 K^3(\tau)}[d\tau^2 + (g^5)^2] + [(g^3)^2 + (g^4)^2]~\cosh^2\left(\frac{\tau}{2}\right) + \nonumber\\
         &~~~~~~~~~~~~~~~~~~~~~~+ [(g^1)^2 + (g^2)^2]~\sinh^2\left(\frac{\tau}{2}\right)\bigg]
\end{align}

\noindent where the six dimensional metric $ds_6^2$ is known as the deformed conifold.  The coordinate $\tau$ is the energy scale in the dual gauge theory, and the deformed conifold is equivalent to the conifold in the UV as shown in appendix~\ref{app:conifold}.

The one-forms, $g^i$, are
\begin{align}\label{eq:gi}
g^1 &=
\frac{1}{ \sqrt{2}}\big[- \sin\theta_1 d\phi_1  -\cos\psi\sin\theta_2 d\phi_2 + \sin\psi d\theta_2\big],\nonumber \\  
g^2 &= \frac{1}{\sqrt{2}}\big[ d\theta_1- \sin\psi\sin\theta_2 d\phi_2-\cos\psi d\theta_2\big], \nonumber \\  
g^3 &= \frac{1}{\sqrt{2}} \big[-\sin\theta_1 d\phi_1+\cos\psi\sin\theta_2 d\phi_2-\sin\psi d\theta_2\big],\nonumber \\  
g^4 &= \frac{1}{ \sqrt{2}} \big[ d\theta_1+\sin\psi\sin\theta_2 d\phi_2+\cos\psi d\theta_2 \big],   \nonumber\\  
g^5 &= d\psi + \cos\theta_1 d\phi_1+ \cos\theta_2 d\phi_2.
\end{align}

\noindent As we will be interested in the dual IR gauge theory, we expand the various functions in the KS background around $\tau = 0$:
\begin{align}\label{eq:KSfunctions}
  h(\tau) &= 2^{2/3}\epsilon^{-8/3}(g_s M \alpha')^2 \int_{\tau}^{\infty} dy \frac{y~\coth~y -1 }{\sinh^2 y} (\sinh(2y) - 2y)^{1/3}\nonumber\\
          &= h_0 - \frac{h_0}{2^{1/3}3^{4/3}I_0}\tau^2 + \dots \nonumber\\
K(\tau) &= \frac{(\sinh(2\tau) - 2\tau)^{1/3}}{2^{1/3}\sinh\tau} = K_0 - \frac{1}{5 \cdot 2^{2/3}3^{1/3}}\tau^2 + \dots\nonumber\\
f(\tau) &= \frac{\tau~\coth\tau - 1}{2\sinh\tau}(\cosh\tau - 1) = \frac{\tau^3}{12} + \dots\nonumber\\
k(\tau) &= \frac{\tau~\coth\tau - 1}{2\sinh\tau}(\cosh\tau - 1) = \frac{\tau}{3} + \frac{\tau^3}{180} + \dots \nonumber\\
F(\tau) &= \frac{\sinh\tau - \tau}{2\sinh\tau} = \frac{\tau^2}{12} + \dots, \nonumber\\
l(\tau) &= f(\tau)(1 - F(\tau)) + k(\tau)F(\tau) = \frac{\tau^3}{9} + \dots
\end{align}

\noindent with
\begin{align}\label{eq:h0K0}
  h_0 &= \left(\frac{2^{1/3}g_s M \alpha'}{\epsilon^{4/3}}\right)^2 I_0 \nonumber\\
I_0 &= \int_{0}^{\infty} dy \frac{y~\coth y -1 }{\sinh^2 y} (\sinh(2y) - 2y)^{1/3} \sim 0.71805 \nonumber\\
  K_0 &= \left(\frac{2}{3}\right)^{1/3}
\end{align}


\subsection{The D3-brane Hamiltonian and \texorpdfstring{$k$}{k}-string Tensions in 3+1}\label{KStension}
To calculate the $d=3+1$ $k$-string tension, we consider a solution of the classical action, Eq. (\ref{eq:Spclassical}) with $p=3$
\begin{align}\label{eq:D3classical}
  S_{3} &=  -\mu_3 \int d^{4}\zeta e^{-\Phi}\sqrt{\mathcal{M}} +\mu_3 \int \left(\frac{1}{2} C_0\wedge \mathcal{F}\wedge \mathcal{F}+ C_2 \wedge \mathcal{F} + C_4\right),
\end{align}

\noindent for an electrically charged probe D3-brane sitting at $\tau = 0$, with world volume parameters $\zeta = (t, x, \theta, \phi)$, and in temporal gauge, $A_t = 0$:
\begin{align}\label{eq:KSclassicalscalars}
   X &= (x^0, x^1, x^2, x^3, \theta_1, \theta_2, \phi_1, \phi_2, \psi, \tau)\nonumber\\
     &= (t, x, 0, 0, \theta, \theta, \phi, -\phi, \psi=\mbox{constant}, 0)&\mbox{scalar fields} \\
   \label{eq:KSclassicalvectors}
   F  &= F_{tx} dt \wedge dx = \dot{A}_x dt \wedge dx &\mbox{$U(1)$ gauge fields} \\
   \label{eq:KSclassicalfermions}
   \Theta &= 0&\mbox{fermion fields}
\end{align}

\noindent We expect this configuration to be dual to an $SU(M)$ $k$-string in $d=3+1$ in the IR. The constant $\psi$ will be determined via minimization of the Hamiltonian.

Plugging this solution into Eq.~\ref{eq:gi}, we calculate
\begin{align}\label{eq:gsimplified}
g^3 &= -\frac{1}{\sqrt{2}}[2\cos^2\frac{\psi}{2}~\sin\theta~d\phi + \sin\psi~d\theta], \nonumber\\
g^4 &= \frac{1}{\sqrt{2}}[2\cos^2\frac{\psi}{2} d\theta -\sin\psi~\sin\theta~d\phi], \nonumber\\
g^5 &= 0,
\end{align}

\noindent which along with the limiting behavior of the function in Eqs.~\ref{eq:KSfunctions}, are all that is necessary to calculate the pullbacks of the metric and the only non-vanishing field, $C_2$
\begin{align}\label{eq:KSginduced}
  ds^2_{D_3} &= g_{ab}d\zeta^a d\zeta^b = h_0^{-1/2}(-dt^2 + dx^2) + \frac{2}{R}(d\theta^2 + \sin^2\theta d\phi^2) \\
   \label{eq:KSC2induced}
   C_2 &= \frac{M\alpha'}{2}(\psi + \sin\psi)\sin\theta d\theta\wedge d\phi
\end{align}

\noindent where the scalar curvature for $g_{ab}$ is
\begin{align}
  \label{eq:KSscalarR}
  R &= g^{bd}R_{bd} \equiv g^{bd}R^a_{~bad} = \frac{2 \sec^2 \frac{\psi}{2}}{b g_s M \alpha'},~~~b = 2^{2/3} 3^{-1/3} I_0^{1/2} \approx 0.933 \\
  R^a_{~bcd} &= \Gamma^a_{~bd,c} - \Gamma^a_{~bc,d} + \Gamma^a_{ce}\Gamma^e_{~bd}-\Gamma^a_{~de}\Gamma^e_{~bc}.
\end{align}

\noindent As $B_{ab} = 0$ on the D3-brane for our solution, we have
\begin{align}
\mathcal{F}_2 &= B_2 + 2\pi\alpha' F\nonumber\\
              &= 2\pi\alpha' F_{tx}dt \wedge dx.  
\end{align}

On our way to calculating the dynamics for our classical probe D3-Brane with the D3-brane action, Eq.~(\ref{eq:DpbraneAction}) with $p=3$, it is first helpful to construct
\begin{align}
  \mathcal{M}_{ab} &=   g_{ab} + \mathcal{F}_{ab} =
        \left(\begin{array}{l l l l}
         -h_0^{-1/2} & 2 \pi\alpha' F_{tx} & 0 & 0  \\
         -2\pi\alpha' F_{tx} & h_0^{-1/2} & 0 & 0\\
         0 & 0 & \frac{2}{R}& 0\\
         0 & 0 & 0 & \frac{2}{R}\sin^2\theta
         \end{array}\right)\nonumber.
\end{align}

\noindent With this, the D3-brane action becomes
\begin{align}\label{eq:DBI3}
    S_3 &= -\frac{\mu_3}{g_s} \int d^{4}\zeta ~ \sqrt{\mathcal{M}} + \mu_3\int e^{\mathcal{F}}\wedge \sum_{q}C_q \nonumber\\
              &= \int d^{4}\zeta \Biggl[-\frac{\mu_3}{g_s}\sqrt{(h_0^{-1} - (2 \pi \alpha' F_{tx})^2)\left(\frac{4 \sin^2\theta}{R^2}\right)} + \nonumber\\
                 &~~~~~~~~~~~~~~~~~~~+ \mu_3 (2 \pi \alpha' F_{tx})\frac{M \alpha'}{2} \sin\theta(\psi + \sin\psi)\Bigg] \nonumber\\
              &= \int dtdx \mathcal{L},  
\end{align}

\noindent where in the last step, we have integrated out the trivial angular dependence:
\begin{align}\label{eq:KSLagrangian0}
   \mathcal{L} &= -2\pi\alpha' M \mu_3 h_0^{-1/2}\left(2 b\sqrt{1-E_x^2}\cos^2{\frac{\psi}{2}} - E_x (\psi + \sin\psi)\right) \\
   \label{eq:KSEx}
   E_x &= 2 \pi \alpha' \sqrt{h_0} F_{tx}.
\end{align}

We set the only component of the conjugate momentum of the gauge field $A^m$ to a constant value, $D$:
\begin{equation}\label{eq:KSD}
D = \frac{\partial \mathcal{L}}{\partial F_{tx}} = \frac{\partial \mathcal{L}}{\partial \dot{A}_{x}} = \mbox{constant},
\end{equation}

\noindent solve for $F_{tx} = \dot{A}_x$
\begin{equation}\label{eq:KSFtx}
F_{tx} = \dot{A}_x =  \frac{\frac{D\pi}{M}-\frac{1}{2}(\psi+\sin\psi)}
     {2 \pi \alpha'  h_0^{1/2}\sqrt{b^2\cos^4{\frac{\psi}{2}}+\left(\frac{D\pi}{M}-\frac{1}{2}(\psi+\sin{\psi})\right)^2}}, 
\end{equation} 

\noindent and apply the Legendre transformation, Eq.~(\ref{eq:LegTransform}), leaving us with the Hamiltonian density:
\begin{equation}\label{eq:KSHamiltonianDensity}
  \mathcal{H} = \frac{M}{2 \pi ^2 \alpha '
   h_0^{1/2}} \sqrt{b^2 \cos ^4\left(\frac{\psi }{2}\right)+\left(\frac{D\pi 
   }{M}-\frac{1}{2} (\psi +\sin{\psi })\right)^2}.
\end{equation}

Minimizing with respect to $\psi$ gives the condition for the classical value for $\psi = \psi_0$: 
\begin{align}\label{eq:psimincondition}
   \psi_0 - \frac{2 D \pi }{M} &= (b^2-1)\sin\psi_0
\end{align}

\noindent whereupon inserting this into Eq.~(\ref{eq:KSHamiltonianDensity}) gives the minimized, classical Hamiltonian density:
\begin{equation}\label{eq:Hamiltonian}
  \mathcal{H}_{min} = \frac{b M}{2 \pi ^2\alpha 'h_0^{1/2}} \sqrt{\cos ^2{\frac{\psi _0}{2}}
   \left(1+(b^2-1) \sin ^2{\frac{\psi
   _0}{2}}\right)}.
\end{equation}

Integrating this over the remaining spatial coordinate $x$ results in the minimized classical energy
\begin{align}\label{eq:hamprop}
  E &= \int dx \mathcal{H}_{min} \nonumber\\
    &= L \mathcal{H}_{min}
\end{align}

\noindent where $L$ is a large distance in the $x$-direction.  

Making the identification
\begin{equation}\label{eq:KSDintermsofk}
   D = k - \frac{M}{2}
\end{equation}

\noindent and comparing the classical, minimized D3-brane energy, Eq.~(\ref{eq:hamprop}), with the $k$-string energy, Eq.~(\ref{eq:kstringenergy}), we identity $\mathcal{H}$ with the $k$-string tension, $T_k$
\begin{align}\label{eq:KSkstringTension}
  \mathcal{H}_{min} &= T_k 
\end{align}

If we are to make the approximation $b \approx 1$, as in~\cite{Herzog:2001fq}, the constraint,  Eq.~(\ref{eq:psimincondition}), becomes
\begin{equation}
     \psi_0 \approx 2\frac{k\pi}{M} - \pi 
\end{equation}

\noindent and we acquire an approximate sine law for the $k$-string tension, Eq.~(\ref{eq:KSkstringTension})
\begin{equation}
T_k \sim b M \sin \frac{k \pi}{M}.
\end{equation}

If we are to solve for the tension exactly, we must solve the transcendental Eqs.~(\ref{eq:psimincondition}) and~(\ref{eq:KSkstringTension}).
Fig.~\ref{fig:KSkstringtension} compares this exact $k$-string tension to both the sine law and Casimir law for various values of $M$, including the large $M$ limit.  Notice the expected phenomenon of $k$-ality exhibited in all models: the $k$-string tension vanishes for $k=0$ and $k=M$. Also, notice how the Klebanov Strassler $k$-string tension always lies in between the Casimir and Sine Laws, and that it approaches both simultaneously for \emph{small} values of $M$.

Summarizing this analysis, we found the gauge/gravity correspondence to manifest itself as a dual description between SU($M$) $k$-strings and D3-branes endowed with  electric flux in the Klebanov-Strassler background.  Our calculation
found approximate agreement with the sine law and casimir laws for the $k$-string tension. 

\begin{figure}[htbp]
\addtocontents{lof}{\protect\vspace{\li}}
  \begin{center}$
  \begin{array}{cc}
\includegraphics[width=0.5\columnwidth]{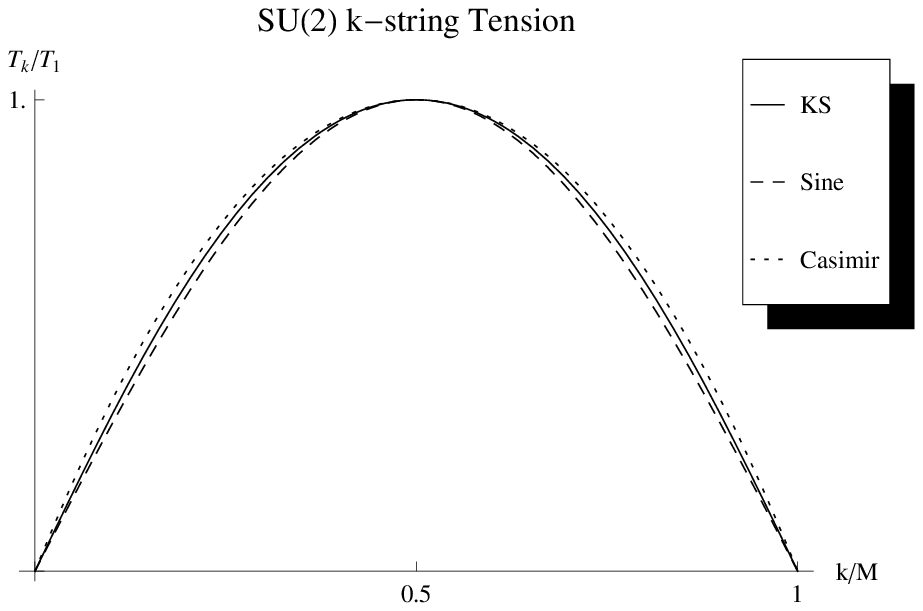}&
\includegraphics[width=0.5\columnwidth]{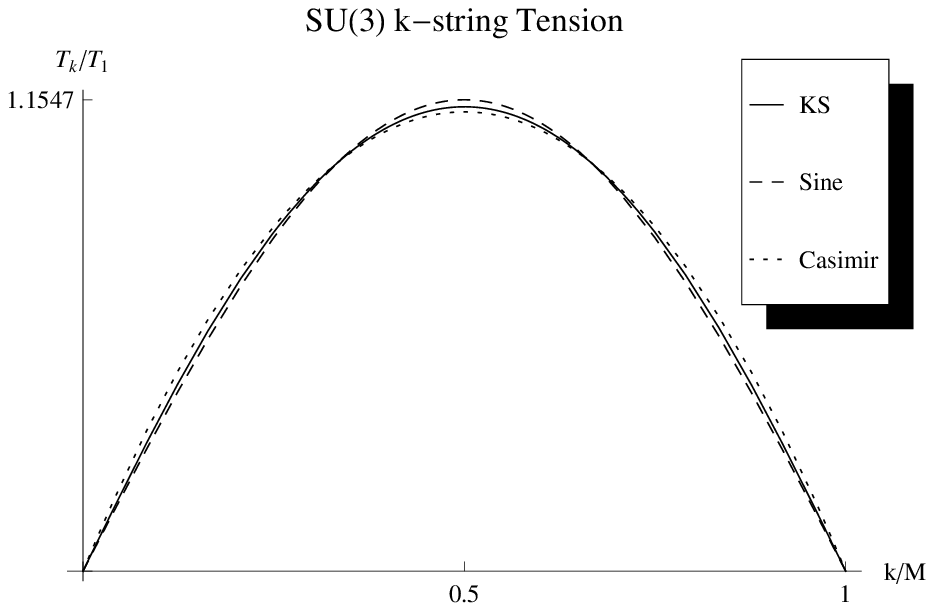}\\
 \includegraphics[width=0.5\columnwidth]{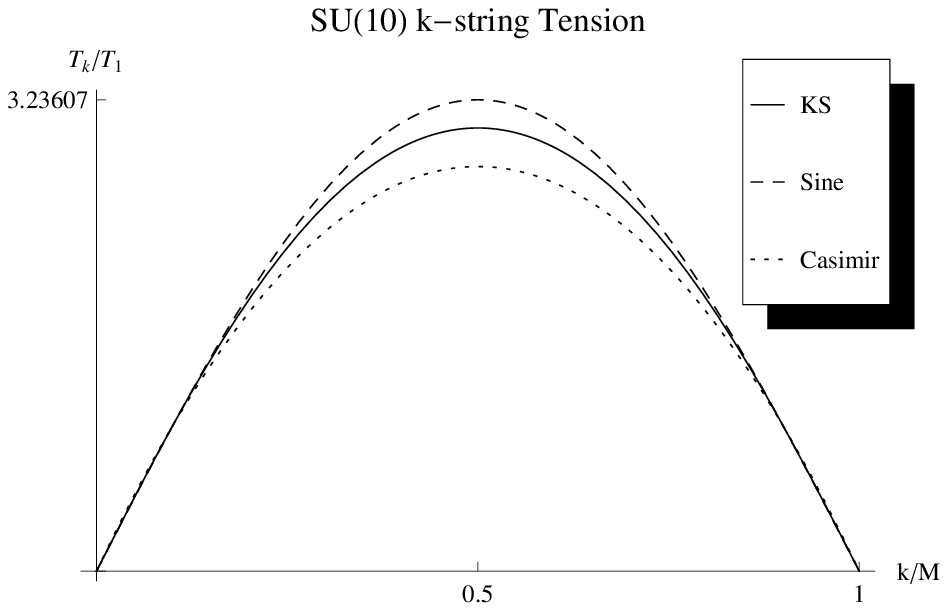}& \includegraphics[width=0.5\columnwidth]{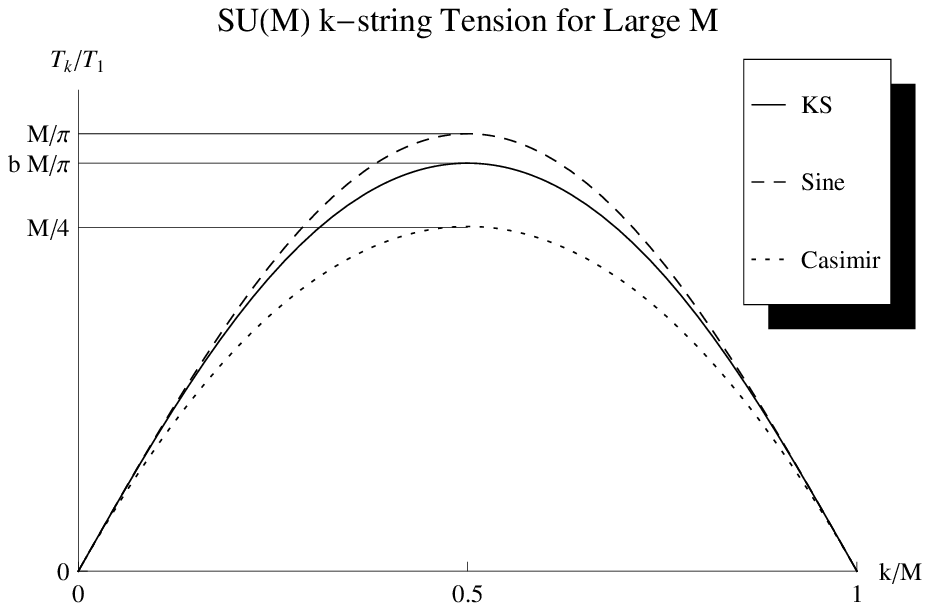}
 \end{array}$
 \end{center}
\caption{Exact solutions for the Klebanov Strassler (KS) $k$-string in units of the $k=1$ tension, compared to the sine and Casimir laws. The large $M$ plot was made for $M=300$.  Plots with $M >> 300$ look very nearly identical to this plot.  The peaks for the $k$-string tension in the large $M$ limit are denoted on the $y$-axis.} 
  \label{fig:KSkstringtension}
\end{figure}

\section{Duality with the Cvetic, Gibbons, L\"u, and Pope Background: \texorpdfstring{$k$}{k}-string Tension in \texorpdfstring{$2+1$}{2+1}}\label{CGLPtension}
The calculation in this section will parallel the calculation in the previous section.  After briefly reviewing the background of Cvetic, Gibbons, L\"u, and Pope (CGLP)~\cite{Cvetic:2001ma}, we will calculate the dual 2+1 $k$-string tension.  In $2+1$, there is more to compare with in the lattice and Hamiltonian communities.  We will see that the $k$-string tension calculated as a supergravity dual of a D4-brane in the CGLP background aligns with the tension of antisymmetric quark representations of $2+1$ $k$-strings in both the lattice and Hamiltonian communities. This section is a summary of work previously published in~\cite{Doran:2009pp}.
\subsection{Review of the CGLP Background}
In type IIA supergravity, a source of D2-branes and $N$ coincident fractional D2-branes with fluxes~\cite{Cvetic:2001ma,Herzog:2002ss,Cvetic:2000mh}
\begin{align}\label{eq:CGLPH3}
H_3 &= d B_2 = \frac{m}{l}a^2 u_1 h dr \wedge X_2 + \frac{m}{l}b^2 u_2 h dr \wedge J_2 + \frac{m}{l}a b^2 u_3X_3,~~~C_1 = 0 \\
\label{eq:CGLPF4}
F_4 &= d C_3 =  m g_s^{-1} \biggl[a b^2 u_3~\epsilon_{ijk}~\mu^i~h dr \wedge D\mu^j \wedge J^k + a^2b^2u_2 X_2 \wedge J_2 + \frac{1}{2}b^4 u_1 J_2 \wedge J_2\biggr] +\nonumber\\
&~~~~~~~~~~+g_s^{-1} dx^0 \wedge dx^1 \wedge dx^2 \wedge dH^{-1}.
\end{align}

\noindent results in the CGLP Einstein frame solution
\begin{align}\label{eq:CGLP}
  ds_{10}^2 &= H^{-5/8}dx^{\alpha}dx^{\beta}\eta_{\alpha\beta} + H^{3/8}ds_7^2,\\
  \label{eq:dilaton}
  e^{\Phi} &= g_s H^{1/4}
\end{align}

\noindent where $\eta_{\alpha\beta}$ is $\mathbb{R}^{1,2}$, and
\begin{align}\label{eq:metric7d}
  ds_7^2 &= l^2[h^2 dr^2 + a^2(D\mu^i)^2 + b^2 d\Omega_4^2], \\
  \label{eq:identities}
  X_2 &\equiv \frac{1}{2}\epsilon_{ijk}\mu^iD\mu^i\wedge D\mu^k,~~~J_2 \equiv \mu^i J^i,~~~ X_3 \equiv dX_2 = dJ_2
\end{align}

\noindent where the radial coordinate, $r=1$ to $\infty$.

In the above, $l$, $m$, and $g_s$ are constants, and $a,b,h,u_i$ and $H$ are functions of $r$~\cite{Cvetic:2001ma,Herzog:2002ss}.
\begin{align}\label{eq:hab}
  h^2 &= (1 - r^{-4})^{-1},~~~a^2 = \frac{1}{4}r^2(1 - r^{-4}),~~~b^2 = \frac{1}{2}r^2 \\
  u_1 &= r^{-4} + P(r)r^{-5}(r^4 -1)^{-1/2},~~~~u_2 = -\frac{1}{2}(r^4 - 1)^{-1} + P(r)r^{-1}(r^4 -1)^{-3/2}, \nonumber\\
  u_3 &= \frac{1}{4}r^{-4}(r^4-1)^{-1} - \frac{3r^4 -1}{4r^5(r^4 - 1)^{3/2}} P(r) \\
  P(r) &= \int_1^r \frac{d\rho}{\sqrt{\rho^4 - 1}} \\
  H(r) &= \frac{m^2}{2l^6}\int_r^\infty \rho (2 u_2(\rho) u_3(\rho) - 3 u_3(\rho))d\rho.
\end{align}

\noindent The parameter $l$ is similar to $\epsilon$ in the deformed conifold~\cite{Herzog:2002ss,Candelas:1989js,Klebanov:2000hb}.

The differential element $D\mu^i$ is
\begin{align}
  D\mu^i = d\mu^i + \epsilon_{ijk}A^j \mu^k
\end{align}

\noindent where the $\mu^i$ are coordinates on a unitless $\mathbb{R}^3$ constrained  to a unit $S^2$ surface, $\mu^i\mu^i = 1$.

The $A^i$ are $SU(2)$ Yang-Mills instanton one forms living on the $S^4$
\begin{align}\label{eq:Asu2}
  A^1 &= \cos\psi d\chi + \cos\theta d\phi \nonumber\\
  A^2 &= \cos\psi~\sin\chi d\theta -\cos\chi~\sin\theta d\phi \nonumber\\
  A^3 &= \cos\psi~\sin\chi~\sin\theta d\phi + \cos\chi d\theta
\end{align}
\noindent  and compose an anti-symmetric $SU(2)$ Yang-Mills two form, $J^i$,
\begin{align}\label{eq:Ji}
  J^i &= J^i_{~\overline{\alpha\beta}}\hat{e}^{\overline{\alpha}}\hat{e}^{\overline{\beta}} =  dA^i + \frac{1}{2}\epsilon^i_{~jk}A^j \wedge A^k, \\
  \hat{e}^{\overline{\alpha}} &= (d\psi,~\sin\psi d\chi,~\sin\psi\sin\chi d\theta,~\sin\psi \sin\chi\sin\theta d\phi)
\end{align}

\noindent which satisfies the algebra of the unit quaternions,
\begin{align}
J^{i~~\overline{\gamma}}_{~\overline{\alpha}}J^j_{~\overline{\gamma\beta}}&= -\delta^{ij}\delta_{\overline{\alpha\beta}} + \epsilon^{ij}_{~~k}J^k_{~\overline{\alpha\beta}},
\end{align}

\noindent whose solution in terms of Eqs.~(\ref{eq:Asu2}) and (\ref{eq:Ji}) is
\begin{align}
   J^1_{~12} &= J^1_{~34} = J^2_{~13} = J^2_{~42} = J^3_{~14} = J^3_{~23} = -1
\end{align}

We now solve Eqs.~(\ref{eq:CGLPH3}) and (\ref{eq:CGLPF4}) for $B_2$ and $C_3$.
In calculating $B_2$, the identities in Eq.~(\ref{eq:identities}) are very helpful.  Using these, we solve for $B_2$, up to a total derivative, to be
\begin{align}\label{eq:CGLPB2}
  lB_2 &= m\left(\int_1^r f_1(u)du\right)X_2 + m\left(\int_1^r f_2(r)du\right)J_2 \nonumber\\
       f_1(u) &= a^2(u) u_1(u)h(u) ,~~~f_2(u) = b^2(u)u_2(r)h(u).
\end{align}

\noindent Notice that this vanishes when $r=1$.  This will be important for our ensuing calculations as it is where we will position our probe D4-brane.
We choose the following solution for $C_3$:
\begin{align}\label{eq:CGLPC3}
   C_3 &= -\frac{x^2 dr\wedge dx^0 \wedge dx^1}{g_s H(r)^2} + \frac{3m}{8g_s}\xi(\psi)d\Omega_3 + \frac{m}{2 g_s}u_2(r)b(r)^2a(r)^2\epsilon_{ijk}\mu^iD\mu^j \wedge J^k\\
  \xi(\psi) &= \int_0^{\psi} \sin^3u~du,~~~d\Omega_3 \equiv \sin^2\chi~\sin\theta~d\chi\wedge d\theta \wedge d\phi,
\end{align}

\noindent which is the same as that chosen in~\cite{Herzog:2002ss}, up to a total derivative.

The constant $m$ is proportional to the number $N$ of stacked fractional D2-branes that the background describes, which is also the number $N$ of colors for the dual supersymmetric $SU(N)$ gauge theory~\cite{Herzog:2002ss}.  Using the Dirac quantization condition~\cite{Herzog:2002ss}
\begin{align}
   \int_{S^4} F_4 &= 8\pi^3 \alpha'^{3/2} N,
\end{align}

\noindent and the $r \to 1$ limiting behavior of $F_4$
\begin{align}
   F_4 &\to \frac{3}{8} d\Omega_4 + \frac{7 l^4}{16 g_s I_0^2}dx^0\wedge dx^1 \wedge dx^2 \wedge dr.
\end{align}

\noindent we can calculate the proportionality constant: $m=8\pi\alpha'^{3/2} g_s N$.

Also, we must mention that there is another CGLP solution. To acquire the other solution, $d\Omega_4^2$ can be substituted for a metric over $\mathbb{C~P}^2$. We will work only with the $4$-sphere, $d\Omega_4^2$.

\subsection{A Coordinate Transformation}
The CGLP metric, Eq.(\ref{eq:CGLP}), has a horizon at $r=1$.  When we apply the coordinate transformation
\begin{align}\label{eq:coordinatetransformation}
  \tau = \sqrt{r -1 }.
\end{align}

\noindent the horizon is now located at $\tau=0$ and the metric becomes
\begin{align}
  ds_7^2 &= l^2\left(f(\tau)^2d\tau^2 + a^2(D\mu^i)^2 + b^2 d\Omega_4^2 \right) \nonumber\\
   f(\tau) &= \frac{2 \tau}{\sqrt{1 - (1 + \tau^2)^{-4}}}.
\end{align}

\noindent Notice that $f(\tau)$ is finite as $\tau \to 0$.  Here are the expansions of all the aforementioned relevant functions in the $\tau \to 0$ limit:
\begin{align}
a(r(\tau)) &= \tau + O(\tau^3),~~~ b(r(\tau)) = \frac{1}{\sqrt{2}}(1 + \tau^2), \nonumber\\
u_1(r(\tau)) &= \frac{3}{2} - 7\tau^2 + O(\tau^4), ~~~u_2(r(\tau)) = -\frac{1}{4} + \frac{7}{10}\tau^2 + O(\tau^4),\nonumber\\
u_3(r(\tau)) &= -\frac{1}{4} + \frac{7}{5}\tau^2 + O(\tau^4),~~~f(\tau) = 1 + \frac{5}{4} \tau^2 + O(\tau^4),\nonumber\\
H(r(\tau)) &= H_0 - H_2\tau^2 + O(\tau^4)),~~~H_0 = \frac{m^2}{l^6}I_0,~~~H_2 = \frac{m^2}{l^6}\frac{7}{16}, \nonumber\\
I_0 &\equiv \int_1^\infty \rho (2 u_2(\rho) u_3(\rho) - 3 u_3(\rho))d\rho \approx 0.10693\dots
\end{align}

One can use these expansions to show that $B_2$ and $C_3$ become, under the coordinate transformations Eq.~(\ref{eq:coordinatetransformation}),
\begin{align}\label{eq:CGLPB2tau}
  B_2 &= -\frac{m}{8l}\tau J_2 + O(\tau^3) \\
  \label{eq:CGLPC3tau}
  C_3 &= -\frac{2 x^2 \tau d\tau\wedge dx^0\wedge dx^1}{g_s H(r(\tau))^2} + \frac{3m}{8 g_s}\left( \xi(\psi)d\Omega_3 - \frac{1}{6}\tau^2 \epsilon_{ijk}\mu^iD\mu^j\wedge J^k\right) + O(\tau^4).
\end{align}

\subsection{The D4-brane Hamiltonian and the 2+1 \texorpdfstring{$k$}{k}-string Tension}

We now outline the calculation of the CGLP $k$-string tension originally presented in~\cite{Herzog:2002ss}.  We use the version of the CGLP metric that was used there, which is conformally related to the Einstein Frame CGLP metric, Eq.~(\ref{eq:CGLP}), by $G_{\mu\nu} \to H_0^{1/8}G_{\mu\nu}$:
\begin{align}\label{eq:CGLPstringframe}
   ds^2 = H_0^{1/8}( H^{-3/8} dx_3^2 + H^{5/8} ds_7^2) 
\end{align}

\noindent which coincides with the string frame metric at $r=1$ $(\tau = 0)${\linespread{1.0}\footnote{The full string frame metric, $ds^2 = H^{-1/2}dx_3^2 + H^{1/2}ds_7^2$,  was used in~\cite{Doran:2009pp}}}.  

Considering the classical D4-brane action, Eq. (\ref{eq:Spclassical}) with $p=4$:
\begin{align}
\label{eq:D4classical}
S_{4} &=  -\mu_4 \int d^{5}\zeta e^{-\Phi}\sqrt{\mathcal{M}} +\mu_4 \int \left(\frac{1}{2} C_1\wedge \mathcal{F}\wedge \mathcal{F}+ C_3 \wedge \mathcal{F} \right),
\end{align}

\noindent we label the world volume coordinates of the probe D4-brane as
\begin{align}
   \zeta^a &= (t, x, \chi,\theta,\phi),
\end{align}

\noindent and investigate the classical solution for an electrically charged D4-brane sitting at the source $\tau = 0$, in temporal gauge $A_t = 0$:
\begin{align}\label{eq:CGLPclassicalscalars}
   X^{\mu} &= (x^0, x^1, x^2, \psi, \chi, \theta, \phi, \mu^1, \mu^2, \mu^3, \tau) \nonumber\\
&=(t, x, 0, \psi=\psi_0, \chi, \theta, \phi, 0,0,1, 0)&\mbox{scalar fields}\\
  \label{eq:CGLPclassicalvectors}
  F &= F_{tx} dt \wedge dx = \dot{A}_x dt \wedge dx &\mbox{$U(1)$ gauge fields}\\
 \label{eq:CGLPclassicalfermions}
\Theta &= 0 &\mbox{fermion fields}.
\end{align}

We again expect this to be dual to an IR $k$-string.  The 11 scalar fields, $X^\mu$, are really 10 independent scalar fields, as the $\mu^i$ fields are constrained to $(\mu^i)^2 = 1$.  Direct substitution of this solution into Eqs. (\ref{eq:CGLPB2tau}), (\ref{eq:CGLPC3tau}), and (\ref{eq:CGLPstringframe}) leads to the pullbacks to the D4-brane:
\begin{align}\label{eq:CGLPpullbacks}
  B_2 &= 0,~~~C_3 = C_3^{(0)} \equiv \frac{3m}{8g_s}\xi(\psi)d\Omega_3,\nonumber\\
ds_{D_4}^2 &= g_{ab}d\zeta^{a}d\zeta^{b} = H_0^{-1/2} (-dt^2 + dx^2) + \frac{6}{R}d\Omega_3^2
\end{align}

\noindent where the scalar curvature is
\begin{align}\label{eq:CGLPRscalar}
   R &= \frac{12}{H_0^{1/2}l^2}\csc^2\psi_0.
\end{align}

\noindent Since $B_2 = 0$, $\mathcal{F}$ becomes simply
\begin{align}\label{eq:CGLPcurlyF}
  H_0^{1/2} \mathcal{F} = H_0^{1/2}(2\pi\alpha' F) \equiv E_x~dt \wedge dx
\end{align}

Now we have all the pieces necessary to calculate the D4-brane action.  Proceeding as in the previous section, we simplify the D4-brane action, Eq. (\ref{eq:D4classical}), by integrating over the angular coordinates $\chi$, $\theta$, and $\phi$:
\begin{align}\label{eq:Lagrangian}
   S_4 &= \int dt \mathcal{L},\\
   \mathcal{L} &= -\alpha N \sqrt{1 - E_x^2}\sin^3\psi~ + q N E_x \xi(\psi) \nonumber\\
   \alpha &= \frac{l^3}{2\sqrt{2}\pi m\alpha'},~~~q = \frac{3}{2^{3/2} I_0^{1/2}}\alpha
\end{align}

\noindent where $L$ is the periodic length of the probe D4-brane's $x$-direction, which has been integrated out.  Because of the periodicity in the gauge field $F$, the conjugate momentum to $A_x$ is quantized to an integer $k$~\cite{Herzog:2002ss}:
\begin{align}
   2\pi\alpha' H_0^{1/2}\frac{\partial\mathcal{L}}{\partial E} = \frac{\partial\mathcal{L}}{\partial \dot{A_x}} = k.
\end{align}

Performing the Legendre transformation 
\begin{align}\label{eq:CGLPHamiltonianDensity}
   \mathcal{H} &= \frac{\partial\mathcal{L}}{\partial \dot{A_x}} \dot{A_x} - \mathcal{L}
\end{align}

\noindent we find the Hamiltonian density to be
\begin{align}
   \mathcal{H} &= \alpha N \sqrt{\sin^6\psi + \frac{q^2}{\alpha^2}\left(\frac{4 k}{3 N} - \xi\right)^2},
\end{align}

\noindent Minimization with respect to $\psi = \psi_0$ results in the condition
\begin{align}\label{eq:CGLPmincon}
  \frac{4 k}{3 N} = \xi(\psi_0) + 3\frac{\alpha^2}{q^2}\sin^2\psi_0\cos\psi_0
\end{align}
\noindent and the minimized Hamiltonian density
\begin{align}\label{eq:CGLPMinimumHamiltonianDensity}
   \mathcal{H}_{min} &= T_k = \alpha N \sin^2\psi_0 \sqrt{\sin^2\psi_0 + (3\alpha/q)^2\cos^2\psi_0}\nonumber\\
   \alpha/q &\approx 0.3083
\end{align}

\noindent where $T_k$ is the $k$-string tension.  Note that the parameter $k$ is once again interpreted as the parameter $k$ in $k$-strings.  The tension, Eq.~(\ref{eq:CGLPMinimumHamiltonianDensity}), and minimization condition, Eq.~(\ref{eq:CGLPmincon}), form a transcendental equation which can be solved numerically for given $k$ and $N$.

A similar calculation in the Maldacena-Nastase (MNa) background~\cite{Maldacena:2001pb}, leads one to a sine law~\cite{Liu:2010i}:
\begin{align}\label{eq:MNaTension}
T\sim N\sin \frac{\pi k}{N}.
\end{align}

\noindent Table~\ref{tab:comparetensions} compares the tension calculated from the CGLP background Eq.~(\ref{eq:CGLPMinimumHamiltonianDensity}) to the sine-law, Eq.~(\ref{eq:MNaTension}), Casimir law, and various results from the lattice calculations and Hamiltonian formulation.

Table~\ref{tab:comparetensions} shows that CGLP $k$-strings seem to be closely related to the anti-symmetric representation, as predicted, while at the same time, are closer to a Casimir law than a sine law.  We also see that MNa $k$-strings, which follow a sine law, seem to align better with the anti-symmetric representation than the symmetric representation.  Following the work of Gomis and Passerini~\cite{Gomis:2006im,Gomis:2006sb}, one can expect certain backgrounds to be dual to $k$-strings in particular representations.  Further tension calculations, in various backgrounds, are necessary to test this. 

\begin{table}[htbp]
\addtocontents{lot}{\protect\vspace{\li}}
\centering
\caption{Comparison of $T_k/T_f$ from various methods.}  
\label{tab:comparetensions}
\begin{tabular}{|c|c|c|c|c|c|c|}
\hline
$Group$ & $k$ &  CGLP & MNa(Sine) & Casimir & lattice & Karabali-Nair  \\
\hline\hline
\multirow{2}{*}{$SU(4)$} & \multirow{2}{*}{2}  & \multirow{2}{*}{1.310} & \multirow{2}{*}{1.414}  & \multirow{2}{*}{1.333} &  1.353(A) & 1.332(A)   \\
&&&&& 2.139(S) & 2.400(S)\\
\hline\hline
$SU(5)$ & 2 &  1.466 & 1.618 & 1.5 & 1.528* & 1.529*  \\
\hline\hline
\multirow{5}{*}{$SU(6)$} & \multirow{2}{*}{2} & \multirow{2}{*}{1.562} & \multirow{2}{*}{1.732} & \multirow{2}{*}{1.6} & 1.617(A) & 1.601(A) \\
\cline{6-7}
&&&&& 2.190(S) & 2.286(S)\\
\cline{2-7}
& \multirow{3}{*}{3} & \multirow{3}{*}{1.744} & \multirow{3}{*}{2.0} & \multirow{3}{*}{1.8} & 1.808(A) & 1.800(A) \\
\cline{6-7}
&&&&&3.721(S) & 3.859(S) \\
\cline{6-7}
&&&&&2.710(M) & 2.830(M)\\
\hline\hline
\multirow{3}{*}{$SU(8)$} & 2 & 1.674 & 1.848 & 1.714 & 1.752* & 1.741*\\
\cline{2-7}
&3 & 2.060 & 2.414 & 2.143 & 2.174* & 2.177*\\
\cline{2-7}
&4 & 2.194 & 2.613 & 2.286 & 2.366* & 2.322*\\
\hline 
\multicolumn{7}{p{0.8\columnwidth}}{Note: $T_k$ is $k$-string tension and $T_f$ is the fundamental string tension, i.e., $k=1$.  CGLP data is calculated from the transcendental Eqs.~(\ref{eq:CGLPMinimumHamiltonianDensity}) and (\ref{eq:CGLPmincon}).  Sine and Casimir data is calculated from Eq.~(\ref{eq:sinelawandCasimirlaw}).  Table from~\cite{Doran:2009pp}.}\\
\multicolumn{7}{l}{S = symmetric, calculated directly from~\cite{Karabali:2007mr}.} \\
\multicolumn{7}{l}{A = antisymmetric, calculated directly from~\cite{Karabali:2007mr}.}\\
\multicolumn{7}{l}{M = mixed, calculated directly from~\cite{Karabali:2007mr}.}\\
\multicolumn{7}{l}{* = antisymmetric, quoted directly from~\cite{Bringoltz:2008nd}.}
\end{tabular}
\end{table}


\Chapter{ONE LOOP CORRECTIONS TO THE K-STRING ENERGY \texorpdfstring{$\newline$}{~} VIA GAUGE/GRAVITY DUALITIES}\label{newresults}

We now investigate the one loop quantum corrections to the D$p$-brane energies found in Chapter~\ref{oldresults}.  We find these corrections to be dual to the L\"uscher term for $k$-strings in the IR of the corresponding gauge theory.  In $2+1$ space-time dimensions, we have lattice gauge theory data to compare with, where we find good agreement.

\section{Approximation Technique for the One Loop Energy}
Employing the same techniques as in~~\cite{Doran:2009pp,PandoZayas:2008hw,PandoZayas:2003yb,Bigazzi:2004ze,Bigazzi:2002gw}, we will now fluctuate around the classical solutions of the previous chapter and calculate the one loop corrections to the classical energy, defined in Eq. (\ref{eq:deltaEgeneral}).   As this is a field theory of scalar fields, vector fields, and fermions, let us first continue our analysis started in section~\ref{generalEformula} of the free energy of scalar field fluctuations, $\delta\varphi$, now in $p+1$ dimensions:
\begin{align}
    \delta S &= \frac{1}{2}\int d^{p+1}\zeta ((\nabla \delta\varphi)^2 + m^2 \delta\varphi^2)
\end{align}

\noindent whose equations of motion give
\begin{align}
   0 &= (-\nabla^2 + m^2)\delta\varphi \nonumber\\
     &= (\partial_t^2 - \partial_i^2 + m^2)\delta\varphi \nonumber\\
     &= (-\omega^2 + p_i^2 + m^2)\delta\varphi \nonumber\\
   \label{eq:EQMvarphi}
\Rightarrow \omega^2 &= p_i^2 + m^2. 
\end{align}

Now, moving to imaginary time, $t = i\tau$, and Wick rotating to a Euclidean action
\begin{align}
    \delta S_E &= \frac{1}{2}\int d\tau d^p\zeta~((\partial_\tau\delta\varphi)^2 + (\partial_i\delta\varphi)^2 + m^2 \delta\varphi^2 \nonumber\\
        &= \frac{1}{2}\int d\tau d^p\zeta~\delta\varphi (-\partial_\tau^2 - \partial_i^2 + m^2)\delta\varphi + \mbox{surface term}
\end{align}

\noindent we can calculate the free energy of the fluctuations:
\begin{align}
  \delta E  &= -\frac{1}{\beta}\log\int D\delta\varphi e^{-\delta S_E} \nonumber\\
     &= -\frac{1}{\beta}\log \left(\det (-\partial_\tau^2 - \partial_i^2 + m^2) \right)^{-1/2}.
\end{align} 

\noindent  Guided by our eventually goal of a gauge dual description of $k$-strings, with fixed quark sources, we impose vanishing boundary conditions~\cite{PandoZayas:2003yb,PandoZayas:2008hw,Doran:2009pp}: 
\begin{align}\label{eq:vanishingBCs}
\delta\varphi = \sin(n\pi \tau/\beta)\sin(n_1 \pi x^1/L_1) \cdots \sin(n_p \pi x^p/L_p),
\end{align}
\noindent where $p_i = n_i \pi/L_i$ with no $i$ sum, which allows us to perform the functional determinate:
\begin{align}
   \delta E &=\frac{1}{2\beta}\log \prod_{n,p_i}\left(\frac{\pi^2 n^2}{\beta^2} + p_i^2 +m^2\right) \nonumber\\
   &= \frac{1}{2\beta}\sum_{n,\omega} \log\left(\frac{\pi^2 n^2}{\beta^2} + \omega^2\right)
\end{align}

\noindent where we have identified $\omega^2 = p_i^2 +m^2$ from the equations of motion of the action, Eq. (\ref{eq:EQMvarphi}). Investigating large times $\beta$, we take the continuum limit, $\sum_n \to \int dn$:
\begin{align}
    \delta E &=  \frac{1}{2\beta}\sum_{\omega} \int dn \log\left( \frac{\pi^2 n^2}{\beta^2} + \omega \right) \nonumber\\
           &= \frac{1}{4\pi} \sum_{\omega} \sqrt{\omega} \int du~ u^{-1/2} \log(u\omega + \omega) \nonumber\\
           &= \frac{1}{4\pi} \sum_{\omega} \sqrt{\omega} \int du~ u^{-1/2} \left[ - \frac{\partial}{\partial v} (u \omega + \omega)^{-v} \right]_{v = 0}
\end{align}

\noindent where we have used 
\begin{align}
   \log x &= - \frac{\partial}{\partial v} x^{-v} |_{v = 0}.
\end{align}

\noindent Moving $\partial/\partial v$ out of the integral, we calculate:
\begin{align}
  \delta E        &= -\frac{1}{4\pi} \sum_{\omega} \sqrt{\omega} \left[ \frac{\partial}{\partial v} \int_0^{\infty} du~u^{-1/2} (u \omega + \omega)^{-v} \right]_{v=0} \nonumber\\
           &=  -\frac{1}{4\pi} \sum_{\omega} \sqrt{\omega} \left[\frac{\partial}{\partial v}\left(\omega^{-v}\int_0^\infty du \frac{u^{\frac{1}{2} - 1}}{(u+1)^{\frac{1}{2} + v - \frac{1}{2}}}\right)\right]_{v=0} \nonumber\\
           &\to -\frac{1}{4\pi} \sum_{\omega} \sqrt{\omega} \left[ \frac{\partial}{\partial v} \left(\omega^{-v} B\left(\frac{1}{2}, v - \frac{1}{2}\right)\right)\right]_{v=0}\nonumber\\
           &= -\frac{1}{4\pi} \sum_{\omega} \sqrt{\omega} (- 2\pi) \nonumber\\
           &= \frac{1}{2} \sum_{\omega}\omega  
\end{align}

\noindent where we have used the regularization procedure (see App.~\ref{app:AnalyticContinuation})
\begin{align}
   B(x,y) &= \int_0^\infty du \frac{u^{x - 1}}{(u+1)^{x+y}}, ~~~\mbox{Re}(x) >0,~\mbox{Re}(y)>0, \nonumber\\  
    &\to B(x,y) = \frac{\Gamma(x)\Gamma(y)}{\Gamma(x+y)}.
\end{align}

\noindent We see that the free energy of the scalar field fluctuations is given by the canonical formula for the free energy of a harmonic oscillator.

Motivated by this calculation, and inspired by \cite{PandoZayas:2003yb,Bertoldi:2004rn,Bigazzi:2004ze,Bigazzi:2002gw}, we define the energy of one loop corrections to the D-brane energy, Eq. (\ref{eq:deltaEgeneral}), as:
\begin{align}\label{eq:oneloopenergy}
   \delta E &\equiv \delta E_b + \delta E_f \nonumber\\
   & = \frac{1}{2}\sum \omega_b -\frac{1}{2}\sum \omega_f,
\end{align}

\noindent where the $\omega_b$'s ($\omega_f$'s) are the eigenmodes of the equations of motion of the fluctuations $\delta X$ and $\delta A$ ($\delta \Theta$):
\begin{equation}
   X^\mu = X^\mu_{(0)} + \delta X^\mu,~~~A^{m} = A^{m}_{(0)} + \delta A^{m},~~~\Theta = 0 + \delta \Theta, \nonumber\tag{\ref{eq:fluctuate}}
\end{equation}

\noindent derived from $\delta S_p$: the fluctuation of the Dp-brane action, Eq. (\ref{eq:DpbraneAction}), from its classical value, Eq. (\ref{eq:Spclassical}):
\begin{align}
  S_{p} &= S_{p}^{(0)} + \delta S_{p} + \mathcal{O}(\delta^3), \nonumber\tag{\ref{eq:expandaction}}
\end{align}

\noindent where $\delta S_p$ splits up into its bosonic and fermionic parts:
\begin{align}
   \label{eq:deltaSbPlusdeltaSf}
  \delta S_{p} &= \delta S_{p}^b[\delta X,\partial \delta X, \partial \delta A]  + \delta S_p^f[\delta\Theta,\partial\delta\Theta].
\end{align}

The topology of the Dp-branes we will investigate is $R^{1,1} \times S^{p-2}$, and in both cases we will find bosonic eigenmodes, derived from $\delta S_p^b$, of the form:
\begin{align}
   \omega &= \sqrt{p_x^2 +m^2 +f(p_x,\Omega_{p-2})}
\end{align}

\noindent where $m$ is the mass of the oscillation and $f(p_x,\Omega_{p-2})$ is a function of $p_x = n \pi/L$, the momentum along the spatial $R^{1,1}$ direction, and $\Omega_{p-2}$, the degrees of freedom associated with the $S^{p-2}$.  As we expect the propagator for massive modes to be exponentially suppressed by a factor $e^{-m L}$ , we do not expect massive modes to contribute to the L\"uscher term for large quark separation $L$~\cite{PandoZayas:2008hw,Doran:2009pp}.  Our calculation for the one loop bosonic energy will support this claim.  Based on this, we will assume that the fermionic one loop energy will not contribute to the L\"uscher term, as all fermionic eigenmodes, derived from $\delta S_p^f$, are all found to be massive.

We proceed to find the form of the bosonic and fermionic actions for the fluctuations, Eq. (\ref{eq:deltaSbPlusdeltaSf}), their resulting equations of motion, their eigenmodes, and finally, the energy of the fluctuations.  Section~\ref{deltaEKS} summarizes this calculation for a probe D3-brane ($p=3$) in the KS background, which is dual to a $d = 3+1$ $k$-string. Section~\ref{deltaECGLP} summarizes the calculation of a probe D4-brane ($p=4$) in the CGLP background, which is dual to a $d=2+1$ $k$-string.  The calculations are given explicitly in appendices~\ref{app:BosonicFluctuations} and~\ref{app:FermionicFluctuations}, both of which can be written succinctly in the formula:
\begin{align}
  \delta E{(d,p)} &= -\frac{\pi(d + p - 3)}{24 L} + \beta^d\nonumber\tag{\ref{eq:Luscherterm}}.
\end{align}

\noindent where $\beta^d$ is constant with respect to large quark separation $L$.  Putting this together with Eq.~(\ref{eq:DpbraneEnergy}), we find the total energy for a Dp-brane embedded in a supergravity background takes the form
\begin{align}
    E(k,d,p) &= T_k L + -\frac{\pi(d + p - 3)}{24 L} + \beta^d
\end{align} 

\noindent for large $L$.  This is the same form the energy for $k$-strings takes, Eq.~(\ref{eq:kstringenergy}), supporting our proposed correspondence, Eq.~(\ref{eq:gaugegravity}).  The next two sections summarize the one loop energy calculations from the explicit calculations given in App.~\ref{app:BosonicFluctuations}.  

\section{One Loop Energy of a D3-brane in the KS Background}\label{deltaEKS}We now discuss fluctuations of the classical D3-brane solution in the KS background.  We first discuss fluctuations of the bosons about the classical solution, Eqs. (\ref{eq:KSclassicalscalars}) and  (\ref{eq:KSclassicalvectors}), and then we discuss fluctuations of the fermions about the classical solution, Eq. (\ref{eq:KSclassicalfermions}).  This section is a summary of work previously published in~\cite{PandoZayas:2008hw}.

\subsection{Bosonic Fluctuations}\label{KSbf}
Keeping the D3-brane parametrization in Eq. (\ref{eq:KSclassicalscalars}) held fixed to:
\begin{equation}
 X^0 = t,~~~X^1 = x,~~~\theta_p \equiv \frac{1}{2}(\theta_1 + \theta_2) = \theta,~~~\phi_m \equiv \frac{1}{2}(\phi_1 - \phi_2) = \phi,
 \end{equation}

\noindent we fluctuate the remaining bosonic fields in the following way
\begin{align}\label{eq:KSbosonicfluctuations}
   \theta_m &\equiv \frac{1}{2}(\theta_1 - \theta_2) = \delta\theta_m,~~~\phi_p \equiv \frac{1}{2}(\phi_1 + \phi_2) = \delta\phi_p \nonumber\\
   X^2 &= \delta X^2,~~~X^3 = \delta X^3 \nonumber\\
  \psi &= \psi_0 + \delta\psi,~~~\tau = \tau_0 + \delta\tau,\nonumber\\
  F &= F_{tx} dt \wedge dx + \partial_a \delta A_b d\zeta^a \wedge d\zeta^b
\end{align}

\noindent leading to the bosonic part of the fluctuation of the D3-brane action

\begin{align}\label{eq:KS2ndorderaction}
  \delta S_3^b = -\int d^4\zeta\sqrt{g^{(\mbox{eff})}}&\Big{\{}c_X\sum_{i=2,3}\nabla^a \delta X^i\nabla_a \delta X^i  + c_A \left[\frac{1}{16\pi}\delta F^{ab}\delta F_{ab} + \delta A_{a}j^{a}\right] + \nonumber\\ \qquad &+c_{\tau}[\nabla^{a}\delta\tau\nabla_{a}\delta\tau + m_{\tau}^2\delta\tau^2 + \nabla^{a}\Psi\nabla_{a}\Psi - R\Psi^2] \\ & \qquad \qquad + \mbox{Total Derivatives}\nonumber\Big{\}}.
\end{align}

\noindent Here we notice that the linear fluctuations vanish, up to total derivatives, signifying that we are truly fluctuating around a classical solution.  The covariant derivative, $\nabla_{a}$, is with respect to an effective metric, $g^{(\mbox{eff})}_{ab}$, on the D3-brane
\begin{align}\label{eq:KSgeff}
 ds^2 &= g^{(\mbox{eff})}_{ab}d\zeta^{a}d\zeta^{b} = g_{xx}(-dt^2 + dx^2) + \frac{2}{R}(d\theta^2 + \sin^2\theta d\phi^2).
\end{align}  

The Euler-Lagrange equations for the bosonic fields derived from the action, Eq.~(\ref{eq:KS2ndorderaction}), take the form:
\begin{align}\label{eq:KSXeqm}
   &\nabla^2 \delta X^i = 0,~~~i = 2,3\\
   \label{eq:KStaueqm}
   &\nabla^2\delta\tau - m_{\tau}^2\delta\tau + \frac{c_A}{2 c_{\tau}}Q_{\tau}  \csc\theta \delta F_{\theta\phi} = 0 \\
   \label{eq:KSchap5Psieqm}
   &\nabla^2\Psi + R\Psi + \frac{c_A}{2c_{\tau}} Q_{\Psi} \delta F_{tx} = 0 \\
   \label{eq:KSAeqm}
   &\nabla^{a}\delta F_{ab} - 4\pi j_{b} = 0.
\end{align}

\noindent In Eq. (\ref{eq:KSchap5Psieqm}), we see the field $\Psi \equiv \delta\psi + 2\cos\theta \delta\phi_p$ is tachyonic.  This is remedied by the gauge fixing procedure for $\delta A_a$ described in App.~\ref{KSbosoniceigenvalues}.  After this and then solving the rest of the equations with Eqs.~(\ref{eq:solA}), (\ref{eq:solPsi}), (\ref{eq:soltau}), and (\ref{eq:solX}), the problem reduces to the eigenvalue problem
 
\begin{align}\label{eq:KSbeigenproblem}
\mbox{{\scriptsize $\omega^2 \left(\begin{array}{l}
             \tilde{\Psi}\\
             \tilde{X^2}\\
             \tilde{X^3}\\
             \tilde{\tau}\\
             \tilde{A_{\theta}} \\
             \tilde{A_{\phi}}
          \end{array}
   \right)$}} &= \mbox{\scriptsize{$
                      \left(
                      \begin{array}{c c c c c c}
                      \omega_1^2 & 0 & 0 & 0 & 0 & 0\\
                      0 & \omega_2^2 & 0 & 0 & 0 & 0\\
                      0 & 0 & \omega_2^2 & 0 & 0 & 0\\
                      0 & 0 & 0 & \omega_2^2 + g_{xx}m_{\tau}^2 & -g_{xx}Q_\tau\frac{c_A}{2c_\tau}\sqrt{l(l+1)} & 0\\
                      0 & 0 & 0 & -g_{xx}Q_\tau\frac{8\pi}{R}\sqrt{l(l+1)} & \omega_2^2 & 0 \\
                      0 & 0 & 0 & -g_{xx}Q_\tau\frac{8\pi}{R}\sqrt{l(l+1)} & 0 & \omega_2^2
                      \end{array}
                      \right)$}}
   \mbox{{\scriptsize $
            \left(\begin{array}{l}
             \tilde{\Psi}\\
             \tilde{X^2}\\
             \tilde{X^3}\\
             \tilde{\tau}\\
             \tilde{A_{\theta}}\\
             \tilde{A_{\phi}}
          \end{array}
   \right)$}}
\end{align}

\noindent where
\[
  \omega_1^2 = p_x^2 + g_{xx}m_{\Psi}^2 \qquad \mbox{and} \qquad
  \omega_2^2 = p_x^2 + g_{xx}\frac{R}{2}l(l+1).
\]

\noindent  The six eigenvalues of Eq.~(\ref{eq:KSbeigenproblem}) are
\begin{align}
  \omega^2 &= \left\{ \begin{array}{l}
                       p_x^2 + g_{xx}m_{\Psi}^2 \\
                       p_x^2 + g_{xx}\frac{R}{2}l(l+1)~\mbox{3-fold degenerate} \\
                       p_x^2 + \mu_{\pm}^{2}(l,\psi_0)
                       \end{array}
              \right.
\end{align}

\noindent where
\begin{align}
  \mu_{\pm}^{2}(l,\psi_0) &= g_{xx}\frac{R}{2}l(l+1)(1 + f_{\pm}(l,\psi_0)\ge 0, \\
   f_{\pm}(l,\psi_0) &= \frac{f_1(\psi_0)}{l (l+1)}\left(1 \pm \sqrt{1 + \frac{f_2(\psi_0)}{f_1^2(\psi_0)}l(l+1)}\right).
\end{align}

\noindent  The effective mass $\mu_\pm^2(l,\psi_0)$ is greater than zero except in the case $\mu_-^2(l=0,\psi_0) = 0$.

The energy of these bosonic fluctuations is calculated in App.~\ref{KSBosonicOneLoopEnergy} to be
\begin{align}\label{eq:KSboneloop}  
\delta E_b    &= -\frac{(d+p-3)\pi}{24L} + \beta^3_b
\end{align}

\noindent where for the present case of a D3-brane in the KS background, $p=3$ and $d=4$.   The term $-\frac{(d+p-3)\pi}{24L}$ is due to the massless modes, and we identify it with the L\"uscher term for a $k$-string in $d=3+1$.  The function $\beta^3_b = \beta^3_b(k,M)$, given by Eq.~(\ref{eq:KSbeta3}), is due to the massive modes and is constant of large quark separation $L$.
\subsection{Fermionic Fluctuations}\label{KSff}
Fluctuating around the classical solution $\Theta = 0 + \delta\Theta$ of the probe D3-brane, the action for fermionic fluctuations in Eq. (\ref{eq:deltaSbPlusdeltaSf}) becomes:
\begin{align}
  \delta S_3^f &= \frac{\mu_p}{2g_s} \int d^4\zeta \sqrt{\mathcal{M}} \delta\overline{\Theta} [(\mathcal{M}^{-1})^{ab}\Gamma_{a}\partial_{b} + M_1 + M_2 + M_3 ] \delta\Theta
\end{align}

\noindent which easily gives the Euler-Lagrange equations:
\begin{align}\label{eq:KSchap5Diraceq}
  [(\mathcal{M}^{-1})^{ab}\Gamma_{a}\partial_{b} + M_1 + M_2 + M_3 ] \delta\Theta = 0.
\end{align}

\noindent The mass matrices, $M_1$, $M_2$, and $M_3$, are found in Eqs.~(\ref{eq:KSM1}), (\ref{eq:KSM2}), and (\ref{eq:KSM3}), and we see that all the fermionic fields are massive.  In App.~\ref{KSfermionsexplicit} we show how Eq. (\ref{eq:KSchap5Diraceq}) can be simplified to the eigenvalue problem
\begin{align}
   \omega \tilde{\Theta}_1 = \mathcal{H}^{(f)}_1 \tilde{\Theta}_1 \\
   \omega \tilde{\Theta}_2 = \mathcal{H}^{(f)}_2 \tilde{\Theta}_2
\end{align}

\noindent where $\tilde{\Theta}_i$ are eight component spinors acted on by the $8 \times 8$ matrices $\mathcal{H}^{(f)}_i$, Eqs.~(\ref{eq:KSHf1final}) and~(\ref{eq:KSHf2final}), which have the same, eight massive eigenmodes
\begin{align}
   \omega &= \left\{\begin{array}{l}
                    \pm\sqrt{c_{10}(p,l) + \sqrt{c_8(p,l)} \pm \sqrt{c_{9+}}(p,l)} \\
                    \pm\sqrt{c_{10}(p,l) - \sqrt{c_8(p,l)} \pm \sqrt{c_{9-}}(p,l)}
                    \end{array}
                    \right.
\end{align} 

Regularization of these eigenmodes proves to be quite a monumental task.  As they are massive, we do not expect them to contribute to the L\"uscher term as the propagators for modes of mass $m$ go as $e^{-mL}$~\cite{PandoZayas:2008hw,Doran:2009pp}.   In fact, we found in the detailed analysis of section \ref{KSbf} that the massive bosons in the KS background contributed only a constant energy at large $L$.  From this evidence, we estimate that these fermions, upon regularization, will contribute such a constant to the KS one loop energy.  We find then the total one loop energy of D3-brane in the KS background to be
\begin{align}\label{eq:OneLoopFinal}
     \delta E(d,p) &= -\frac{\pi(d + p - 3)}{24 L} + \beta^d
\end{align}

\noindent with $d=4$ and $p=3$, and where $\beta^d$ is the sum of the bosonic constant energy, given by Eq.~(\ref{eq:KSbeta3}), and the contribution of the fermions.
\section{One Loop Energy of a D4-brane in the CGLP Background}\label{deltaECGLP}
In a parallel calculation to the one in the previous section, we now discuss fluctuations of the classical D4-brane solution in the CGLP background.  We first discuss fluctuations of the bosons about the classical solution, Eqs. (\ref{eq:CGLPclassicalscalars}) and  (\ref{eq:CGLPclassicalvectors}), and then we discuss fluctuations of the fermions about the classical solution, Eq. (\ref{eq:CGLPclassicalfermions}).  This section is a summary of work previously published in~\cite{Doran:2009pp}.

\subsection{Bosonic Fluctuations}\label{CGLPbf}
As shown explicitly in App.~\ref{app:CGLPfluctuations}, the fluctuations
\begin{align}
   x^2(\zeta) &= 0 + \delta x^2(\zeta),~~~\psi(\zeta) = \psi_0 + \delta\psi(\zeta), \nonumber\\
      \tau(\zeta) &= \tau_0 + \delta\tau(\zeta),~~~\mu^i(\zeta) = \mu^i_0 + \delta\mu^i(\zeta), \nonumber\\
  F &= \frac{E}{2\pi\alpha' H_0^{1/2}}dt\wedge dx + \partial_a \delta A_b d\zeta^a \wedge d\zeta^b
\end{align}
\noindent lead to the action for bosonic fluctuations of a D4-brane in the CGLP background:
\begin{align}\label{eq:CGLPfullbosonicaction}
  \delta S_4^b &= -\int \sqrt{-\det(g^{(eff)})}d^5\zeta\biggl\{c_x \nabla_a\delta x^2\nabla^a \delta x^2 + c_{\psi} \left[\nabla_a\delta\psi\nabla^a \delta\psi - \frac{R}{2} \delta\psi^2\right] \biggr. + \nonumber\\
            &~~~+ c_{\tau}\left[\nabla_a\delta\tau\nabla^a \delta\tau + m_{\tau}^2(\chi,\theta)\delta\tau^2\right]+ c_A\left[\frac{1}{16 \pi} \delta F^{ab} \delta F_{ab} + j^a \delta A_a \right] + \nonumber\\  &~~~ \biggl. + \mbox{total derivatives} \biggr\},
\end{align}

\noindent noticing that as in the KS calculation, the linear fluctuations vanish, up to total derivatives, signifying that we are truly fluctuating around a classical solution.  The covariant derivative, $\nabla_{a}$, is with respect to an effective metric, $g^{(\mbox{eff})}_{ab}$, on the D4-brane
\begin{align}
   ds^2 &= g^{(eff)}_{ab} d\zeta^a d\zeta^b = \frac{1}{g_{xx}}(-dt^2 + dx^2) + \frac{6}{R}d\Omega_3^2, 
\end{align}
\noindent where $R$ is the same scalar curvature, Eq.~(\ref{eq:CGLPRscalar}), as the induced metric. 

The equations of motion of the action, Eq.~(\ref{eq:CGLPfullbosonicaction}), are quite difficult to solve.  To simplify the problem, we recall the important physical features that were found for a D3-brane in the KS background in section~\ref{KSBosonicOneLoopEnergy}~\cite{PandoZayas:2008hw}.  There we found the massless modes, from which the L\"uscher term was derived, to be independent of the angular degrees of freedom.  Inspired by these results, we propose that in the current case of a D4-brane in the CGLP background, we can integrate out the spherical degrees of freedom, $\chi$, $\theta$, and $\phi$, and still have the same number of massless modes as before and as a result, the same L\"uscher term as would be calculated from the full five dimensional theory.

To proceed with this integration, we consider the fluctuations to be independent of the $S^3$ variables,
\begin{align}
   \delta X^{\mu} &= \delta X^{\mu}(t,x),~~~\delta A_{a} = \delta A_{a}(t,x)
\end{align}

\noindent and we integrate out the $S^3$ from the action, Eq. (\ref{eq:CGLPfullbosonicaction}). This results in an effective action
\begin{align}\label{eq:CGLPS2bosoniceff}
   \delta S^b_{4eff} &= - V_3 \int dt~dx \biggl\{c_x \nabla_m\delta x^2\nabla^m \delta x^2 + c_{\psi} \left[\nabla_m\delta\psi\nabla^m \delta\psi - \frac{R}{2} g_{xx} \delta\psi^2\right] \biggr. + \nonumber\\
            &~~~+ c_{\tau}\left[\nabla_m\delta\tau\nabla^m \delta\tau + m_{\tau e}^2\delta\tau^2\right]+ c_A\left[\frac{1}{g_{xx}16 \pi} \delta F^{mn} \delta F_{mn} + g_{xx} j^m \delta A_m \right] + \nonumber\\
            &~~~ + \frac{c_A}{16 \pi}(\nabla_m \delta A_{\chi}\nabla^m \delta A_{\chi} + 2\nabla_m \delta A_{\theta}\nabla^m \delta A_{\theta} + I_1 \nabla_m \delta A_{\phi}\nabla^m \delta A_{\phi}) + \nonumber\\
            &~~~ + \biggl.\mbox{total derivatives} \biggr\},
\end{align}

\noindent whose equations of motion are
\begin{align}
   &(-\partial_t^2 + \partial_x^2) \delta x^2 = 0 \\
   &(-\partial_t^2 + \partial_x^2) \delta \psi + \frac{R}{2}g_{xx} \delta\psi + \frac{c_A g_{xx} Q_{\psi}}{2 c_{\psi}} \delta F_{tx} = 0 \\
   &(-\partial_t^2 + \partial_x^2)\delta\tau - m_{\tau e}^2 \delta \tau = 0 \\
   &(-\partial_t^2 + \partial_x^2) \delta A_i = 0,~~~i = \chi,\theta,\phi \\
   \label{eq:CGLPFmn}
   &\partial_m \delta F^{mn} = 4\pi g_{xx}^2 j^n
\end{align}

\noindent which have the following eigenmodes
\begin{align}\label{eq:CGLPbosonomegas}
  \omega^2 &= \left\{\begin{array}{l}
                 p^2~~~\mbox{4 fold degenerate}  \\
                 p^2 + m_{\tau e}^2 \\
                 p^2 + m_{\psi}^2
                 \end{array}
              \right..
\end{align}

The calculation for the one loop correction to the bosonic $k$-string energy, $\delta E_b$, follows similarly to the KS calculation in App.~(\ref{KSBosonicOneLoopEnergy}); the result for large quark separation $L$ being
\begin{align}\label{eq:CGLPbosononeloop}
  \delta E_b &= -\frac{(d+p-3)\pi}{24 L} - \frac{1}{4}(m_{\tau e} + m_{\psi})
\end{align}

\noindent  where $p=4$, $d=3$, and the L\"uscher term is $\frac{(d+p-3)\pi}{24L} = -\frac{\pi}{6L}$, which is the same as the expected value, as $N$ increases, for lattice calculations done in~\cite{Bringoltz:2008nd}.  We see that the L\"uscher term is composed of $d-2 =1$ massless mode, $\delta x^2$, and $p-1 = 3$ massless modes, $\delta A_i$.  This results in the same formula for $d$ and $p$ that we found in the KS case, Eq. (\ref{eq:KSboneloop}).
\subsection{Fermionic Fluctuations}\label{CGLPff}
As in the previous section, we investigate $S^3$ independent solutions for the fluctuations about the classical solution
\begin{align}
   \Theta = 0 + \delta\Theta(t,x)
\end{align}

\noindent which after integrating out the spherical degrees of freedom results is the effective action for fermionic fluctuations:
\begin{align}
  \delta S^{f}_{eff} \propto \int dt dx \delta\overline{\Theta}\Gamma_{D_4}'((\mathcal{M}^{-1})^{mn}\Gamma_m\partial_n + M_f) \delta\Theta,~~~ m,n = t,x. 
\end{align}

\noindent where the mass matrix, $M_f$, is given by Eq.~(\ref{eq:CGLPMf}).  The Euler equations of this effective action can be solved by Fourier transform;
\begin{align}\
  \Gamma_{D_4}'(i(\mathcal{M}^{-1})^{mn}\Gamma_m p_n + M_f) \delta\Theta=0,~~~ m,n = t,x~~~
  p_t& = -\omega,~~~p_x = p,
\end{align}

\noindent and, after using the constraint, Eq. (\ref{eq:Thetaconstraint}), can be reorganized into sixteen equations
\begin{align}
  \omega \delta\Theta &= H_f \delta\Theta
\end{align}

\noindent where $H_f$ is the $16 \times 16$ matrix in Eq.~(\ref{eq:CGLPHf}), with massive eigenmodes:
\begin{align}\label{eq:CGLPwf1}
  \omega &= \pm\sqrt{p^2 + \alpha_{1} \pm \alpha_{2}} \\
  \label{eq:CGLPwf2}
  \omega &= \pm\sqrt{p^2 + \alpha_{3} \pm \alpha_{4}} \\
  \label{eq:CGLPwf3}
   \omega &= \left\{\begin{array}{l}
                    \pm\sqrt{\alpha_{7}(p) + \alpha_{5}(p) \pm \alpha^{+}_{6}(p)} \\
                    \pm\sqrt{\alpha_{7}(p) - \alpha_{5}(p) \pm \alpha^{-}_{6}(p)}
                    \end{array}
                    \right.,
\end{align}

From the regularization procedure used in App.~\ref{MassiveModes}, we see that the first two sets of fermionic eigenmodes, eq.~(\ref{eq:CGLPwf1}) and eq.~(\ref{eq:CGLPwf2}), will contribute a constant to the fermionic energy.  The remaining eigenmodes, Eq.~(\ref{eq:CGLPwf3}), prove to be very difficult to regulate from their very complicated $p$-dependence.  We assume that they will not contribute to the L\"uscher term, as they are massive, and the propagators will go as $e^{-mL}$, $m$ being the mass of the eigenmode and $L$ the large quark separation. This leaves us with the same, succinct formula for the one loop corrections to the D4-brane energy in the CGLP background, as was found in the KS case, Eq.~({\ref{eq:OneLoopFinal}), where now $d=3$, $p=4$, and $\beta^d = \beta^4$ includes both bosonic and fermionic contributions, independent of large quark separation, $L$.

\Conclusion
\vskip -\li
Superstring theory allows for an attempt at unifying gravity  with the gauge forces, strong and electroweak.  Gauge/gravity dualities from superstring theory have been found to be useful tools to make gauge theory calculations, most notably applicable in the low energy regime.  It is interesting to see how close gauge theory calculations from these gauge/gravity dualities can come to more direct methods of gauge theory calculations.  In this thesis, specific gauge theory objects, known as $k$-strings were investigated. These are colorless combinations of strongly coupled quark-antiquark pairs which give rise to flux tubes of gauge flux.  The most common calculation of this configuration found in the literature is the energy, which consists of a tension term, $T_k L$, and a Coulombic $\alpha/L$ correction, where $L$ is the length between quark-antiquark pairs.  Lattice gauge theory, direct Hamiltonian analysis, and string theory using gauge/gravity dualities are methods used to calculate the energy of $k$-strings. 

  This thesis reviewed string theory, how gauge/gravity dualities emerge from string theory, and how they can be used to calculate the $k$-string tension and make a direct comparison with lattice gauge theory and Hamiltonian results.  Specifically, the objects dual to $k$-strings are D$p$-branes embedded in confining supergravity backgrounds from low energy superstring theories.  Branes embedded into two different backgrounds were investigated in detail: a D$3$-brane embedded in the Klebanov Strassler (KS) background, which a dual to a $3+1$ $k$-string, and a D$4$-brane embedded in the Cvetic, Gibbons, L\"u, and Pope (CGLP) background, which is dual to a $2+1$ $k$-string.

In the KS case, the tension term was found to interpolate nicely between the competing models in lattice gauge theory and Hamiltonian methods: the sine law and Casimir law. We were able to briefly touch on the problem of quark representations in the CGLP case.  We found, by direct comparison to tension calculations in lattice gauge theory and Hamiltonian methods, that a D$4$-brane in the CGLP background was more likely dual to an anti-symmetric $k$-string representation. Both the tensions found in the KS and CGLP cases were of classical calculations on the supergravity side.

The main result of this thesis was to go beyond these classical energy calculations on the string theory side, and calculate the one loop quantum corrections, which were found to be dual to the L\"uscher term on the gauge theory side.    A succinct formula for the L\"uscher term was found in both cases, KS and CGLP, which in the $2+1$ case reproduced the expected value, as $N$ increases, from a lattice gauge theory calculation.


\appendix


\Appendix{REGULARIZATION OF INFINITE SUMS}
\label{app:AnalyticContinuation}

\section{The Riemann Zeta Function}
The Riemann zeta Function is defined as ~\cite{Elizalde:1995a}
\begin{align}\label{eq:ZetaDefSGreater1}
   \zeta(s) &= \sum_{n=1}^{\infty} n^{-s}~~~~~~\mbox{Re}(s) > 1,
\end{align}

\noindent where Re$(s)$ stand for the real part of $s$.  Through analytic continuation, one can define the $\zeta$-function to have finite values for all Re$(s) \ne 1$.  An example, particularly useful for Eq.~(\ref{eq:abdefinition}), is
\begin{align}
  \zeta(-1) &= -\frac{1}{12}
\end{align}

\noindent which we can use to regulate the associated infinite sum by replacing it with the zeta function:
\begin{align}\label{eq:SumofAllPositiveIntegersRegularization}
   \sum_{n=1}^{\infty} n \to \zeta(-1) = -\frac{1}{12},
\end{align}

Let us proceed to explicitly calculate the analytic continuation of the Riemann zeta function to values for Re$(s) < 1$.
Consider~\cite{z1}
\begin{align}
   \Gamma(s)\zeta(s) &= \int_0^{\infty} dt~t^{s-1} e^{-t} \sum_{n=1}^{\infty} n^{-s} \nonumber\\
                     &= \int_0^{\infty}  \sum_{n=1}^{\infty} dt~t^{s-1} e^{-t} n^{-s}~~~~~\mbox{Re}(s) > 1,
\end{align}

\noindent apply the coordinate shift $t \to nt$, and reorganize the result:
\begin{align}\label{eq:GammaZeta}
 \Gamma(s)\zeta(s) &= \int_0^{\infty} dt~t^{s-1} \sum_{n=1}^{\infty} e^{-nt} \nonumber\\
                    &= \int_0^{\infty} dt~t^{s-1} \left( \sum_{n=0}^{\infty} e^{-nt}  - 1\right) \nonumber\\
                    &= \int_0^{\infty} dt~t^{s-1} \left( \frac{1}{1 - e^{-t}} - 1 \right) \nonumber\\
                    &= \int_0^{\infty} dt \frac{t^{s-1}}{e^{t} - 1}
\end{align}

\noindent This leads us to the integral representation of the Riemann zeta function~\cite{Elizalde:1995a}:
\begin{align}\label{eq:ZetaIntegralRepSGreater1}
   \zeta(s) &= \frac{1}{\Gamma(s)}\int_0^{\infty} dt \frac{t^{s-1}}{e^{t} - 1}
\end{align}

We wish to pull out the pole structure of Eq.~(\ref{eq:GammaZeta}).  As it will soon be shown, the poles in this integral come from the interval $0 < t < 1$ and so we first break up the integral
\begin{align}\label{eq:GammaZeta1}
 \Gamma{(s)}\zeta{(s)} &= \int_0^{1} dt \frac{t^{s-1}}{e^{t} - 1} + \int_1^{\infty} dt \frac{t^{s-1}}{e^{t} - 1}
\end{align}

\noindent Looking at the power expansion
\begin{align}\label{eq:expansion}
   \frac{1}{e^t - 1} &= \frac{1}{t} - \frac{1}{2} + \frac{t}{12} - \frac{t^3}{120} + \dots,
\end{align}

\noindent we add and subtract these first few terms from the left integrand in Eq.~(\ref{eq:GammaZeta1}).

\begin{align}\label{eq:GammaZeta2}
\Gamma(s)\zeta(s) &= \int_0^1 dt~t^{s-1}\left(\frac{1}{e^t - 1} - \frac{1}{t} + \frac{1}{2} - \frac{t}{12} + \frac{t^3}{120} \right) + \int_0^1 dt~t^{s-1}\left(\frac{1}{t} - \frac{1}{2} + \frac{t}{12} - \frac{t^3}{120} \right) + \nonumber\\
                   &~~~~~~~~~~~~~~~~~~~+\int_1^{\infty} dt \frac{t^{s-1}}{e^{t} - 1},
\end{align}

Rearranging, we acquire
\begin{align}\label{eq:GammaZeta2b}
  \Gamma(s)\zeta(s) &= I_1(s) +  I_2(s) + \int_0^1 dt~t^{s-1}\left(\frac{1}{t} - \frac{1}{2} + \frac{t}{12} - \frac{t^3}{120} \right),
\end{align}

\noindent where 
\begin{align}
   I_1(s) &\equiv \int_0^1 dt~t^{s-1}\left(\frac{1}{e^t - 1} - \frac{1}{t} + \frac{1}{2} - \frac{t}{12} + \frac{t^3}{120} \right) \\
   I_2(s) &\equiv \int_1^{\infty} dt \frac{t^{s-1}}{e^{t} - 1},
\end{align}

\noindent and for Re$(s) > 1$, we can perform the integrals leading to yet another representation of $\zeta(s)$, still \emph{precisely} equivalent to the other two: Eqs.~(\ref{eq:ZetaDefSGreater1}) and~(\ref{eq:ZetaIntegralRepSGreater1}):
\begin{align}\label{eq:ZetaACRep}
   \zeta(s) &\equiv \frac{1}{\Gamma(s)}\left(I_1(s) + \frac{1}{s-1} - \frac{1}{2s} + \frac{1}{12(s+1)} - \frac{1}{120(s+3)} + I_2(s)\right) \nonumber\\
            &   \mbox{Re}(s) > 1.
\end{align}

Now as $\zeta(s)$ is not yet defined for Re$(s) \le 1$, we are free to \emph{analytically continue} its definition to Eq.~(\ref{eq:ZetaACRep}) for Re$(s) > -4$:
\begin{align}\label{eq:ZetaACRepSGreaterNegative4}
   \zeta(s) &\equiv \frac{1}{\Gamma(s)}\left(I_1(s) + \frac{1}{s-1} - \frac{1}{2s} + \frac{1}{12(s+1)} - \frac{1}{120(s+3)} + I_2(s)\right) \nonumber\\
            &   \mbox{Re}(s) > -4.
\end{align}

The key point here is that for Re$(s) > 1$, $\zeta(s)$ can be given by either Eq.~(\ref{eq:ZetaDefSGreater1}) \emph{or} Eq.~(\ref{eq:ZetaACRep}), but for $-4 < $ Re$(s) \le 1$, $\zeta(s)$ is only regular when defined by Eq. (\ref{eq:ZetaACRepSGreaterNegative4}).  We must be sure to understand that
\begin{align}
   \zeta(s) \ne \sum_{n=1}^{\infty}n^{-s}~~~~~~~\mbox{Re}(s) \le 1.
\end{align}

\noindent but instead we can replace the infinite sum with its associated zeta function, calling the procedure regularization:
\begin{align}
   \sum_{n=1}^{\infty}n^{-s} \to \zeta(s)
\end{align}

\noindent where for Re$(s) > -4$, the definition of $\zeta(s)$ is given by Eq.~(\ref{eq:ZetaACRepSGreaterNegative4}).  To analytically continue to more negative values of Re$(s)$, we must merely include more terms from the expansion, Eq.~(\ref{eq:expansion}), in Eq.~(\ref{eq:GammaZeta2}).

Looking at Eq.~(\ref{eq:ZetaACRepSGreaterNegative4}), we see that we still must define what we mean by $\Gamma(s)$ when Re$(s) \le 0$.  By the same techniques that gave us Eq.~(\ref{eq:GammaZeta2b}) from Eq.~(\ref{eq:GammaZeta2}), we find
\begin{align}
   \Gamma(s) &\equiv \int_0^{\infty} dt~t^{s-1} e^{-t} \nonumber\\
             &= I_3(s) + I_4(s) + \sum_{n=0}^3\frac{(-1)^n}{n!} \frac{1}{s+n}
             &\mbox{Re}(s) > 0
\end{align} 

\noindent where
\begin{align}
    I_3(s) &= \int_0^1 dt~t^{s-1} \left(e^{-t} - \sum_{n=0}^3\frac{(-t)^n}{n!} \right) \\
    I_4(s) &= \int_1^\infty dt~t^{s-1} e^{-t}
\end{align}

We notice, similar to before, that $I_3(s)$ is finite for Re$(s) > -4$ and $I_4(s)$ is finite for all $s$, allowing us to analytically continue $\Gamma(s)$ to the region Re$(s) > -4$:
\begin{align}\label{eq:GammaAC}
   \Gamma(s) &= I_3(s) + I_4(s) + \sum_{n=0}^3\frac{(-1)^n}{n!} \frac{1}{s+n} &\mbox{Re}(s) > -4
\end{align}

Now that we have all the tools necessary, let's calculate an example: the previously stated value
\begin{align}
   \zeta(-1) = -\frac{1}{12}.
\end{align}

\noindent From Eqs.~(\ref{eq:ZetaACRepSGreaterNegative4}) and~(\ref{eq:GammaAC}) we can calculate the limit
\begin{align}
   \zeta(-1) &= \lim_{s \to -1} \frac{s +1}{s + 1} \zeta(s) \nonumber\\
             &= \lim_{s \to -1} \frac{(s+1)(I_1(s) + I_2(s) + \frac{1}{s-1} - \frac{1}{2s}  + \frac{1}{12(s+1)} - \frac{1}{120(s+3)})}{(s+1)\Gamma(s)}  \nonumber\\
             &= \lim_{s \to -1} \frac{(s+1)(I_1(s) + I_2(s)) + \frac{s+1}{s-1} - \frac{s+1}{2s}  + \frac{s+1}{12(s+1)} - \frac{s+1}{120(s+3)}}{(s+1)(I_3(s) + I_4(s) + \sum_{n=0}^3\frac{(-1)^n}{n!}\frac{1}{s+n})} \nonumber\\
             &= \lim_{s \to -1} \frac{(s+1)(I_1(s) + I_2(s)) + \frac{s+1}{s-1} - \frac{s+1}{2s}  + \frac{s+1}{12(s+1)} - \frac{s+1}{120(s+3)}}{(s+1)(I_3(s) + I_4(s)) + \frac{s+1}{s} + \frac{-(s+1)}{s+1} + \frac{s+1}{2!(s+2)} + \frac{-(s+1)}{3!(s+3)}} \nonumber\\
             &= \frac{0 + \frac{1}{12}}{0 -1} \nonumber\\
     \zeta(-1)&= -\frac{1}{12}
\end{align}

\noindent which allows us to regularize the sum as in Eq.~(\ref{eq:SumofAllPositiveIntegersRegularization}).  We can apply this same regularization procedure to the following sums:
\begin{align}
     \sum_{n=1}^{\infty} 1 &\to \zeta(0) = -\frac{1}{2} \\
     \sum_{n=1}^{\infty} n^{-s} &\to \zeta(s) = 0, ~~~s =-2,-4,-6,\dots\\
     \sum_{n=1}^{\infty}n^3 &\to \zeta(-3) = \frac{1}{120} \\
     \sum_{r=\frac{1}{2},\frac{3}{2},\cdots}^\infty r &\to \frac{1}{24}.
\end{align}

\noindent As a final note, the Riemann zeta function, analytically continued to the entire complex plane, $s \in \mathbb{C}$, has a single pole at $s=1$, around which it has the expansion~\cite{Elizalde:1995a}
\begin{align}\label{eq:zetaexpansion}
 \zeta(s) &= \frac{1}{s-1} + \gamma + \gamma_1(s-1) + \gamma_2(s-1)^2 +\dots \\
   \gamma_k &= \lim_{n\to\infty}\left[\sum_{\nu=1}^\infty \frac{(\log \nu)^k}{\nu} - \frac{1}{k+1}(\log n)^{k+1} \right],
\end{align} 

\noindent where $\gamma \approx 0.577216$, Euler's constant.

\section{Regularization of Infinite Sums Using the Riemann zeta Function}
We now show a few examples of regularization of infinite sums, pertinent to this thesis. 

\subsection{Massive Modes}\label{MassiveModes}
\noindent Here we apply the regularization methods of~\cite{PandoZayas:2008hw,Doran:2009pp} for infinite sums resulting from massive modes, generally of the form:
\begin{align}\label{eq:Hurwitz}
  \sum_{n=1}^{\infty}\left(\frac{n^2\pi^2}{L^2} +\mu^2\right)^{-s} = \mu^{-2s}\sum_{n=1}^{\infty}\left(1 +\left(\frac{n\pi}{\mu L}\right)^2\right)^{-s}.
\end{align} 

We do this in two distinct ways, showing that, in the large $L$ limit and at least in the case of interest, $s = -1/2$, both schemes give
\begin{align}\label{eq:MassiveModeRegularization}
    \sum_{n=1}^{\infty}\left(\frac{n^2\pi^2}{L^2} +\mu^2\right)^{-s} \to -\frac{\mu^{-2s}}{2}.
\end{align} 

\noindent Our first method is as in~\cite{PandoZayas:2008hw,Doran:2009pp} where we use a cutoff $\{\frac{\mu L}{\pi}\}$ which is the largest integer less than $\frac{\mu L}{\pi}$:
\begin{align}\label{eq:cutoff}
  \mu^{-2s}\sum_{n=1}^{\infty}\left(1 +\left(\frac{n\pi}{\mu L}\right)^2\right)^{-s}~ &\longrightarrow ~\mu^{-2s}\sum_{n=1}^{\{\frac{\mu L}{\pi}\}}\left( 1 + \left(\frac{n\pi}{\mu L}\right)^2\right)^{-s}
\end{align}

\noindent where the right hand side may now be expanded in a binomial series for all $s \in \mathbb{C}$:
\begin{align}\label{eq:cutoffresult}
   \mu^{-2s}\sum_{n=1}^{\{\frac{\mu L}{\pi}\}}\left( 1 + \left(\frac{n\pi}{\mu L}\right)^2\right)^{-s} &= \mu^{-2s} \sum_{n=1}^{\{\frac{\mu L}{\pi}\}} \sum_{q=0}^{\infty}\binom{-s}{q} \left(\frac{n\pi}{\mu L}\right)^{2q} \nonumber\\
       &= \mu^{-2s} \sum_{q=0}^{\infty}\binom{-s}{q}\left(\frac{\pi}{\mu L}\right)^{-2q} \sum_{n=1}^{\{\frac{\mu L}{\pi}\}}n^{2q} \nonumber\\
       &\approx\mu^{-2s} \sum_{q=0}^{\infty}\binom{-s}{q}\left(\frac{\pi}{\mu L}\right)^{-2q} \sum_{n=1}^{\infty}n^{2q} \nonumber\\
       &\longrightarrow\mu^{-2s} \sum_{q=0}^{\infty}\binom{-s}{q}\left(\frac{\pi}{\mu L}\right)^{-2q} \delta^0_q \zeta(0) \nonumber\\
       &= -\frac{\mu^{-2s}}{2}.
\end{align}

\noindent In this calculation, we have used $\zeta$-function regularization, and the approximation $\{\frac{\mu L}{\pi}\} \to \infty$, corresponding to large $L$ quark separation.

Next, we show the equivalence, in the large $L$ limit, between this result and that found using the regularization scheme from~\cite{Bertoldi:2004rn,Doran:2009pp}.  First, we split the sum up
\begin{align}\label{eq:HurwitzExpanded}
   \sum_{n=1}^{\infty}\left(\frac{n^2\pi^2}{L^2} +\mu^2\right)^{-s} &= -\frac{1}{2}\mu^{-2s} + \frac{1}{2}\left(\frac{\pi}{L}\right)^{-2s}\sum_{n \in \mathcal{Z}} \left(n^2 + \left(\frac{\mu L}{\pi}\right)^2\right)^{-s} \nonumber\\
 &= -\frac{1}{2}\mu^{-2s} + 2\frac{L^{s+\frac{1}{2}}\mu^{-s + \frac{1}{2}}}{\sqrt{\pi}\Gamma(s)} \sum_{n=1}^{\infty}n^{s-\frac{1}{2}}K_{s-\frac{1}{2}}(2 n \mu L) + \nonumber\\ &~~~~~~~~~~~~~~~~~~+\frac{1}{2}\frac{L\mu^{-2s +1}}{\sqrt{\pi}\Gamma(s)}\Gamma{\left(s-\frac{1}{2}\right)}, 
\end{align}

\noindent and regulate by dropping the last term{\linespread{1.0}\footnote{This term diverges in the case of interest, the $s\to -1/2$ limit.  However, it was argued in~\cite{Bertoldi:2004rn} that this term could be subtracted off as a Casimir energy.}}.  Taking the large $L$ limit then gives us 
\begin{align}\label{eq:Bertoldiresult}
   \sum_{n=1}^{\infty}\left(\frac{n^2\pi^2}{L^2} +\mu^2\right)^{-s} &\longrightarrow-\frac{\mu^{-2s}}{2}\left[1 - 2 \frac{ (\mu L)^{s+\frac{1}{2}}}{\Gamma(s)}\sum_{n=1}^{\infty}n^{s-\frac{1}{2}}e^{-2 n \mu L}\left((n \mu L)^{-\frac{1}{2}} + \mathcal{O}(n L)^{-\frac{3}{2}}\right)  \right] \nonumber\\
      &= -\frac{\mu^{-2s}}{2}\left[1 - \frac{2\mu L}{\Gamma(s)}\sum_{n=1}^{\infty}e^{-2 n \mu L}\left((n\mu L)^{s-1} + \mathcal{O}(nL)^{s-2}\right)\right] 
\end{align}

\noindent The sum in Eq. (\ref{eq:Bertoldiresult}) is exponentially suppressed even without further regularization in the large $L$ limit and for the case of interest $s=-1/2$, where we recover the same result as with the cut off method, Eq.~(\ref{eq:cutoffresult}).

\subsection{Regularization of Two Sphere Eigenmodes}\label{TwoSphereRegularization}

As our next example, we regularize the following infinite sum
\begin{align}\label{eq:S2eigensum}
    \tilde{f}_\zeta(s) &\equiv \sum_{l=1}^\infty (2 l+1)[l(l+1)]^{-s} = 2 f_{\zeta}(s;1,1,0) + f_{\zeta}(s;1,0,0)
\end{align}

\noindent where~\cite{Elizalde:1995a}
\begin{align}
   f_{\zeta}(s;a,b,c) &\equiv \sum_{l=1}^\infty l^{-s+b}(l+a)^{-s+c} 
\end{align}

\noindent The function $f_{\zeta}(s;a,b,c)$ can be regularized using a binomial expansion and the Riemann zeta function to
\begin{align}\label{eq:fzetaRegularized}
   f_{\zeta}(s;a,b,c) \to  &\sum_{\nu=0}^\infty \frac{\Gamma(1-s+c)a^\nu}{\nu!\Gamma(1-s-\nu+c)}\left[\zeta(2s+\nu-b-c) - \sum_{l=1}^{[a]}l^{-2s-\nu+b+c}\right] + \nonumber\\
         &~~~~~+ \sum_{l=1}^{[a]}l^{-2s+b+c}(1 + a l^{-1})^{-s+c} .
\end{align}

\noindent where the cutoff $[a]$ is the integer part of $a$.  We use this to regularize the sum, Eq.~(\ref{eq:S2eigensum}), for $s \in \frac{\mathbb{Z}}{2}$:
\begin{align}\label{eq:ftildezetaregularized}
   \tilde{f}_\zeta(s)  &\to \sum_{\nu=0}^\infty\frac{\Gamma(1-s)}{\nu! \Gamma(1-s-\nu)}(2\zeta(2s +\nu-1) + \zeta(2s + \nu) - 3) + 3 \cdot 2^{-s}\nonumber\\
        &= 2\sum_{\nu=2s-1}^\infty\frac{\Gamma(1-s)\zeta(\nu)}{(\nu + 1 -2s)! \Gamma(s-\nu)} + \sum_{\nu = 2s}^\infty\frac{\Gamma(1-s)\zeta(\nu)}{(\nu-2s)! \Gamma(1+s-\nu)} + \nonumber\\ 
&~~~~~~~~~~~~~~~~~~~~~~~~~~~~~~~~~~~~~~~~ 3\left[2^{-s} - \sum_{\nu=0}^\infty\frac{\Gamma(1-s)}{\nu!\Gamma(1-s-\nu)} \right]  \nonumber\\       &=\sum_{\nu=2s}^\infty\zeta(\nu)\left[\frac{2\Gamma{(1-s)}}{(\nu+1-2s)!\Gamma(s-\nu)} +\frac{\Gamma{(1-s)}}{(\nu-2s)!\Gamma(1+s-\nu)}\right] + \nonumber\\
 &~~~~~+2\zeta(2s-1) +    3\left[2^{-s} - \sum_{\nu=0}^\infty\frac{\Gamma(1-s)}{\nu!\Gamma(1-s-\nu)} \right] \nonumber\\
&=-\sum_{\nu=2s}^\infty\frac{\Gamma{(1-s)}~(\nu-1)\zeta(\nu)}{(\nu +1-2s)!\Gamma(1+s-\nu)} + \nonumber\\
&~~~~~+2\zeta(2s-1) +    3\left[2^{-s} - \sum_{\nu=0}^\infty\frac{\Gamma(1-s)}{\nu!\Gamma(1-s-\nu)} \right] \nonumber\\
      &=-\sum_{\nu=2s,\nu\ne 1}^\infty\frac{\Gamma{(1-s)}~(\nu-1)\zeta(\nu)}{(\nu +1-2s)!\Gamma(1+s-\nu)} -\left(\frac{\Gamma{(1-s)}~(\nu-1)\zeta(\nu)}{(\nu +1-2s)!\Gamma(1+s-\nu)}\right)_{\nu=1} \nonumber\\
&~~~~~+2\zeta(2s-1) +    3\left[2^{-s} - \sum_{\nu=0}^\infty\frac{\Gamma(1-s)}{\nu!\Gamma(1-s-\nu)} \right]
\end{align}

\noindent The second term, with $\nu =1$, is only there for $s \le \frac{1}{2}$.  In this term, the factor $(\nu-1)$ cancels with the pole in the expansion of $\zeta(\nu)$,  Eq. (\ref{eq:zetaexpansion}), leaving us with:
\begin{align}
\tilde{f}_\zeta(s)  &\to -\sum_{\nu=2s,\nu\ne 1}^\infty\frac{\Gamma{(1-s)}~(\nu-1)\zeta(\nu)}{(\nu +1-2s)!\Gamma(1+s-\nu)} -\frac{\Gamma{(1-s)}}{(2-2s)!\Gamma(s)} \nonumber\\
&~~~~~+2\zeta(2s-1) +    3\left[2^{-s} - \sum_{\nu=0}^\infty\frac{\Gamma(1-s)}{\nu!\Gamma(1-s-\nu)} \right]
\end{align}

This formula is particularly useful for $s=-\frac{1}{2}$, the sum over spherical eigenmodes, where we have~\cite{PandoZayas:2008hw}:
\begin{align}\label{eq:SphericalEigenmodeRegularization}
  \tilde{f}_\zeta\left(-\frac{1}{2}\right) &\to-\sum_{\nu=-1,\nu\ne 1}^\infty\frac{\Gamma{(\frac{3}{2})}~(\nu-1)\zeta(\nu)}{(\nu +2)!\Gamma(\frac{1}{2}-\nu)} -\frac{\Gamma{(\frac{3}{2})}}{3!\Gamma(-\frac{1}{2})} \nonumber\\
&~~~~~+2\zeta(-2) +    3\left[\sqrt{2} - \sum_{\nu=0}^\infty\frac{\Gamma(\frac{3}{2})}{\nu!\Gamma(\frac{3}{2}-\nu)} \right] \nonumber\\
&= 2\zeta(-1) +\frac{1}{4} \zeta(0) - \sum_{\nu=2 }^\infty\frac{\Gamma{(\frac{3}{2})}~(\nu-1)\zeta(\nu)}{(\nu +2)!\Gamma(\frac{1}{2}-\nu)}  + \frac{1}{24} + \nonumber\\
&~~~~~~+ 0 + 3[\sqrt{2} - \sqrt{2}] \nonumber\\
  &=  -\frac{1}{6} -\frac{1}{8}  - \Gamma{\left(\frac{3}{2}\right)(0.0170335)}  + \frac{1}{24} \nonumber\\
            &\approx -0.265096
\end{align} 

We can extend this regularization method to more complicated infinite sums as in Eq. (\ref{eq:KSdeltaEbnsumperformed}) where we have a sum of the form
\begin{align}
    \tilde{g}_\pm(s) &\equiv \sum_{l=1}^\infty(2l+1) [l(l+1)]^{-s}[1 + f_\pm(l)]^{-s},\\
    f_\pm(l) &= \frac{f_1}{l(l+1)}\left(1 \pm \sqrt{1 + \frac{f_2}{f_1^2}l(l+1)}\right)
\end{align}

\noindent for $s = -\frac{1}{2}$, and where $f_2$ and $f_1$ are constant with respect to $l$.  We consider only cases as in the KS calculation, Figs.~\ref{fig:KS_fpluslkM} and~\ref{fig:KS_fminuslkM}, where $f_-(l) <1$ for all $l \ge 1$, but $f_+(1) > 1$ definitely only for $l >1$.  After a series of careful binomial expansions, and a zeta function regularization, as in Eq. (\ref{eq:ftildezetaregularized}), we find for $s = -\frac{1}{2}$:
\begin{align}\label{eq:EmbeddedSquareRootRegularization}
\tilde{g}_\pm\left(-\frac{1}{2}\right) &\to \sum_{w=0}^\infty \sum_{q=0}^{[\frac{w}{2}]}\sum_{n=0}^q \binom{q}{n}\frac{\Gamma{(\frac{3}{2})}f_2^n f_1^{w-2n}}{\Gamma{(\frac{3}{2}-w)}}\frac{\tilde{f}_\zeta(w-n-\frac{1}{2})}{(2q)!\Gamma{(w+1-2q)}}+ \nonumber\\
&~~\pm \sum_{w=1}^\infty\sum_{q=0}^{[\frac{w-1}{2}]}\frac{\Gamma{(\frac{3}{2})}f_1^w}{\Gamma{(\frac{3}{2}-w)}}\frac{g_\zeta(q,w)}{(2q+1)!\Gamma(w-2q)}
+ h_\pm
\end{align}

\noindent with
\begin{align}\label{eq:gzeta}
g_\zeta(q,w) &=  \left\{\begin{array}{ll}\sum_{n=0}^\infty\binom{q+\frac{1}{2}}{n}\left(\frac{f_2}{f_1^2}\right)^{q-n+\frac{1}{2}}\left[\tilde{f}_\zeta(w+n-q-1) + \right.& \\
\left.~~~~~~~~~~~~~~~-\sum_{l=1}^{\{l_f\}}\frac{(2l+1)}{[l(l+1)]^{w+n-q-1}}\right]  + & 0 < \frac{k}{M} < 1\\
~~~~~~~~+  \sum_{l=1}^{\{l_f\}}\frac{(2l+1)\left(1 + \frac{f_2}{f_1^2}l(l+1)\right)^{q + \frac{1}{2}}}{[l(l+1)]^{w-\frac{1}{2}}} &   \\     \sum_{n=0}^\infty\binom{q+\frac{1}{2}}{n}\left(\frac{f_2}{f_1^2}\right)^n\tilde{f}_\zeta\left(w-n-\frac{1}{2}\right) & k=0,M
                                  \end{array}
                           \right. 
\end{align}

\noindent and
\begin{align}
h_+ &= \left\{\begin{array}{ll}
                  2\sqrt{2}\left(\sqrt{1 + f_+(1)} - \sum_{w=0}^\infty \binom{\frac{1}{2}}{w}f_+^w(1)\right) &f_+(1) > 0 \\
                  0 & f_+(1) < 0,
                 \end{array}
        \right.  \\
h_- &= 0,
\end{align}

\noindent where $\{l_f\}$ is the largest integer less than $l_f = \frac{1}{2}\left(-1 + \sqrt{1+\frac{4f_1^2}{f_2}}\right)$.


\Appendix{ANTI-DE SITTER SPACE}
\label{app:AdSspace}
\addtocontents{toc}{\protect\vspace{-\li}} 

This appendix is based on a summary found in~\cite{Maldacena:1997re}.  Anti-de Sitter space of dimension $d$ ($AdS_d)$ is defined by the hyperboloid:
\begin{equation}\label{eq:hyperboloid}
-(X^{-1})^2 -(X^0)^2 + (X^1)^2 + \cdots + (X^{d-2})^2 + (X^{d-1})^2 = -L^2
\end{equation}

\noindent embedded in $R^{2,d-1}$:
\begin{equation}\label{eq:embedding}
ds^2 = -(dX^{-1})^2 -(dX^0)^2 + (dX^1)^2 + \cdots + (dX^{d-2})^2 + (dX^{d-1})^2
\end{equation}
\noindent where the constant $L$ is known as the $AdS$ radius~\cite{Maldacena:1997re}. 

We shall henceforth use the following simplifier notation:
\begin{align}
X^2 &= X_i X^i = -(X^0)^2 + (X^1)^2 + \cdots + (X^{d-2})^2 \nonumber\\
dX^2 &= dX_i dX^i = -(dX^0)^2 + (dX^1)^2 + \cdots + (dX^{d-2})^2 \nonumber
\end{align}

\noindent which casts the defining Eqs.~(\ref{eq:hyperboloid}) and~(\ref{eq:embedding}) for $AdS_d$ space into:
\begin{align}\label{eq:hyperboloidcompact}
-L^2 &= -(X^{-1})^2 + X^2 + (X^{d-1})^2 \\
\label{eq:embeddingcompact}
ds^2 &= -(dX^{-1})^2 + dX^2 + (dX^{d-1})^2
\end{align}
 
We now set out to find the form of the $AdS_d$ metric, which is given by the intersection of the embedding metric, Eq.~(\ref{eq:embeddingcompact}), with the hyperboloid, Eq.~(\ref{eq:hyperboloidcompact}).  This computation is made easier if we change coordinates to:
\begin{align}\label{eq:coordinatechange}
r &= X^{-1} + X^{d-1} \nonumber\\
x^i &= \frac{X^i L}{r} \nonumber\\
q &= X^{-1} - X^{d-1}
\end{align}
This coordinate change casts the hyperboloid, Eq.~(\ref{eq:hyperboloidcompact}), into the form:
\begin{equation}
q = \frac{L^2}{r} + \frac{r}{L^2}x^2
\end{equation}
giving for the differential element $dq$:
\begin{equation}\label{eq:dq}
dq = -\frac{L^2}{r^2}dr + 2\frac{r}{L^2}x_i dx^i + \frac{x^2}{L^2}dr.
\end{equation}

Applying the coordinate change, Eq.~(\ref{eq:coordinatechange}), to the embedding metric, Eq.~(\ref{eq:embeddingcompact}), we have:
\begin{align}
ds^2 &= -(dX^{-1})^2 + (dX^{d-1})^2 + dX^2 \nonumber\\
     &= -dr~dq + \left(\frac{x^i}{L}dr + \frac{r}{L}dx^i\right)^2 \nonumber\\
     &= -dr~dq + \frac{x^2}{L^2}dr^2 + \frac{r^2}{L^2}dx^2 + 2\frac{r}{L^2}x_i dx^i dr 
\end{align}

\noindent whereupon plugging Eq.~(\ref{eq:dq}) in for the differential element $dq$ leaves us with the common form for the $AdS_d$ metric:
\begin{equation}\label{eq:AdScommon1}
ds^2 = \frac{L^2}{r^2}dr^2 + \frac{r^2}{L^2}dx^2
\end{equation}
Switching variables again to $z = \frac{L^2}{r}$ gives another very common form of the $AdS_d$ metric:
\begin{equation}\label{eq:AdScommon2}
ds^2 = \frac{L^2}{z^2}\left(dx^2 + dz^2\right)
\end{equation}


\Appendix{THE CONIFOLD AND THE DEFORMED CONIFOLD}
\label{app:conifold}
\section{The Conifold}

The conifold is described in four dimensional complex space $\mathbb{C}^4$ by:
\begin{equation}\label{eq:conifold}
   \sum_{i=1}^4 z_i^2 = 0
\end{equation}

\noindent and is said to have a singularity where all $z_i = 0$.  The conifold's base is given by its intersection with a real eight-sphere~\cite{Candelas:1989js}:
\begin{equation}\label{eq:foursphere}
   \sum_{i=1}^4 |z_i|^2 = constant
\end{equation}

One such base that result from this intersection is the Einstein manifold $T^{1,1}$, which is topologically equivalent to $S^2$ $\times$ $S^3$~\cite{Candelas:1989js}. The Einstein metric for the $T^{1,1}$ is:
\begin{equation}\label{eq:basemetric}
ds_5^2 = \frac{1}{9}(g^5)^2 + \frac{1}{6}\sum_
{i=1}^4 (g^i)^2
\end{equation}

\noindent where the 1-forms, $g^i$ are~\cite{PandoZayas:2008hw,Klebanov:2000hb}
\begin{align}
g^1 &= \frac{1}{\sqrt{2}}[-\sin\theta_1 d\phi_1 - \cos\psi \sin\theta_2 d\phi_2 + \sin\psi d\theta_2] \nonumber\\
g^2 &= \frac{1}{\sqrt{2}}[d\theta_1 - \sin\psi \sin\theta_2 d\phi_2 - \cos\psi d\theta_2]\nonumber\\
g^3 &= \frac{1}{\sqrt{2}}[-\sin\theta_1 d\phi_1 + \cos\psi \sin\theta_2 d\phi_2 - \sin\psi d\theta_2] \nonumber\\
g^4 &= \frac{1}{\sqrt{2}}[d\theta_1 + \sin\psi \sin\theta_2 d\phi_2 + \cos\psi d\theta_2]\nonumber\\
g^5 &= d\psi + \cos\theta_1 d\phi_1 + \cos\theta_2 d\phi_2\nonumber\tag{\ref{eq:gi}}
\end{align}

\noindent and the coordinates have the ranges: $0 \le \theta_i < \pi$, $0 \le \phi_i < 2\pi$, and $0 \le \psi < 4\pi$~\cite{Candelas:1989js}.
\begin{figure}
\addtocontents{lof}{\protect\vspace{\li}}
 \centering
 \subfigure[The conifold has a singular tip in both the $S^3$ and the $S^2$ vanish there.]{\label{fig:conifold}\includegraphics[width = 0.4\columnwidth]{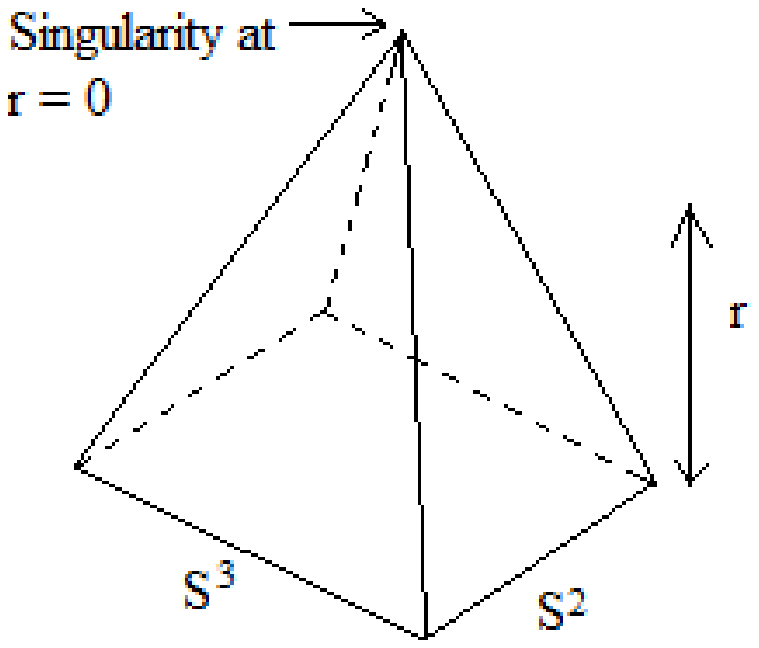}}\qquad
 \subfigure[The tip of the deformed conifold is singular only the in the $S^2$ direction: the $S^3$ is ``blown up" to size $\epsilon$ here.]{\label{fig:deformedconifold}\includegraphics[width=0.4\columnwidth]{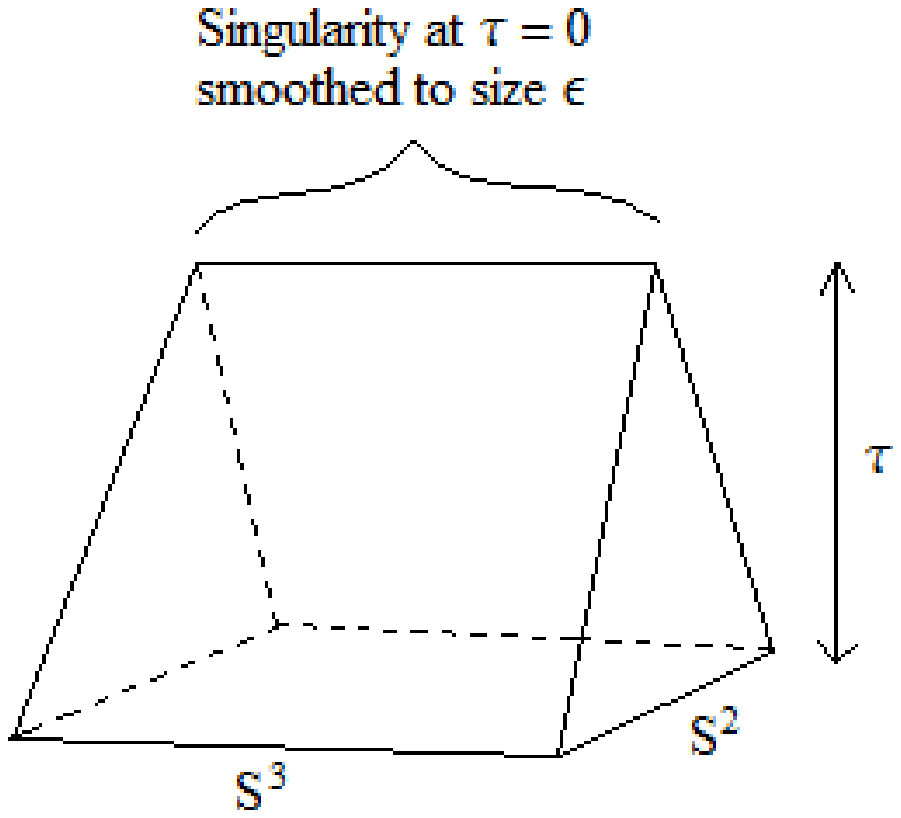}}
 \caption{The conifold and deformed conifold both have a base which is an $S^3$ $\times$ $S^2$. The main difference is at the tip: the conifold is singular in both the $S^3$ and $S^2$ directions, where as the deformed conifold is only singular in the $S^2$ direction.}
 \label{fig:conifoldanddeformed}
\end{figure}
Augmenting the base with the ``height" direction, $r$, the full metric for our conifold is~\cite{Candelas:1989js}:
\begin{equation}\label{eq:confmetric}
ds_6^2 =  dr^2 + r^2 ds_5^2
\end{equation}

\noindent Figure~\ref{fig:conifold} shows how one can think about a conifold as a generalization of a three dimensional pyramid, homeomorphic to a cone.  One edge of the base is thought of as an $S^3$ and the other edge as an $S^2$.  The base of the conifold, then, is topologically $S^3$ $\times$ $S^2$.  The tip of the pyramid corresponds to $r=0$ in the conifold, where a singularity exists as both the $S^3$ and the $S^2$ shrink to zero size~\cite{kachru}.

\section{The Deformed Conifold}

The Klebanov-Strassler background actually arises from the \emph{deformed conifold}.  The conifold is deformed by ''blowing up" the tip to size $\epsilon$:
\begin{equation}\label{eq:deformedconifold}
   \sum_{i=1}^4 z_i^2 = \epsilon^2,
\end{equation}
\noindent smoothing out the singularity as depicted in Fig.~\ref{fig:deformedconifold}~\cite{Klebanov:2000hb}.  The tip of the deformed conifold is located where the new deformed conifold coordinate $\tau = 0$~\cite{kachru}. In the end, the deformed conifold is described by the metric~\cite{Klebanov:2000hb,PandoZayas:2008hw}:
\begin{align}\label{eq:deformedconfmetric}
ds_6^2 = \frac{1}{2}\epsilon^{4/3}K(\tau) \bigg[\frac{1}{3 K^3(\tau)}[d\tau^2 + (g^5)^2] &+ [(g^3)^2 + (g^4)^2]~\cosh^2\left(\frac{\tau}{2}\right) + \nonumber\\
         &+ [(g^1)^2 + (g^2)^2]~\sinh^2\left(\frac{\tau}{2}\right)\bigg]
\end{align}

\noindent where the new function $K(\tau)$ is given by:
\begin{equation}\label{eq:Ktau}
K(\tau) = \frac{(\sinh(2\tau) - 2\tau)^{1/3}}{2^{1/3}\sinh\tau}
\end{equation}

Comparing Eqs.~(\ref{eq:conifold}) and~(\ref{eq:deformedconifold}), we would expect the deformed conifold to become the conifold as the infinitesimal parameter $\epsilon \to 0$.  The conifold metric, Eq.~(\ref{eq:confmetric}), does in fact arise from the deformed conifold metric, Eq.~(\ref{eq:deformedconfmetric}), if we take the limit~\cite{Ohta:1999we}:

\begin{equation}\label{eq:limit}
\frac{1}{\epsilon}, \tau \to \infty~~\mbox{as}~~~~ r^3\left(\frac{2}{3}\right)^{2/3}  = ~\epsilon^2 \cosh\tau~~ \mbox{remains fixed}
\end{equation}
In summary, the deformed conifold is mapped to the regular conifold through the coordinate $\tau$ at infinity.  This corresponds to the UV in the dual gauge theory.  Here we shrink the $S^3$, which had been ``blown up" to size $\epsilon$ here, back to zero size.  We do all of this all while keeping the conifold coordinate $r(\epsilon,\tau)$ fixed. 

Stacking $M$ D5-branes and and $N$ D3-branes at the tip of the deformed conifold, $\tau = 0$, gives rise to an $SU(N + M) \times SU(N)$ dual gauge theory~\cite{Klebanov:2000hb,Gubser:1998fp,Klebanov:2000nc,Klebanov:1999rd}.  We wrap two of the dimensions of the D5-branes around the $S^2$, which shrinks to zero size at the tip as in Fig.~\ref{fig:deformedconifold}, creating $M$ \emph{fractional} D3-branes at $\tau=0$. It has been shown in the literature~\cite{Klebanov:2000hb,Klebanov:2000nc}, that for objects located at $\tau=0$, the dual, IR $SU(N +M) \times SU(N)$ gauge theory becomes an $SU(M)$ gauge theory in the IR.


\Appendix{BOSONIC FLUCTUATIONS}
\label{app:BosonicFluctuations}
This appendix is based on the calculations published previously in~\cite{PandoZayas:2008hw,Doran:2009pp,Stiffler:2009ma}.  Expanding the Dp-brane action 
\begin{align}
  S_{p}& =  -\mu_p \int d^{p+1}\zeta e^{-\Phi}\sqrt{\mathcal{M}} +\mu_p \int \sum_n C_n\wedge e^\mathcal{F} + \nonumber\\
       &~~~+ \frac{\mu_p}{2}\int d^{p+1}\zeta e^{-\Phi}\sqrt{\mathcal{M}}~\mathcal{L}_p^f(\Theta)\nonumber\tag{\ref{eq:DpbraneAction}}
\end{align}

\noindent to second order in the fluctuations, $\delta X^\mu$, $\delta A^m$, and $\delta \Theta$
\begin{equation}
   X^\mu = X^\mu_{(0)} + \delta X^\mu,~~~A^{m} = A^{m}_{(0)} + \delta A^{m},~~~\Theta = 0 + \delta \Theta, \nonumber\tag{\ref{eq:fluctuate}}
\end{equation}

\noindent results in
\begin{align}
  S_{p} &= S_{p}^{(0)} + \delta S_{p} + \mathcal{O}(\delta^3), \nonumber\tag{\ref{eq:expandaction}}\\ 
  \delta S_{p} &= \delta S_{p}^b[\delta X,\partial \delta X, \partial \delta A]  + \delta S_p^f[\delta\Theta,\partial\delta\Theta] \nonumber\tag{\ref{eq:deltaSbPlusdeltaSf}},
\end{align}

\noindent where $ \delta S_{p}^b[\delta X,\partial \delta X, \partial \delta A]$ is the action for the bosonic fluctuations.  The matrix $\mathcal{M}_{ab}$ expands as
\begin{align}
    \mathcal{M}_{ab} &= \mathcal{M}^{(0)}_{ab} + \delta\mathcal{M}_{ab} \nonumber\\
                     &= (g_{ab}^{(0)} + \mathcal{F}_{ab}^{(0)}) + (\delta g_{ab} + \delta\mathcal{F}_{ab})
\end{align}

\noindent and following formula to expand the square root is useful
\begin{align}\label{eq:SqrtExpansion}
           \sqrt{\mathcal{M}} &\equiv \sqrt{-\det(\mathcal{M}_{ab})} \nonumber\\
 &= \sqrt{-\det(\mathcal{M}^{(0)}_{ab} + \delta\mathcal{M}_{ab})} \nonumber\\
         &= \sqrt{\mathcal{M}^{(0)}}(1 + \Delta)
\end{align}
\noindent where
\begin{align}\label{eq:Delta}
 \Delta &\equiv  \frac{1}{2}(\mathcal{M}_{(0)}^{-1})^{ab} \delta\mathcal{M}_{ab} + \frac{1}{8} [(\mathcal{M}_{(0)}^{-1})^{ab} \delta\mathcal{M}_{ab}]^2 + \nonumber\\
                             &-\frac{1}{4}[ (\mathcal{M}_{(0)}^{-1})^{ab} \delta\mathcal{M}_{bc} (\mathcal{M}_{(0)}^{-1})^{cd}\delta\mathcal{M}_{da}] + \mathcal{O}(\delta\mathcal{M}_{ab}^3).
\end{align}

For branes probing each background, KS and CGLP, we will first show in each case that $\delta S_p^b$ is, to lowest order, quadratic in the fluctuations, up to total derivatives, confirming our previous result that we are fluctuating around a classical solution.  Then we will find the bosonic eigenvalues, $\omega_b$, for the equations of motion of the fluctuations, and use them to calculate the bosonic contribution 
\begin{align}\label{eq:deltaEb}
   \delta E_b = \frac{1}{2}\sum \omega_b
\end{align}

\noindent to the one loop energy:
\begin{align}
   \delta E &\equiv \frac{1}{2}\sum \omega_b -\frac{1}{2}\sum \omega_f. \nonumber\tag{\ref{eq:oneloopenergy}}
\end{align}

\section{D3-brane Bosonic Fluctuations in the KS Background}\label{app:KSfluctuations}
Utilizing Eq.~(\ref{eq:SqrtExpansion}), the action for bosonic fluctuations of a D3-brane in the KS background becomes
\begin{align}
    \delta S_3^b &=  -\mu_3 \int d^4\zeta\sqrt{\mathcal{M}^{(0)}}~\Delta +  \mu_3 \int (\delta C_2 \wedge \mathcal{F}^{(0)} + C_2^{(0)} \wedge \delta\mathcal{F} + \delta C_2 \wedge \delta\mathcal{F}^{(0)})
\end{align}
\noindent where $\Delta$, $\delta C_2$, and $\delta\mathcal{F}$ depend on the bosonic fluctuations
\begin{align}\label{eq:bosonicfluctuations}
   \theta_m &\equiv \frac{1}{2}(\theta_1 - \theta_2) = \delta\theta_m,~~~\phi_p \equiv \frac{1}{2}(\phi_1 + \phi_2) = \delta\phi_p \nonumber\\
   X^2 &= \delta X^2,~~~X^3 = \delta X^3 \nonumber\\
  \psi &= \psi_0 + \delta\psi,~~~\tau = \tau_0 + \delta\tau,\nonumber\\
  F &= F_{tx} dt \wedge dx + \partial_a \delta A_b d\zeta^a \wedge d\zeta^b
\end{align}

\noindent where we will be careful to take the $\tau_0 \to 0$ limit at the appropriate time to avoid singularities.  In the above, we have maintained the following parameterization of the D3-brane
\begin{equation}
 X^0 = t,~~~X^1 = x,~~~\theta_p \equiv \frac{1}{2}(\theta_1 + \theta_2) = \theta,~~~\phi_m \equiv \frac{1}{2}(\phi_1 - \phi_2) = \phi.
 \end{equation}

\noindent We will gauge fix $\delta A_a$ to be in temporal gauge, $\delta A_t = 0$, when we analyze the equations of motion.

With these explicit fluctuations, we have to second order in the fluctuations
\begin{align}
 \delta \mathcal{F} &= \delta B_2 + 2\pi \alpha' \partial_a \delta A_b d\zeta^a \wedge d\zeta^b \nonumber\\
        &= \frac{g_s M\alpha'}{6}\delta\tau(g_{(0)}^3 \wedge g_{(0)}^4 + \delta g^3 \wedge g_{(0)}^4 + g_{(0)}^3 \wedge \delta g^4)
\end{align}

\noindent where $g_{(0)}^i$ are the values of $g^i$, Eq.~(\ref{eq:gi}), evaluated at the classical solution, Eq.~(\ref{eq:KSclassicalscalars}). We see that in this case, $\delta g^i$ need only to be expanded to first order in the fluctuations, Eq.~(\ref{eq:bosonicfluctuations}), for $\delta \mathcal{F}$ to be expanded to second order in the fluctuations.  We can similarly expand $\delta C_2$ and $\Delta$ to second order in the fluctuations.

\subsection{Explicit Action for Bosonic Fluctuations}\label{KSbosonicExplicitAction}
The action for bosonic fluctuations of a probe D3-brane in the KS background, after some simplification, takes the form
\begin{align}\label{eq:2ndorderaction}
  \delta S_3^b = -\int d^4\zeta\sqrt{g^{(\mbox{eff})}}&\Big{\{}c_X\sum_{i=2,3}\nabla^a \delta X^i\nabla_a \delta X^i  + c_A \left[\frac{1}{16\pi}\delta F^{ab}\delta F_{ab} + \delta A_{a}j^{a}\right] + \nonumber\\ \qquad &+c_{\tau}[\nabla^{a}\delta\tau\nabla_{a}\delta\tau + m_{\tau}^2\delta\tau^2 + \nabla^{a}\Psi\nabla_{a}\Psi - R\Psi^2] \\ & \qquad \qquad + \mbox{Total Derivatives}\nonumber\Big{\}}.
\end{align}

\noindent Here we notice that the linear fluctuations vanish, up to total derivatives, signifying that we are truly fluctuating around a classical solution.  The covariant derivative, $\nabla_{a}$, is with respect to an effective metric, $g^{(\mbox{eff})}_{ab}$, on the D3-brane
\begin{align}\label{eq:geff}
 ds^2 &= g^{(\mbox{eff})}_{ab}d\zeta^{a}d\zeta^{b} = g_{xx}(-dt^2 + dx^2) + \frac{2}{R}(d\theta^2 + \sin^2\theta d\phi^2).
\end{align}

\noindent This effective metric has the same topology, $R^{1,1} \times S^2$, and scalar curvature as the induced metric:
\begin{align}
  R &= \frac{8}{9b^3M \alpha' g_sf_2(\psi_0)},~~~b\approx 0.933 \nonumber\tag{\ref{eq:KSscalarR}}
\end{align}

\noindent now written in terms of the newly defined
\begin{align}
    \label{eq:KSf2}
    f_2(\psi_0) &= \frac{4 \cos^2\frac{\psi_0}{2}}{9 b^2}.
\end{align}

\noindent The field $\Psi$ is a combination of the fields $\delta\psi$, and $\delta\phi_p$
\begin{align}\label{eq:KSPsiFieldDef}
   \Psi \equiv \delta\psi + 2\cos\theta \delta\phi_p,
\end{align}

\noindent and contains all the contributions of $\delta\psi$ and $\delta\phi$ to the quadratic action suggesting a redundancy in the fields. We discuss this below.
\noindent The covariantly conserved $U(1)$ gauge current is given by
\begin{align}
    j^a &= (-Q_{\Psi} \nabla_x \Psi, ~Q_{\Psi} \nabla_t \Psi,~ -Q_{\tau} \csc\theta \nabla_{\phi} \delta\tau,~ Q_{\tau} \csc\theta\nabla_{\theta} \delta\tau ),
\end{align}

\noindent The various constants in the previous few equations are
\begin{align}\label{eq:constants}
    g_{xx} &= \frac{1}{f_3(\psi_0) \sqrt{h_0}},~~~c_A = 32 \sqrt{h_0} \pi^3 \alpha'^2 c_X = \frac{128 \pi^3 \alpha'^3}{b g_s M} c_\tau = 2 g_s^{-1} \sqrt{f_3(\psi_0)}, \nonumber\\
m_{\tau}^2 &=\frac{8}{9 b^3 g_s M \alpha'}\frac{f_1(\psi_0)}{f_2(\psi_0)} ,~~~Q_{\tau} = \frac{1}{54 b^4 g_s M \pi^2 \alpha'^2 f_2(\psi_0)},~~~Q_{\Psi} = \frac{\sqrt{h_0}f_3^{3/2}(\psi_0)}{8b\pi^2 \alpha'}, \nonumber\\
f_1(\psi_0) &= \frac{7}{9} + \frac{10 - 8b^2}{20}f_2(\psi_0),~~~ f_3(\psi_0) =1 + b^2 \tan^2\frac{\psi_0}{2}.
\end{align}

The Euler-Lagrange equations for the bosonic fields derived from the action, Eq.~(\ref{eq:2ndorderaction}), take the form:
\begin{align}\label{eq:Xeqm}
   &\nabla^2 \delta X^i = 0,~~~i = 2,3\\
   \label{eq:taueqm}
   &\nabla^2\delta\tau - m_{\tau}^2\delta\tau + \frac{c_A}{2 c_{\tau}}Q_{\tau}  \csc\theta \delta F_{\theta\phi} = 0 \\
   \label{eq:Psieqm}
   &\nabla^2\Psi + R\Psi + \frac{c_A}{2c_{\tau}} Q_{\Psi} \delta F_{tx} = 0 \\
   \label{eq:Aeqm}
   &\nabla^{a}\delta F_{ab} - 4\pi j_{b} = 0.
\end{align}

Observe that we found no field equation for $\delta \theta_m$ and that a field redefinition absorbs $\delta \phi_p$ in $\Psi$.  This is consistent with the way we arrived at the D3 brane through the D5 brane of the KS background via a deformed conifold where the base is an $S^3\times S^2$ and the diffeomorphism gauge is fixed. The $\tau \rightarrow 0$ limit shrinks the $S^2$ and yields $M$ fractional D3 branes.  From this point of view, $\theta_m$ and $\phi_p$ were already fixed  and the absence of any fluctuations of these fields is equivalent to there being no residual gauge freedom in fixing the coordinates.  One might wonder whether the absence of field equations for $\theta_m$ and $\phi_p$ could be related to a degenerate coordinate choice.  Indeed by following ~\cite{Bigazzi:2004ze}, and recalculating the Lagrangian after applying the following coordinate transformation
\begin{align}
  W \equiv \theta_m \cos\phi_p \nonumber\\
  Z \equiv \theta_m \sin\phi_p
\end{align}
we again find no field equations for $\delta W$ or $\delta Z$, up to  total derivatives.

\subsection{Bosonic Eigenvalues}\label{KSbosoniceigenvalues}
We now set out to solve the bosonic Eqs.~(\ref{eq:Xeqm})~-~(\ref{eq:Aeqm}).
Notice that the composite field $\Psi$ looks like a tachyon with an electric source $\delta F_{tx}$.  With the definition of the Riemann curvature tensor
\begin{equation}\label{eq:Riemann}
  ?R^a_bcd? \delta A^b = [\nabla_{c}, \nabla_{d}] \delta A^{a},
\end{equation}

\noindent we can cast the $U(1)$ gauge field Eqs.~(\ref{eq:Aeqm}) into the following form:
\begin{align}\label{eq:Aeqmsimplified}
4\pi j_{b} &= \nabla^{a}\delta F_{ab} = \nabla^{a}(\nabla_{a} \delta A_{b} - \nabla_{b} \delta A_{a}) \nonumber\\
                       &= \nabla^{a}\nabla_{a}\delta A_{b} - \nabla_{b} \nabla_{a} \delta A^{a} - ?R^a_cab?\delta A^{c} \nonumber\\
                       &= \nabla^{a}\nabla_{a}\delta A_{b} - \nabla_{b} \nabla_{a} \delta A^{a} - R_{cb}\delta A^{c},
\end{align}
\noindent where we have used
\begin{align}
    R_{cb} \equiv R^{a}_{~cab}
\end{align}
We can further simplify these three equations by noticing that the Ricci tensor has only two non-vanishing components:
\begin{align}\label{eq:RicciComponents}
R_{\theta\theta} &= 1,~~~R_{\phi\phi} = \sin^2(\theta),~~~\mbox{all others zero}.
\end{align}

Working in the temporal gauge, $\delta A^{t} = 0$, the Gauss's law constraint is identified in Eq.~(\ref{eq:Aeqmsimplified}) as
\begin{align}\label{eq:U1constraint1}
          \nabla_t \nabla_{a} \delta A^{a} = -4\pi g_{xx} Q_{\Psi} \nabla_x \Psi.
\end{align}

We try the following ansatz for $\delta A_{a}$, $\Psi$, and $\delta\tau$:
\begin{align}
   \label{eq:solA}
   \delta A_j &= \int dp~d\omega \sum_{l=0}^{\infty}\sum_{m=-l}^{m=l} \tilde{A_j}^{(lm)} (p,\omega)~e^{i(px - \omega t)}~Y_j^{(lm)}(\theta,\phi),~~j = x,\theta,\phi  \\
   \label{eq:solPsi}
   \Psi &= \int dp~d\omega \sum_{l=0}^{\infty}\sum_{m=-l}^{m=l} \tilde{\Psi}^{(lm)}(p,\omega)~e^{i(px - \omega t)}~Y_{(lm)}(\theta,\phi) \\
   \label{eq:soltau}
   \delta\tau &= \int dp~d\omega \sum_{l=0}^{\infty}\sum_{m=-l}^{m=l} \tilde{\tau}^{(lm)} (p,\omega)~e^{i(px - \omega t)}~Y_{(lm)}(\theta,\phi),
\end{align}

\noindent where the $Y_j^{(lm)}(\theta,\phi)$ are
\begin{align}\label{eq:vectorYlm}
Y_x^{(lm)} &\equiv Y_{(lm)}(\theta,\phi)\nonumber\\
Y_{\theta}^{(lm)} &\equiv \frac{\csc{\theta}}{\sqrt{l(l+1)}}\partial_{\phi} Y_{(lm)}(\theta,\phi)\nonumber\\
Y_{\phi}^{(lm)} &\equiv \frac{-\sin{\theta}}{\sqrt{l(l+1)}}\partial_{\theta} Y_{(lm)}(\theta,\phi),
\end{align}

\noindent and $Y_{\theta}^{(lm)}$ and $Y_{\phi}^{(lm)}$ are \emph{vector} spherical harmonics which satisfy the eigenvalue equation
\begin{align}\label{eq:vectorYlmeigen}
   \hat{L}^2 Y_j^{(lm)} &= [-l(l+1) + 1] Y_j^{(lm)}, j=\theta,\phi
\end{align}

\noindent where
\begin{align}
   \hat{L}^2 &= \frac{1}{\sin{\theta}}\partial_{\theta}\sin{\theta}\partial_{\theta} + \frac{1}{\sin^2{\theta}}\partial_{\phi}^2.
\end{align}

Using an ansatz with $\tilde{A_{\theta}} = \tilde{A_{\phi}}$, the Gauss law constraint Eq.~(\ref{eq:U1constraint1}) becomes simply
\begin{align}\label{eq:U1constraint2}
    \nabla_x \nabla_t\delta A_x = -4\pi g_{xx}^2 Q_{\Psi} \nabla_x \Psi.
\end{align}

\noindent This can be used to simplify the $b=x$ component of Eq.~(\ref{eq:Aeqmsimplified}) to the non-dynamical form
\begin{align}
   \hat{L}^2 \tilde{A_x} = 0
\end{align}

\noindent which means $l=0$ for the coupled fields $\delta A_x$ and $\Psi$.  The equation of motion for $\Psi$, Eq.~(\ref{eq:Psieqm}), then becomes the eigenvalue equation
\begin{align}\label{eq:beigenproblem1}
   [\omega^2 - p^2 - m_{\Psi}^2] \tilde{\Psi} = 0
\end{align}
where
\begin{equation}\label{eq:Psimass}
   m_{\Psi}^2 = 4\pi g_{xx}^2 Q_{\Psi}^2 \frac{c_A}{2 c_{\tau}} - R
               = \frac{4}{b^3 M\alpha' g_s}\left(1-b^2 + \frac{2}{9 f_2(\psi_0)}\right)   \end{equation}

\noindent is always positive:
  \[\frac{2.78}{g_s M \alpha'} \le m_{\Psi}^2 < \infty.\label{eq:Psimassrange}
\]

Next, the solution Eqs.~(\ref{eq:solA}) and~(\ref{eq:soltau}) for the coupled fields $\delta A_{\theta}$, $\delta A_{\phi}$, and $\delta\tau$, reduce the $b=\theta,\phi$ components of Eq.~(\ref{eq:Aeqmsimplified}) and Eq.~(\ref{eq:taueqm}) to the eigenvalue problem
\begin{align}\label{eq:beigenproblem2}
&[\omega^2 - p^2 - g_{xx}\frac{R}{2}l(l+1)] \tilde{A_i} +g_{xx}Q_\tau\frac{8\pi}{R}\sqrt{l(l+1)}\tilde{\tau} = 0,~~~i=\theta,\phi \nonumber\\
&[\omega^2 - p^2 - g_{xx}\frac{R}{2}l(l+1) - g_{xx}m_{\tau}^2]\tilde{\tau} + g_{xx} Q_\tau \frac{c_A}{2 c_\tau}\sqrt{l(l+1)}\tilde{A_{\theta}} = 0.
\end{align}

Finally, the two massless equations~(\ref{eq:Xeqm}) can be solved with
\begin{equation}\label{eq:solX}
   \delta X^j = \int dp~d\omega~\tilde{X^j}_{(lm)}(p,\omega) e^{i(p x - \omega t)}Y_{(lm)}(\theta,\phi)~~~j=2,3
\end{equation}

\noindent yielding two identical eigenvalue problems
\begin{align}\label{eq:beigenproblem3}
   [\omega^2 - p^2 - g_{xx}\frac{R}{2}l(l+1)]\tilde{X^j} = 0,~~~j=2,3.
\end{align}

We now organize Eqs.~(\ref{eq:beigenproblem1}),~(\ref{eq:beigenproblem2}), and~(\ref{eq:beigenproblem3}) into a succinct system of six scalar bosons:
\begin{align}\label{eq:beigenproblem}
\mbox{{\scriptsize $\omega^2 \left(\begin{array}{l}
             \tilde{\Psi}\\
             \tilde{X^2}\\
             \tilde{X^3}\\
             \tilde{\tau}\\
             \tilde{A_{\theta}} \\
             \tilde{A_{\phi}}
          \end{array}
   \right)$}} &= \mbox{\scriptsize{$
                      \left(
                      \begin{array}{c c c c c c}
                      \omega_1^2 & 0 & 0 & 0 & 0 & 0\\
                      0 & \omega_2^2 & 0 & 0 & 0 & 0\\
                      0 & 0 & \omega_2^2 & 0 & 0 & 0\\
                      0 & 0 & 0 & \omega_2^2 + g_{xx}m_{\tau}^2 & -g_{xx}Q_\tau\frac{c_A}{2c_\tau}\sqrt{l(l+1)} & 0\\
                      0 & 0 & 0 & -g_{xx}Q_\tau\frac{8\pi}{R}\sqrt{l(l+1)} & \omega_2^2 & 0 \\
                      0 & 0 & 0 & -g_{xx}Q_\tau\frac{8\pi}{R}\sqrt{l(l+1)} & 0 & \omega_2^2
                      \end{array}
                      \right)$}}
   \mbox{{\scriptsize $
            \left(\begin{array}{l}
             \tilde{\Psi}\\
             \tilde{X^2}\\
             \tilde{X^3}\\
             \tilde{\tau}\\
             \tilde{A_{\theta}}\\
             \tilde{A_{\phi}}
          \end{array}
   \right)$}}
\end{align}

\noindent where
\[
  \omega_1^2 = p_x^2 + g_{xx}m_{\Psi}^2 \qquad \mbox{and} \qquad
  \omega_2^2 = p_x^2 + g_{xx}\frac{R}{2}l(l+1).
\]

\noindent  The six eigenvalues of Eq.~(\ref{eq:beigenproblem}) are
\begin{align}\label{eq:bosoniceigenvalues}
  \omega^2 &= \left\{ \begin{array}{l}
                       p_x^2 + g_{xx}m_{\Psi}^2 \\
                       p_x^2 + g_{xx}\frac{R}{2}l(l+1)~\mbox{3-fold degenerate} \\
                       p_x^2 + \mu_{\pm}^{2}(l,\psi_0)
                       \end{array}
              \right.
\end{align}

\noindent where
\begin{align}\label{eq:muPlusMinus}
  \mu_{\pm}^{2}(l,\psi_0) &= g_{xx}\frac{R}{2}l(l+1)(1 + f_{\pm}(l,\psi_0)\ge 0, \\
   f_{\pm}(l,\psi_0) &= \frac{f_1(\psi_0)}{l (l+1)}\left(1 \pm \sqrt{1 + \frac{f_2(\psi_0)}{f_1^2(\psi_0)}l(l+1)}\right).
\end{align}

\noindent The effective mass $\mu_+^2(l,\psi_0) >0$ for all $l \ge 0$, where as $\mu_-^2(l,\psi_0)>0$ for all $l >0$, as $\mu_-^2(0,\psi_0) = 0$.
\subsection{Bosonic One Loop Energy of a D3-brane in the KS-background}\label{KSBosonicOneLoopEnergy}
Using fixed quark boundary conditions, as in Eq. (\ref{eq:vanishingBCs}),
\begin{align}
   p_x &= \frac{n \pi}{L}
\end{align}
\noindent we calculate the bosonic contribution to the one loop energy, Eq.~(\ref{eq:deltaEb}), to be
\begin{align}\label{eq:KSdeltaEb}
    \delta E_b &= \frac{1}{2}\sum \omega \nonumber\\
           &=\frac{1}{2}\sum_{n=1}^\infty\sum_{l=0}^\infty\sum_{m=-l}^l \left[ \delta_{l}^{~0}\sqrt{\left(\frac{n\pi}{L}\right)^2+g_{xx}m_{\Psi}^2} +3\sqrt{\left(\frac{n\pi}{L}\right)^2 + g_{xx}\frac{R}{2}l(l+1)}+ \right.\nonumber\\
             &\left.~~~~~~~~~~~~~~~~~~~~~~~~~~~~~+\sqrt{\left(\frac{n\pi}{L}\right)^2 + \mu_{-}^{2}(l,\psi_0)} + \sqrt{\left(\frac{n\pi}{L}\right)^2 + \mu_{+}^{2}(l,\psi_0)}~\right].
\end{align}

\noindent Performing the trivial sum over 
$m$ and splitting the $l$ sum into $l=0$ and $l>0$ modes leaves us with $d +p -3 = 4$ massless eigenmodes{\linespread{1.0}\footnote{Three massless modes clearly come from the second term in Eq. (\ref{eq:KSdeltaEb}): two from the Minkowski fields, $\delta X^i$, and one from the $U(1)$ gauge fields, $\delta A_a$.  The other massless mode from the gauge fields comes from the third term as we see from Eq. (\ref{eq:muPlusMinus}) that $\mu_-^2(0,\psi_0) = 0$}} and an infinite tower of massive eigenmodes:
\begin{align}
   \delta E_b &= \frac{1}{2}\sum_{n=1}^\infty\left((d+p-3)\frac{n\pi}{L} + \sqrt{\left(\frac{n\pi}{L}\right)^2 + g_{xx}m_{\Psi}^2} + \sqrt{\left(\frac{n\pi}{L}\right)^2 + g_{xx}m_{\tau}^2}\right)+\nonumber\\
 &~~+\frac{1}{2}\sum_{l=1}^\infty (2 l+1) \sum_{n=1}^\infty\left[ 3\sqrt{\left(\frac{n\pi}{L}\right)^2 + g_{xx}\frac{R}{2}l(l+1)}~+\right.\nonumber\\
           &\left.~~~~~~~~~~~~~~~~~~~~~~~~~~~~+\sqrt{\left(\frac{n\pi}{L}\right)^2 + \mu_{-}^{2}(l,\psi_0)}+ \sqrt{\left(\frac{n\pi}{L}\right)^2 + \mu_{+}^{2}(l,\psi_0)}~\right]
\end{align}

We pause here to reflect on how the massless eigenmodes came about.  There are $d-2 = 2$ massless modes from the Minkowski fields $\delta X^i$ that were not statically fixed to the D-brane parameters.  There were in addition $p-1 = 2$ massless modes from the $U(1)$ gauge fields, $\delta A_a$, after gauge fixing.  This leaves us with the $d-2 + p-1 = d+p-3 = 4$ massless modes we see here.  

Continuing with our calculation of $\delta E_b$, we perform the $n$ sum, in the large $L$ limit, by regulating the massless modes using Eq. (\ref{eq:SumofAllPositiveIntegersRegularization}) and the massive modes using Eq. (\ref{eq:MassiveModeRegularization}), leaving us with
\begin{align}\label{eq:KSdeltaEbnsumperformed}
    \delta E_b &= -\frac{(d + p-3)\pi}{24 L} - \frac{\sqrt{g_{xx}}m_{\Psi}}{4}  - \frac{\sqrt{g_{xx}}m_{\tau}}{4}+\nonumber\\
    &~~~~~~-\frac{1}{4} \sum_{l=1}^\infty(2l+1)\left[3\sqrt{g_{xx}\frac{R}{2}l(l+1)} + \mu_-(l,\psi_0) + \mu_+(l,\psi_0)\right].
\end{align} 
\noindent Notice, that in the large $L$ limit, only the massless modes contribute to the $1/L$ L\"uscher term, the remnants of the massive mode sums all now shown to be constant with respect to $L$.  These remaining sums can be regulated using Eqs. (\ref{eq:SphericalEigenmodeRegularization}) and (\ref{eq:EmbeddedSquareRootRegularization}), resulting in:
\begin{align}
           \delta E_b    &= -\frac{(d+p-3)\pi}{24L} + \beta^3_b(k,M) \\
    \label{eq:KSbeta3}
    \beta^3_b(k,M) &=  - \frac{1}{4} \sqrt{g_{xx}}(m_{\Psi}  + m_{\tau}) -\frac{1}{4}\left[3 \tilde{f}_\zeta\left(-\frac{1}{2}\right)+\tilde{g}_-\left(-\frac{1}{2}\right)+\tilde{g}_+\left(-\frac{1}{2}\right) \right] ,
\end{align}

\noindent where the regularizations of the functions $f_\zeta(s)$  and $\tilde{g}_\pm(s)$, calculated in  App.~\ref{TwoSphereRegularization}, are:
\begin{align}
   \tilde{f}_\zeta(s) &\to -0.265096 \\
\tilde{g}_\pm\left(-\frac{1}{2}\right) &\to \sum_{w=0}^\infty \sum_{q=0}^{[\frac{w}{2}]}\sum_{n=0}^q \binom{q}{n}\frac{\Gamma{(\frac{3}{2})}f_2^n f_1^{w-2n}}{\Gamma{(\frac{3}{2}-w)}}\frac{\tilde{f}_\zeta(w-n-\frac{1}{2})}{(2q)!\Gamma{(w+1-2q)}}+ \nonumber\\
&~~\pm \sum_{w=1}^\infty\sum_{q=0}^{[\frac{w-1}{2}]}\frac{\Gamma{(\frac{3}{2})}f_1^w}{\Gamma{(\frac{3}{2}-w)}}\frac{g_\zeta(q,w)}{(2q+1)!\Gamma(w-2q)}
+ h_\pm \nonumber\tag{\ref{eq:EmbeddedSquareRootRegularization}}
\end{align}

\noindent with
\begin{align}\label{eq:KSgzeta}
g_\zeta(q,w) &=  \left\{\begin{array}{ll}\sum_{n=0}^\infty\binom{q+\frac{1}{2}}{n}\left(\frac{f_2}{f_1^2}\right)^{q-n+\frac{1}{2}}\left[\tilde{f}_\zeta(w+n-q-1) + \right.& \\
\left.~~~~~~~~~~~~~~~-\sum_{l=1}^{\{l_f\}}\frac{(2l+1)}{[l(l+1)]^{w+n-q-1}}\right]  + & 0 < \frac{k}{M} < 1\\
~~~~~~~~+  \sum_{l=1}^{\{l_f\}}\frac{(2l+1)\left(1 + \frac{f_2}{f_1^2}l(l+1)\right)^{q + \frac{1}{2}}}{[l(l+1)]^{w-\frac{1}{2}}} &   \\     \sum_{n=0}^\infty\binom{q+\frac{1}{2}}{n}\left(\frac{f_2}{f_1^2}\right)^n\tilde{f}_\zeta\left(w-n-\frac{1}{2}\right) & k=0,M,
                                  \end{array}
            \right. 
\end{align}
\noindent and
\begin{align}
h_+ &= \left\{\begin{array}{ll}
                  2\sqrt{2}\left(\sqrt{1 + f_+(1)} - \sum_{w=0}^\infty \binom{\frac{1}{2}}{w}f_+^w(1)\right) &f_+(1) > 0 \\
                  0 & f_+(1) < 0,
                 \end{array}
        \right.  \label{eq:KShplus}
h_- &= 0,
\end{align}
\noindent where $\{l_f\}$ is the largest integer less than $l_f = \frac{1}{2}\left(-1 + \sqrt{1+\frac{4f_1^2}{f_2}}\right)$, as graphed versus $k/M$ in Fig.~\ref{fig:KSlf}.

The constant $\beta_3^b$ is constant with respect to $L$, but depends on $k$ and $M$ through $f_2$, $f_1$, $f_+$, $g_{xx} R/2$, and $g_{xx} m_\psi^2$ and the transcendental Eq.~(\ref{eq:psimincondition}).  As seen in Figs.~(\ref{fig:KS_f1f2}), (\ref{fig:KS_fpluslkM}), and (\ref{fig:KS_masses}), these are all symmetric under $k \to k - M$ and thus so is the constant $\beta_3^b$, a symmetry which we indeed expect to be manifest in the $k$-string energy~\cite{PandoZayas:2008hw}.
\begin{figure}[htbp]
\addtocontents{lof}{\protect\vspace{\li}}
  \centering
  \includegraphics[width = 0.7 \columnwidth]{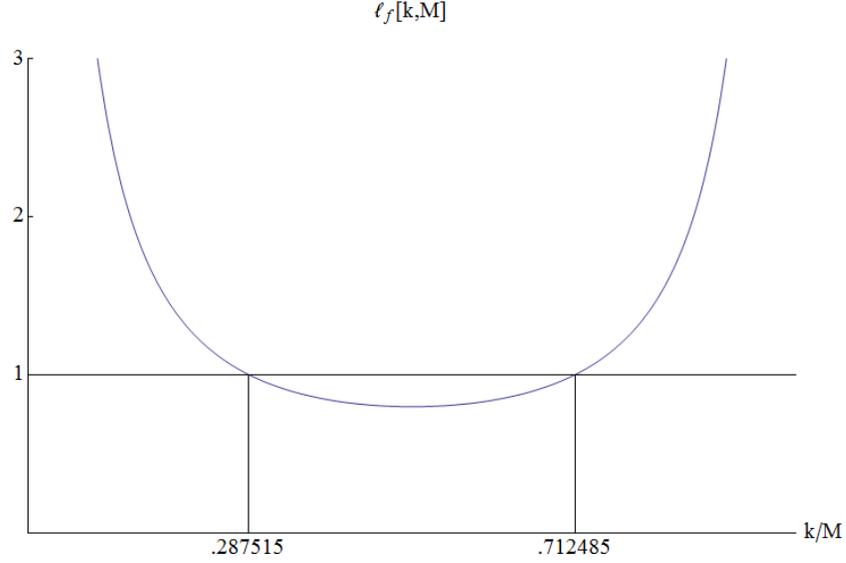}
  \caption{Solution for $l_f = \frac{1}{2}\left(-1 + \sqrt{a + \frac{4f_1^2}{f_2}}\right)$.  We see for $0.288 \lesssim k/M \lesssim 0.712$, $l_f <1$ and Eq.~(\ref{eq:KSgzeta}) simplifies in this case, as the latter two sums vanish.  Also, $l_f$ clearly respects the $k \to M-k$ symmetry.}
\label{fig:KSlf}
\end{figure}

\begin{figure}[htbp]
\addtocontents{lof}{\protect\vspace{\li}}
  \centering
  \includegraphics[width = 0.7\columnwidth]{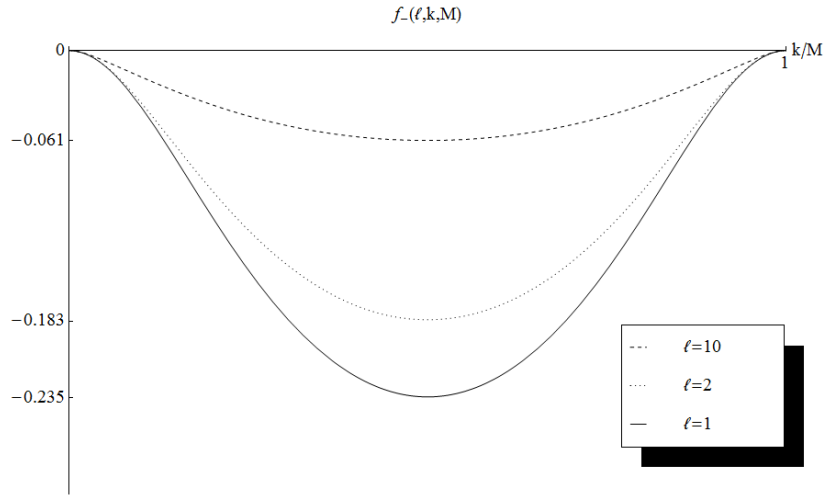}
\caption{$|f_-| < 0$ for all $l \ge 1$, and shrink to zero as $l$ increases.  The $k \to k-M$ symmetry is respected by $f_-$.}
\label{fig:KS_fminuslkM}
\end{figure}

\begin{figure}[htbp]
\addtocontents{lof}{\protect\vspace{\li}}
\centering
 \includegraphics[width = 0.85\columnwidth]{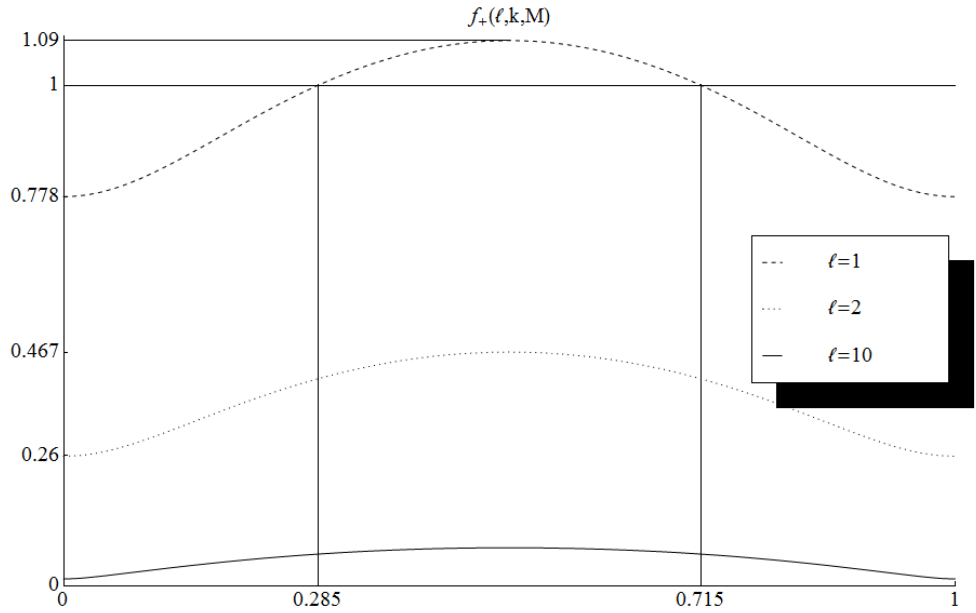}
 \caption{$f_+ >1$ only for $0.285 \lesssim k/M \lesssim 0.715$, where as a result, $h_+ \ne 0$ as seen in Eq.~(\ref{eq:KShplus}).  Also, $f_+$ respects the $k \to k -M$ symmetry.}
\label{fig:KS_fpluslkM}
\end{figure}

\begin{figure}[htbp]
\addtocontents{lof}{\protect\vspace{\li}}
\centering
  \includegraphics[width = 0.85\columnwidth]{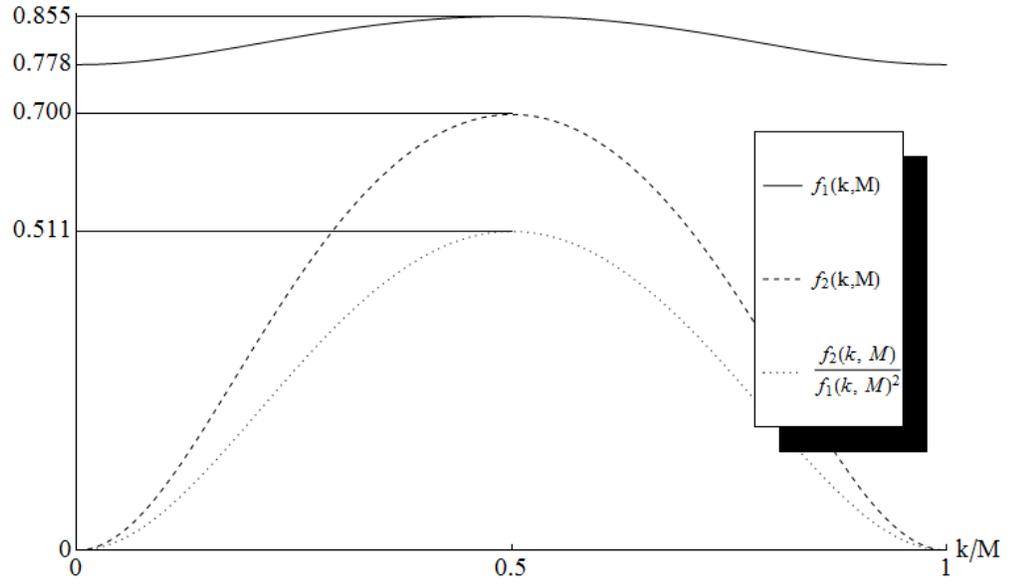}
  \caption{The functions $f_1$, $f_2$, and the ratio $\frac{f_2}{f_1^2}$ all respect the $k \to k-M$ symmetry.}
\label{fig:KS_f1f2}
\end{figure}

\begin{figure}[htbp]
\addtocontents{lof}{\protect\vspace{\li}}
   \centering
  \includegraphics[width = 0.85\columnwidth]{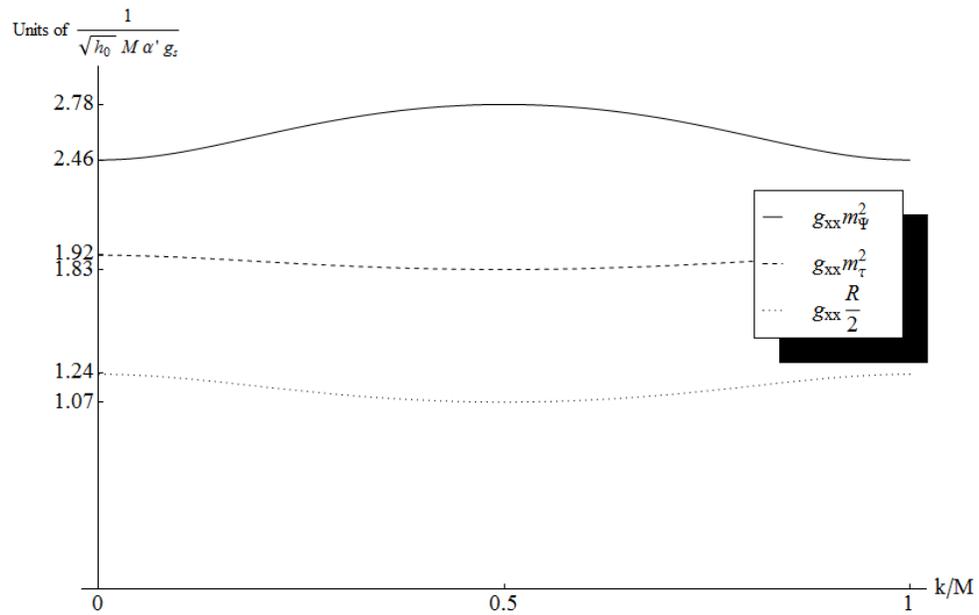}
\caption{These are all finite and respect the $k \to k-M$ symmetry.}
\label{fig:KS_masses}
\end{figure}
\newpage
\section{D4-brane Bosonic Fluctuations in the CGLP Background}\label{app:CGLPfluctuations}
The fluctuations
\begin{align}
   x^2(\zeta) &= 0 + \delta x^2(\zeta),~~~\psi(\zeta) = \psi_0 + \delta\psi(\zeta), \nonumber\\
      \tau(\zeta) &= \tau_0 + \delta\tau(\zeta),~~~\mu^i(\zeta) = \mu^i_0 + \delta\mu^i(\zeta), \nonumber\\
  F &= \frac{E}{2\pi\alpha' H_0^{1/2}}dt\wedge dx + \partial_a \delta A_b d\zeta^a \wedge d\zeta^b
\end{align}
\noindent lead to the action for bosonic fluctuations of a D4-brane in the CGLP background
\begin{align}\label{eq:deltaS4b}
    \delta S_4^b &=  -\mu_4 \int d^5\zeta e^{-\Phi_0}\sqrt{\mathcal{M}^{(0)}}(\Delta-\delta\Phi) +  \mu_4 n\int \bigg(\delta C_3 \wedge \mathcal{F}^{(0)} + C_3^{(0)} \wedge \delta\mathcal{F} + \nonumber\\
   &~~~~~~~~~~~~~~~~~~~~~~~~~~~~~~~~~~~~~~~~~~~~~~~~~~~~~~~~~~~~~~~~~~~~~~~~~+ \delta C_3 \wedge \delta\mathcal{F}^{(0)}\bigg)
\end{align}

\noindent where $\Delta$ is defined through Eq.~(\ref{eq:Delta}).  The $\mu^i_0$ refer to the classical value for the field, specified in Eq.~(\ref{eq:CGLPclassicalscalars}), and the three $\mu^i$ are still constrained by $(\mu^i)^2 = 1$.  In the action, Eq.~(\ref{eq:deltaS4b}), the classical fields with subscript or superscript $(0)$ are as in section~\ref{CGLPtension}, and for the fluctuations we have
\begin{align}\label{eq:CGLPginducedfluctuated}
   \delta g_{ab}d\zeta^a d\zeta^b  &= H_0^{1/2}\frac{l^2}{2}\sin(2\psi_0) d\Omega_3^2 \delta\psi+ \delta\psi^2 H_0^{1/2}\frac{l^2}{2}\cos(2\psi_0) d\Omega_3^2 + \nonumber\\
  &~~~+ \left(H_0^{-1/2} \frac{\partial \delta x^2}{\partial\zeta^a}\frac{\partial\delta x^2}{\partial\zeta^b} + H_0^{1/2}l^2\left(\frac{\partial\delta\tau}{\partial\zeta^a}\frac{\partial\delta\tau}{\partial\zeta^b} + \frac{1}{2}\frac{\partial\delta\psi}{\partial\zeta^a}\frac{\partial\delta\psi}{\partial\zeta^b}\right)\right)d\zeta^a d\zeta^b + \nonumber\\
   &~~~+ \delta\tau^2\left[H_0^{1/2}l^2\left(A^i_{\alpha} A^i_{\beta} d\Omega_3^{\alpha} d\Omega_3^{\beta} + \sin^2\psi_0(1 - \frac{H_2}{4H_0})d\Omega_3^2\right)\right. + \nonumber\\
          &~~~~~~~~~~~~~\left. +\frac{H_2}{2H_0^{3/2}}(-dt^2 + dx^2)\right]\mbox{$i$ = 1,2},
\end{align}

\noindent and
\begin{align}\label{eq:CGLPB2fluctuated}
   \delta B_2 &= \frac{m}{8l}\sin^2\psi_0~\sin\chi~\delta\tau~d\chi \wedge d\theta +\nonumber\\
   &~~~+\frac{m}{8l}\delta\tau~\biggl(\delta\mu^1 \sin^2\psi_0~ \sin^2\chi~\sin\theta~d\theta \wedge d\phi + \delta\mu^2 \sin^2\psi_0~\sin\chi~\sin\theta~d\phi\wedge d\chi + \biggr.\nonumber\\
             &~~~~~~~~~~~~~~~\biggl.+\sin\psi_0~\sin\chi~\sin\theta~\frac{\partial\delta\psi}{\partial\zeta^a}d\zeta^a \wedge d\phi + \delta\psi~\sin(2\psi_0)\sin\chi~d\chi\wedge d\theta\biggr),
\end{align}

\noindent and
\begin{align}\label{eq:CGLPC3fluctuated}
  \delta C_3 &= \frac{3m}{8 g_s}\delta\psi~\sin^3\psi_0~d\Omega_3 +\nonumber\\
  &~~~+\frac{m}{16 g_s}\left(9 \sin^2\psi_0~\cos\psi_0 \delta\psi^2 - \frac{A^1_{\chi}J^1_{\theta\phi} + A^2_{\theta}J^2_{\phi\chi}}{\sin^2\chi~\sin\theta}\delta\tau^2\right)d\Omega_3
\end{align}

As the dilaton depends on $\tau$ through Eq.~(\ref{eq:dilaton}), the fluctuation of the dilaton, to second order in the tau fluctuation, is
\begin{align}\label{eq:dilatonfluctuation}
   \delta\Phi &= - \frac{H_2}{H_0}\delta\tau^2
\end{align}

Using Eq.~(\ref{eq:CGLPginducedfluctuated}) through Eq.~(\ref{eq:dilatonfluctuation}), we calculate the bosonic action, Eq.~(\ref{eq:deltaS4b}), to second order in the fluctuations:
\begin{align}
  \label{eq:S2bosonic}
  \delta S_4^b &= -\int \sqrt{-\det(g^{(eff)})}d^5\zeta\biggl\{c_x \nabla_a\delta x^2\nabla^a \delta x^2 + c_{\psi} \left[\nabla_a\delta\psi\nabla^a \delta\psi - \frac{R}{2} \delta\psi^2\right] \biggr. + \nonumber\\
            &~~~+ c_{\tau}\left[\nabla_a\delta\tau\nabla^a \delta\tau + m_{\tau}^2(\chi,\theta)\delta\tau^2\right]+ c_A\left[\frac{1}{16 \pi} \delta F^{ab} \delta F_{ab} + j^a \delta A_a \right] + \nonumber\\  &~~~ \biggl. + \mbox{total derivatives} \biggr\},
\end{align}

\noindent Here we notice that the linear fluctuations vanish, up to total derivatives, signifying that we are truly fluctuating around a classical solution.  The covariant derivatives are with respect to $g^{(eff)}$, an effective metric on the D4-brane
\begin{align}\label{eq:CGLPgeff}
   ds^2 &= g^{(eff)}_{ab} d\zeta^a d\zeta^b = g_{xx}(-dt^2 + dx^2) + \frac{6}{R}d\Omega_3^2, \nonumber\\
   g_{xx} &= \frac{12^3 A^2 l^6}{H_{min}^2 I_0^2 R^3 m^4},
\end{align}
\noindent where $R$ is the same scalar curvature, Eq.~\ref{eq:CGLPRscalar}, as the induced metric. The $U(1)$ gauge current $j^a$, and $m_{\tau}(\chi,\theta)$ are
\begin{align}
   j^a &= \left(-Q_{\psi} \nabla_x \delta\psi,~Q_{\psi} \nabla_t \delta\psi, ~ Q_{\tau} \frac{\nabla_{\theta}(\sin\theta \delta\tau)}{\sin\chi~\sin\theta},~ -Q_{\tau} \frac{\nabla_{\chi}(\sin\chi~\delta\tau)}{\sin^2\chi}, 0\right), \\
   m_{\tau}^2(\chi,\theta) &= m_{\tau 0}^2 + \frac{R}{6}\csc^2\chi \csc^2\theta,
\end{align}

\noindent and the various constants are
\begin{align}
 c_x &= \frac{\mu_4 R^{3/2}l^3 H_{min}}{48 \sqrt{3} g_s A} = \frac{2 l^4}{I_0 m^2}c_{\psi} = \frac{l^4}{I_0 m^2}c_{\tau} = \frac{l^3}{32 I_0^1/2 \pi^3 m\alpha'^2}c_A, \nonumber\\
  Q_{\tau} &= \frac{3 R}{4 I_0^{1/2} \pi^2 12^2 \alpha'},~~~Q_{\psi} = \frac{9 H_{\min}^3 I_0^{9/4} R^{9/2} m^{11/2}}{8\sqrt{6} A^3 \pi^2 l^{15/2} 12^5 \alpha'} \nonumber\\
  m_{\tau 0}^2 &= \frac{l}{128 m I_0^{3/2}}(43 - 48 I_0) + \frac{35}{96}R.
\end{align}

Applying the variational principle to the action, Eq.~(\ref{eq:S2bosonic}), results in the field equations:
\begin{align}\label{eq:x2}
   &\nabla^2\delta x^2 = 0 \\
   \label{eq:KSPsieqm}
   &\nabla^2\delta\psi + \frac{R}{2}\delta\psi + \frac{c_A Q_{\psi}}{2 c_{\psi}}\delta F_{tx} = 0 \\
   \label{eq:tau}
  &\nabla^2\delta\tau - m_{\tau}^2(\chi,\theta)\delta\tau + \frac{c_A Q_{\tau}}{2 c_{\tau}} \csc\chi \delta F_{\theta\chi} = 0 \\
  \label{eq:FU1}
   &\nabla_{a}\delta F^{ab} - 4 \pi j^{b} = 0,
\end{align}

\noindent where $\nabla_a$ is the covariant derivative compatible with Eq.~(\ref{eq:CGLPgeff}).

The solution to Eq.~(\ref{eq:x2}) is
\begin{align}\label{eq:x2solution}
   \delta x^2 &= \int d\omega dp \sum_{n \ge l \ge |m|} {\tilde{x}}^{(n,l,m)}(p,\omega) e^{i(p x - \omega t)}Y^{nlm}(\chi,\theta,\phi),
\end{align}

\noindent where the $Y^{nlm}(\chi,\theta,\phi)$ are the spherical harmonics on an $S^3$~\cite{Higuchi:1986wu}
\begin{align}
   Y^{nlm}(\chi,\theta,\phi) &= c_{nl}\frac{1}{\sqrt{\sin\chi}} P^{l+1/2}_{n+1/2}(\cos\chi) Y^{(lm)}(\theta,\phi), \nonumber\\
   c_{nl} &= \sqrt{\frac{(n+1)(n+l+1)!}{(n-l)!}},
\end{align}

\noindent and $P^{l}_{n}(x)$ are the associated Legendre polynomials. The $S^3$ spherical harmonics \newline $Y^{nlm}(\chi,\theta,\phi)$ satisfy the eigenvalue problem
\begin{align}
   \tilde{\nabla}^2 Y^{nlm}(\chi,\theta,\phi) &= -n(n+2)Y^{nlm}(\chi,\theta,\phi),
\end{align}

\noindent where $\tilde{\nabla}^2$ is the Laplacian for an $S^3$ whose action on scalar functions such as $Y^{nlm}(\chi,\theta,\phi)$ is explicitly given by
\begin{align}
   \tilde{\nabla}^2 &= \frac{1}{\sin^2\chi}\left(\partial_{\chi}(\sin^2\chi \partial_{\chi}) + \frac{1}{\sin\theta}\partial_{\theta}(\sin\theta \partial_{\theta}) + \frac{1}{\sin^2\theta}\partial^2_{\phi} \right).
\end{align}

With that said, solving Eq.~(\ref{eq:x2}) with the solution Eq.~(\ref{eq:x2solution}) results in the eigenvalue problem
\begin{align}
  &\left[\frac{1}{g_{xx}}(\omega^2 - p^2) - \frac{R}{6}n(n+2) \right]\tilde{x} = 0
\end{align}

The rest of the equations prove quite difficult and require perturbation theory to solve; their solution is not given here.  To simplify the problem, we recall the important physical features that were found for a D3-brane in the KS background in section~\ref{KSBosonicOneLoopEnergy}~\cite{PandoZayas:2008hw}.  There we found the massless modes, from which the L\"uscher term was derived, to be independent of the angular degrees of freedom.  Inspired by these results, we propose that in the current case of a D4-brane in the CGLP background, we can integrate out the spherical degrees of freedom, $\chi$, $\theta$, and $\phi$, and still have the same number of massless modes as before and as a result, the same L\"uscher term as would be calculated from the full five dimensional theory.

To proceed with this integration, we consider the fluctuations to be independent of the $S^3$ variables,
\begin{align}
   \delta X^{\mu} &= \delta X^{\mu}(t,x),~~~\delta A_{a} = \delta A_{a}(t,x)
\end{align}

\noindent and we integrate out the $S^3$ from the action Eq.~(\ref{eq:S2bosonic}). This results in an effective action
\begin{align}\label{eq:S2bosoniceff}
   \delta S^b_{4eff} &= - V_3 \int dt~dx \biggl\{c_x \partial_m\delta x^2\partial^m \delta x^2 + c_{\psi} \left[\partial_m\delta\psi\partial^m \delta\psi - \frac{R}{2} g_{xx} \delta\psi^2\right] \biggr. + \nonumber\\
            &~~~+ c_{\tau}\left[\partial_m\delta\tau\partial^m \delta\tau + m_{\tau e}^2\delta\tau^2\right]+ c_A\left[\frac{1}{g_{xx}16 \pi} \delta F^{mn} \delta F_{mn} + g_{xx} j^m \delta A_m \right] + \nonumber\\
            &~~~ + \frac{c_A}{16 \pi}(\partial_m \delta A_{\chi}\partial^m \delta A_{\chi} + 2\partial_m \delta A_{\theta}\partial^m \delta A_{\theta} + I_1 \partial_m \delta A_{\phi}\partial^m \delta A_{\phi}) + \nonumber\\
            &~~~ + \biggl.\mbox{total derivatives} \biggr\},
\end{align}

\noindent where the indices $m$ and $n$ now sum only over the coordinates $t$ and $x$, and are raised and lowered by the two dimensional Minkowski metric
\begin{align}
   \eta_{mn} d\zeta^m d\zeta^n &= -dt^2 + dx^2,
\end{align}

\noindent and the effective $\delta\tau$ mass, $m_{\tau e}$, the constant $V_3$, and the integral, $I_1$ are
\begin{align}
    m_{\tau e}^2 &= g_{xx}\left(m_{\tau 0}^2 + \frac{R}{6}I_1\right), ~~~V_3 = 2\pi^2 \left(\frac{6}{R}\right)^{3/2},~~~I_1 = \int_0^{\pi} \csc\theta~d\theta.
\end{align}

The equations of motion of the action, Eq.~(\ref{eq:S2bosoniceff}), are
\begin{align}
   &(-\partial_t^2 + \partial_x^2) \delta x^2 = 0 \\
   &(-\partial_t^2 + \partial_x^2) \delta \psi + \frac{R}{2}g_{xx} \delta\psi + \frac{c_A g_{xx} Q_{\psi}}{2 c_{\psi}} \delta F_{tx} = 0 \\
   &(-\partial_t^2 + \partial_x^2)\delta\tau - m_{\tau e}^2 \delta \tau = 0 \\
   &(-\partial_t^2 + \partial_x^2) \delta A_i = 0,~~~i = \chi,\theta,\phi \\
   \label{eq:Fmn}
   &\partial_m \delta F^{mn} = 4\pi g_{xx}^2 j^n
\end{align}
To solve these equations, we move to Fourier space
\begin{align}
  &(\omega^2-p^2)\delta x^2 =0 \\
  &(\omega^2-p^2 + \frac{R}{2}g_{xx}) \delta\psi -i \frac{c_A g_{xx} Q_{\psi}}{2 c_{\psi}} (\omega \delta A_x + p \delta A_t) = 0 \\
   &(\omega^2 - p^2 - m_{\tau e}^2) \delta \tau = 0 \\
   &(\omega^2 - p^2) \delta A_i = 0,~~~i = \chi,\theta,\phi \\
   &p^2\delta A_t + p\omega \delta A_x = -i4\pi g_{xx}^2 Q_{\psi} p \delta\psi \\
   &\omega^2\delta A_x + p\omega \delta A_t = -i4\pi g_{xx}^2 Q_{\psi} \omega \delta\psi,
\end{align}

\noindent and work in temporal gauge, $\delta A_t = 0$, where the $\psi$ field becomes massive instead of tachyonic:
\begin{align}
    &(\omega^2 - p^2 - m_\psi^2)\delta \psi = 0
\end{align}

\noindent with
\begin{align}
   m_{\psi}^2 &= \frac{c_A}{c_{\psi}}2\pi Q_{\psi}^2 g_{xx}^3 -\frac{R}{2}g_{xx}   \nonumber\\
  &= \frac{3 l^4}{4 I_0^2 m^2}\left(3  - \frac{8 I_0}{8 I_0 \cos^2\psi_0 + \sin^2\psi_0}\right) \\
   131 &\lesssim \frac{m^2}{l^4} m_{\psi}^2 \lesssim 141.
\end{align}

The rest of the bosonic equations solve easily and we have the six bosonic eigenmodes:
\begin{align}\label{eq:bosonomegas}
  \omega^2 &= \left\{\begin{array}{l}
                 p^2~~~\mbox{4 fold degenerate}  \\
                 p^2 + m_{\tau e}^2 \\
                 p^2 + m_{\psi}^2
                 \end{array}
              \right..
\end{align}

The calculation for the one loop correction to the bosonic $k$-string energy, $\delta E_b$, follows similarly to the KS calculation in App.~(\ref{KSBosonicOneLoopEnergy}); the result for large quark separation $L$ being
\begin{align}\label{eq:bosononeloop}
  \delta E_b &= -\frac{(d+p-3)\pi}{24 L} - \frac{1}{4}(m_{\tau e} + m_{\psi})
\end{align}

\noindent  where $p=4$, $d=3$, and the L\"uscher term is $\frac{(d+p-3)\pi}{24L} = -\frac{\pi}{6L}$, which is the same as the expected value, as $N$ increases, for lattice calculations done in~\cite{Bringoltz:2008nd}.


\Appendix{FERMIONIC FLUCTUATIONS}
\label{app:FermionicFluctuations}
In this appendix we first lay out the explicit fermionic portion of the D$p$-brane action.  We then show the explicit calculations for the fermionic eigenmodes for the D-branes embedded in the KS and CGLP backgrounds.  First, a quick note on the index convention.  Throughout this thesis, and unless otherwise noted, Latin indices $a,b,c,\dots$ are Dp-brane indices running from $0$ to $p$, Greek indices, $\alpha,\beta,\mu,\nu,\dots$ are 10 dimensional curved indices, and overlined Greek indices, $\overline{\mu},\overline{\nu},\overline{\alpha},\dots$ are 10 dimensional flat indices.  The calculations in this appendix are based on those previously published in~\cite{PandoZayas:2008hw,Doran:2009pp,Stiffler:2009ma}. 

\section{Fermionic Action Definitions}\label{app:FermionicDefinitions}
The fermionic portion of the D$p$-brane action~\cite{Marolf:2003ye,Marolf:2003vf,Marolf:2004jb,Martucci:2005rb}
\begin{align}
  S_{p}& =  -\mu_p \int d^{p+1}\zeta e^{-\Phi}\sqrt{\mathcal{M}} +\mu_p \int \sum_n C_n\wedge \mathcal{F} + \nonumber\\
       &~~~+ \frac{\mu_p}{2}\int d^{p+1}\zeta e^{-\Phi}\sqrt{\mathcal{M}}~\mathcal{L}_p^f(\bar{\Theta},\partial_a\Theta)\nonumber\tag{\ref{eq:DpbraneAction}}
\end{align}

\noindent is the action for a 32 component, Grassmann valued spinor field, $\Theta$:
\begin{align}\label{eq:fermionicaction}
   S_p^f &= \frac{\mu_p}{2}\int d^{p+1}\zeta e^{-\Phi}\sqrt{\mathcal{M}}\mathcal{L}_p^f(\overline{\Theta},\partial_a\Theta)\\
   \mathcal{L}^{f}_p &=  \overline{\Theta} \left(1 -\Gamma_{D_p}\right)\biggl[(\mathcal{M}^{-1})^{ab}\left(\Gamma_a D_b^{(0)} +\Gamma_b W_a\right) - \Delta^{(1)}-\Delta^{(2)}\biggr]\Theta,~~~\mbox{type IIA}\nonumber\tag{\ref{eq:LfIIA}}\\
\mathcal{L}^{f}_{p}&=\overline{\Theta}\biggl[\left(\mathcal{M}^{-1}\right)^{ab}\left(\Gamma _{a}D_{b}^{(0)}-\Gamma_{D_p}^{-1}\Gamma_{b}W_{a}\right) - \Delta^{(1)} + \Gamma_{D_p}^{-1}\Delta^{(2)}\biggr]\Theta,~~~\mbox{type IIB}
\nonumber\tag{\ref{eq:LfIIB}}
\end{align}

\noindent where for both type IIA and IIB we have
\begin{align}\label{eq:Sfcommonpieces}
\mathcal{M}_{ab} &= g_{ab} + \mathcal{F}_{ab},~~~\mathcal{M} \equiv -\det \mathcal{M} \nonumber\\
D_{\alpha}^{(0)} &= \partial_{a} + \frac{1}{4}?\Omega_a^{\overline{\mu\nu}}?\Gamma_{\overline{\mu\nu}} + \frac{1}{4 \cdot 2!} H_{a\nu\rho}\Gamma^{\nu\rho} \nonumber\\
\Delta^{(1)} &= \frac{1}{2}(\Gamma^\mu\partial_\mu\Phi + \frac{1}{2\cdot3!} H_{\mu\nu\rho}\Gamma^{\mu\nu\rho}),
\end{align}

\noindent for type IIA we have
\begin{align}\label{eq:SfIIApieces}
 W_a &= -\frac{1}{8}e^{\Phi}\left(\frac{1}{2}F_{\mu\nu}\Gamma^{\mu\nu} + \frac{1}{4!}\tilde{F}_{\mu\nu\alpha\beta}\Gamma^{\mu\nu\alpha\beta}\right)\Gamma_a \nonumber\\
   \Delta^{(2)} &= \frac{1}{8} e^{\Phi}\left(\frac{3}{2!}F_{\mu\nu}\Gamma^{\mu\nu} - \frac{1}{4!}\tilde{F}_{\mu\nu\alpha\beta}\Gamma^{\mu\nu\alpha\beta}\right)\nonumber\\
   \Gamma_{D_p} &=\frac{\epsilon^{a_1\dots a_{p+1}}\Gamma_{a_1\dots a_{p+1}}}{(p+1)!\sqrt{\mathcal{M}}} (\Gamma^{11})^{p/2+1}\sum_{q \ge 0} \frac{(-1)^q (\Gamma^{11})^q}{q! 2^q}\Gamma^{b_1 \dots b_{2q}}\mathcal{F}_{b_1 b_2} \dots \mathcal{F}_{b_{2q-1}b_{2q}}
\end{align}

\noindent and for type IIB we have
\begin{align}\label{eq:SfIIBpieces}
W_a &= \frac{1}{8}\left[ F_\mu\Gamma^\mu + \frac{1}{3!}\tilde{F}_{\mu\nu\alpha}\Gamma^{\mu\nu\alpha}+ \frac{1}{2\cdot 5!}\tilde{F}_{\mu\nu\alpha\beta\rho}\Gamma^{\mu\nu\alpha\beta\rho}\right]\Gamma_a \nonumber\\
\Delta^{(2)} &= -\frac{1}{2}e^{\Phi}[F_\mu \Gamma^\mu + \frac{1}{2\cdot 3!}\tilde{F}_{\mu\nu\rho}\Gamma^{\mu\nu\rho}] \nonumber\\
\Gamma_{D_p} &= (-1)^{\frac{(p-2)(p-3)}{2}} \frac{\epsilon^{a_1 \dots a_{p+1}}\Gamma_{a_1\dots a_{p+1}}}{(p+1)!\sqrt{-\det M_0}}\sum_q \frac{\Gamma^{\lbrack b_1 \dots b_{2q}\rbrack}}{q!2^q} \mathcal{F}_{b_1 b_2}\cdots\mathcal{F}_{b_{2q-1}b_{2q}} 
\end{align}

\noindent We are using the conventions
\begin{align}
  H_3 &= dB_2,~~~F_p = d C_{p-1},~~~\tilde{F}_3 = F_3 - C_0 H_3,~~~\tilde{F}_4 = F_4 - C_1 \wedge H_3,\nonumber\\
 \tilde{F}_5 &= *\tilde{F}_5 = F_5 + B_2 \wedge F_3.
\end{align}

The spin connection, $\Omega_{a}^{\overline{\mu\nu}}$, is the pull back
\begin{align}
  \Omega_a^{~\overline{\mu\nu}} &= \frac{\partial X^{\mu}}{\partial\zeta^a}\Omega_{\mu}^{~\overline{\mu\nu}}
\end{align}
\noindent where the 10-D spin connection is built out of the frame fields~\cite{Lawrie:1990}
\begin{align}\label{eq:10dSpinConnection}
  ?\Omega_\mu^{\overline{\mu \nu}}? = -?\Omega_\mu^{\overline{\nu \mu}}?  &= \frac{1}{2} ?e_\mu^{\overline{\alpha}}?(?C_{\overline{\alpha}}^{\overline{\mu\nu}}? - ?C^{\overline{\mu\nu}}_{\overline{\alpha}}? - ?C^{\overline{\nu}}_{\overline{\alpha}}^{\overline{\mu}}?) \nonumber\\
       &=
\eta^{\overline{\nu\rho}}e^{\overline{\mu}}_{~\alpha}\left(\partial_{\mu}e_{\overline{\rho}}^{~\alpha} + e_{\overline{\rho}}^{~\nu}\Gamma^{\alpha}_{~\mu\nu}\right)
\end{align}

\noindent where the Christoffel symbol and holonomy elements are
\begin{align}
   \Gamma^{\alpha}_{~\mu\nu} &= \frac{1}{2}G^{\alpha\beta}\left(\partial_{\mu}G_{\beta\nu} + \partial_{\nu}G_{\beta\mu} - \partial_{\beta}G_{\mu\nu}\right) \\
   ?C^{\overline{\alpha}}_{\overline{\mu\nu}}? &= (?e_{\overline{\mu}}^{\alpha}??e_{\overline{\nu}}^\beta? - ?e_{\overline{\mu}}^{\beta}??e_{\overline{\nu}}^\alpha?)\partial_\beta ?e^{\overline{\alpha}}_\alpha?
\end{align}

\noindent with flat indices raised (lowered) by $\eta^{\overline{\mu\nu}}$($\eta_{\overline{\mu\nu}}$) and the curved indices raised (lowered) by $G^{\mu\nu}$ ($G_{\mu\nu}$):
\begin{align}
      e^{\overline{\mu}\mu} &= e_{\overline{\nu}}^{~\mu}\eta^{\overline{\mu\nu}} = e^{\overline{\mu}}_{~\nu}G^{\mu\nu},~~~e_{\overline{\mu}\mu} = e^{\overline{\nu}}_{~\mu}\eta_{\overline{\mu\nu}} = e_{\overline{\mu}}^{~\nu}G_{\mu\nu}
\end{align}

\noindent The frame fields thus frame the 10-D metric from 10-D Minkowski space:
\begin{align}
    G_{\mu\nu} = e^{\overline{\mu}}_{~\mu} e^{\overline{\nu}}_{~\nu} \eta_{\overline{\mu\nu}}
\end{align}

\noindent which is consistent with the definition of the inverse frame fields:
\begin{align}
    ?e_{\overline{\mu}}^\mu??e^{\overline{\mu}}_\nu? &= ?\delta^\mu_\nu?,~~~?e_{\overline{\mu}}^\mu??e^{\overline{\nu}}_\mu? = ?\delta_{\overline{\mu}}^{\overline{\nu}}?.
\end{align}

The 10-D curved $\Gamma^{\mu}$ matrices are framed from the 10-D flat $\Gamma^{\overline{\mu}}$ matrices
\begin{align}\label{eq:10dGammaRelation}
\Gamma^\mu = ?e_{\overline{\mu}}^\mu? \Gamma^{\overline{\mu}},~~~\Gamma_{\mu} = ?e^{\overline{\mu}}_\mu?\Gamma_{\overline{\mu}}
\end{align}

\noindent which satisfy a Clifford algebra:
\begin{align}\label{eq:10dcurvedClifford}
\left\{ \Gamma^{\overline{\mu}}, \Gamma^{\overline{\nu}} \right\} = 2\eta^{\overline{\mu\nu}}.
\end{align}.

The $\Gamma^{a}$ matrices are the pull backs of the curved $\Gamma^\mu$ matrices onto the D$p$-brane:
\begin{align}\label{eq:D3braneGamma}
\Gamma_{a} &= \frac{\partial X^\mu}{\partial \zeta^{a}} \Gamma_\mu \nonumber\\
\Gamma^{a} &= g^{ab}\Gamma_{b}
\end{align}

\noindent with $g_{ab}$ the induced metric.

In type IIA, $\Theta$ is constrained by~\cite{Marolf:2003vf,Marolf:2003ye,Marolf:2004jb,Martucci:2005rb}:
\begin{align}\label{eq:Thetaconstraint}
   \Gamma^{11} \Theta = \Theta,~~~\mbox{type IIA constraint}
\end{align}

\noindent with $\Gamma^{11} = \Gamma^{\overline{0123456789}}$, where  \begin{align}
\Gamma^{\overline{\mu_1\mu_2\dots\mu_n}} &= \frac{1}{n!} \Gamma^{[\overline{\mu}_1}\Gamma^{\overline{\mu}_2}\dots\Gamma^{\overline{\mu}_n]} \nonumber\\
 &= \frac{1}{n!}\left(\Gamma^{\overline{\mu}_1}\Gamma^{\overline{\mu}_2}\dots\Gamma^{\overline{\mu}_n} + \mbox{all positive permutations} - \mbox{all negative permutations}\right) \nonumber\\
&= \left\{
   \begin{array}{l}
\Gamma^{\overline{\mu}_1}\Gamma^{\overline{\mu}_2}\dots\Gamma^{\overline{\mu}_n},~~~\mbox{all $\overline{\mu}_i$ different} \\
     0,~~~\mbox{any two $\overline{\mu}_i$ the same}
   \end{array}
\right.
\end{align}

Our representations for $\Gamma^{\overline{\mu}}$ will be tensor products, such as
\begin{align}
    \Gamma^{\overline{\mu}} = \sigma^a \otimes \sigma^b \otimes \sigma^c \otimes \sigma^d \otimes \sigma^e,
\end{align}
\noindent of the Pauli spin matrices, augmented with the identity:
\begin{align}\label{eq:Paulispin}
   \sigma^0 &= \mbox{{\scriptsize $\left(\begin{array}{c c}
                     1 & 0 \\
                     0 & 1
                     \end{array}
                \right)$}},~~~
   \sigma^1 &= \mbox{{\scriptsize $ \left(\begin{array}{c c}
                       0 & 1 \\
                       1 & 0
                      \end{array}
                \right)$}},~~~
   \sigma^2 &= \mbox{{\scriptsize $\left(\begin{array}{c c}
                      0 & -i \\
                      i & 0
                     \end{array}
               \right)$}},~~~
   \sigma^3 &= \mbox{{\scriptsize $\left(\begin{array}{c c}
                      1 & 0 \\
                      0 & -1
                     \end{array}
               \right)$}} .
\end{align}

\section{D3-brane Fermionic Fluctuations in KS Background}\label{KSfermionsexplicit}
For the KS background at $\tau = 0$, we have
\begin{align}\label{eq:simplifiedSfpieces}
  D_{\alpha}^{(0)} &= \partial_a + \frac{1}{4} ?\Omega_{a}^{\overline{\mu\nu}}?\Gamma_{\overline{\mu\nu} }\nonumber\\
  W_a &= \frac{1}{8 \cdot 3!} F_{\mu\nu\alpha}\Gamma^{\mu\nu\alpha} \Gamma_a \nonumber\\
  \Delta^{(1)} &= 0,~~~~~ \Delta^{(2)} = -\frac{1}{4\cdot 3!} F_{\mu\nu\alpha}\Gamma^{\mu\nu\alpha} \nonumber\\
  \overset{\vee}{\Gamma}_{D_3} &= \frac{\varepsilon^{abcd} \Gamma_{abcd}}{2! 4! \sqrt{\mathcal{M}}} \Gamma^{mn}\mathcal{F}_{mn}
\end{align}
\noindent with the non-zero components of the spin connection given by
\begin{align}\label{eq:KSspinconnection}
  ?\Omega_{\theta}^{\bar{4}\bar{9}}? &= ?\Omega_{\theta}^{\bar{8}\bar{7}}? = \frac{\sin\psi_0}{2},~~~?\Omega_{\theta}^{\bar{5}\bar{9}}? = ?\Omega_{\theta}^{\bar{6}\bar{8}}? =  \sin^2\frac{\psi_0}{2} \nonumber\\
  ?\Omega_{\phi}^{\bar{5}\bar{4}}? &= ?\Omega_{\phi}^{\bar{7}\bar{6}}? = \cos\theta,~~~?\Omega_{\phi}^{\bar{9}\bar{4}}? = ?\Omega_{\phi}^{\bar{7}\bar{8}}? = \sin\theta~\sin^2\frac{\psi_0}{2} \nonumber\\
  ?\Omega_{\phi}^{\bar{5}\bar{9}}? &= ?\Omega_{\phi}^{\bar{6}\bar{8}}? = \frac{1}{2}\sin\theta~\sin\psi_0.
\end{align}

\noindent Fluctuating around the classical solution $\Theta = 0 + \delta\Theta$ of the probe D3-brane, the action for fermionic fluctuations, Eq.~(\ref{eq:fermionicaction}), becomes:
\begin{align}
  \delta S_3^f &= \frac{\mu_p}{2g_s} \int d^4\zeta \sqrt{\mathcal{M}} \delta\overline{\Theta} [(\mathcal{M}^{-1})^{ab}\Gamma_{a}\partial_{b} + M_1 + M_2 + M_3 ] \delta\Theta \\
  \label{eq:KSM1}
  M_1 &= \frac{1}{4} (\mathcal{M}^{-1})^{ab}\Gamma_{a}?\Omega_{b}^{\overline{\mu\nu}}?\Gamma_{\overline{\mu\nu}} \\
  \label{eq:KSM2}
  M_2 &= -\frac{1}{8 \cdot 3!}\overset{\vee}{\Gamma}_{D_3}^{-1} (\mathcal{M}^{-1})^{ab}\Gamma_{b} F_{\mu\nu\alpha}\Gamma^{\mu\nu\alpha} \Gamma_{a} \\
  \label{eq:KSM3}
  M_3 &= -\frac{1}{4 \cdot 3!}\overset{\vee}{\Gamma}_{D_3}^{-1} F_{\mu\nu\alpha}\Gamma^{\mu\nu\alpha}
\end{align}

\noindent which easily gives the Euler-Lagrange equations of a massive fermionic field:
\begin{align}\label{eq:KSDiraceq}
   [(\mathcal{M}^{-1})^{ab}\Gamma_{a}\partial_{b} + M_1 + M_2 + M_3 ] \delta\Theta = 0.
\end{align}

For the probe D3-brane in the KS background, we use the following representation for the 10-D flat $\Gamma^{\overline{\mu}}$ matrices:
\begin{align}\label{eq:KSflatgamma}
  \Gamma^{\overline{0}} &= -i \sigma^1 \otimes \sigma^1 \otimes\sigma^1 \otimes \sigma^1 \otimes \sigma^3,~~~\Gamma^{\overline{1}} = \sigma^1 \otimes \sigma^1 \otimes \sigma^1 \otimes \sigma^2 \otimes \sigma^0 \nonumber\\
  \Gamma^{\overline{2}} &= \sigma^1 \otimes \sigma^3 \otimes \sigma^0\otimes \sigma^0 \otimes \sigma^0,~~~\Gamma^{\overline{3}} = \sigma^2 \otimes \sigma^0 \otimes \sigma^0 \otimes \sigma^0 \otimes \sigma^0 \nonumber\\
  \Gamma^{\overline{4}} &= \sigma^1 \otimes \sigma^1 \otimes \sigma^3\otimes \sigma^0 \otimes \sigma^0,~~~\Gamma^{\overline{5}} = \sigma^1 \otimes \sigma^2 \otimes \sigma^0 \otimes \sigma^0 \otimes \sigma^0 \nonumber\\
  \Gamma^{\bar{6}} &= -\sigma^1 \otimes \sigma^1 \otimes \sigma^1 \otimes \sigma^1 \otimes \sigma^1,~~~\Gamma^{\overline{7}} = \sigma^1\otimes \sigma^1 \otimes \sigma^1 \otimes \sigma^1 \otimes \sigma^2 \nonumber\\
  \Gamma^{\overline{8}} &= \sigma^1 \otimes \sigma^1 \otimes \sigma^2\otimes \sigma^0 \otimes \sigma^0,~~~\Gamma^{\overline{9}} = \sigma^1 \otimes \sigma^1 \otimes \sigma^1 \otimes \sigma^3 \otimes \sigma^0
\end{align}

For our solution of the KS background, we find the object $\overset{\vee}{\Gamma}_{D_3}^{-1}$ can be expressed as:
\begin{align}\label{eq:InvGammav}
  \overset{\vee}{\Gamma}_{D_3}^{-1} &= b^{-1}\cot{\frac{\psi_0}{2}}\Gamma^{\bar{6}}\Gamma^{\bar{7}}.
\end{align}

\noindent and we calculate the pulled back $\Gamma^a$ matrices to be:
\begin{align}\label{eq:Gammpb}
  \Gamma^a &= \left(\begin{array}{l}
                       h_0^{1/4}\Gamma^{\bar{0}} \\
                       h_0^{1/4}\Gamma^{\bar{1}} \\
                       \sqrt{\frac{R}{2}}(\cos{\frac{\psi_0}{2}}\Gamma^{\bar{7}} - \sin{\frac{\psi_0}{2}}\Gamma^{\bar{6}}) \\
                       -\sqrt{\frac{R}{2}}\csc{\theta}(\sin{\frac{\psi_0}{2}}\Gamma^{\bar{7}} + \cos{\frac{\psi_0}{2}}\Gamma^{\bar{6}})
                     \end{array}
                 \right).
\end{align}

We proceed to solve Eq.~(\ref{eq:KSDiraceq}) with a harmonic ansatz for $\delta\Theta$\cite{Kirsch:2006he}:
\begin{align}\label{eq:Thetasoln}
\delta\Theta &= \int dp~d\omega\sum_{l,m}e^{i(p x - \omega t)}\tilde{\Theta}_{lm}(p,\omega) \circ \Phi_{lm}(\theta,\phi)
\end{align}

\noindent where $\Phi_{lm}(\theta,\phi)$ is a 32 component complex spinor, whose components are arbitrary functions of $\theta$ and $\phi$, and $\tilde{\Theta}_{lm}(p,\omega)$ is a 32 component spinor of Grassman valued expansion coefficients.  The \emph{component} product $\circ$ is a commutative operator, defined for N component vectors or spinors as:
\begin{align}\label{eq:compproduct}
   A\circ B &= \left(\begin{array}{l} A^1 \\ A^2 \\ \vdots \\ A^N  \end{array} \right)
                      \circ
               \left(\begin{array}{l} B^1 \\ B^2 \\ \vdots \\ B^N \end{array}\right)
                      \equiv
               \left(\begin{array}{l} A^1 B^1 \\ A^2 B^2 \\ \vdots \\ A^N B^N \end{array} \right).
\end{align}
With the solution Eq.~(\ref{eq:Thetasoln}) for $\delta\Theta$, the Dirac Eq.~(\ref{eq:KSDiraceq}) can be reorganized and expressed as two distinct eigenvalue problems
\begin{align}\label{eq:feigenproblems}
   \omega \tilde{\Theta}_1\circ \Phi_1 = \mathcal{H}^{(f)}_1 \tilde{\Theta}_1 \circ \Phi_1 \\
   \omega \tilde{\Theta}_2\circ \Phi_2 = \mathcal{H}^{(f)}_2 \tilde{\Theta}_2\circ \Phi_2
\end{align}

\noindent where $\tilde{\Theta}_1 \circ \Phi_1$ and $\tilde{\Theta}_2 \circ \Phi_2$ are each eight components of the original 32 component spinor, Eq.~(\ref{eq:Thetasoln}).  The matrices $\mathcal{H}^{(f)}_i$ are \begin{align}\label{eq:Hf1}
   \mathcal{H}^{(f)}_1 &= \mbox{{\tiny $\left(\begin{array}{c c c c c c c c}
                          -p & -c_{3+}\mathcal{O}^{(2)}_+ & c_{2+} & 0 & 0 & i c_+ & 0 & 0 \\
                          -c_{4-}\mathcal{O}^{(1)}_- & p & 0 & c_{1-} & 0 & 0 & 0 & 0 \\
                          -c_{1-} & 0 & p & c_{3-} \mathcal{O}^{(2)}_+ & 0 & 0 & 0 & i c_- \\
                          0 & -c_{2+} & c_{4+}\mathcal{O}^{(1)}_- & -p & 0 & 0 & 0 & 0 \\
                          0 & 0 & 0 & 0 & -p & -c_{3+}\mathcal{O}^{(1)}_+ & -c_{2+} & 0 \\
                          i c_- & 0 & 0 & 0 & -c_{4-}\mathcal{O}^{(2)}_- & p & 0 & -c_{1-} \\
                          0 & 0 & 0 & 0 & c_{1-} & 0 & p & c_{3-}\mathcal{O}^{(1)}_+ \\
                          0 & 0 & i c_+ & 0 & 0 & c_{2+} & c_{4+}\mathcal{O}^{(2)}_- & -p
                          \end{array}
                          \right)$}}
\end{align}

\noindent and
{\footnotesize \begin{align}\label{eq:Hf2}
   \mathcal{H}^{(f)}_2 &= \mbox{{\tiny $\left(\begin{array}{l l l l l l l l}
                          -p & -c_{3+}\mathcal{O}^{(1)}_+ & c_{2+} & 0 & 0 & 0 & 0 & 0 \\
                          -c_{4-}\mathcal{O}^{(2)}_- & p & 0 & c_{1-} & -i c_- & 0 & 0 & 0 \\
                          -c_{1-} & 0 & p & c_{3-} \mathcal{O}^{(1)}_+ & 0 & 0 & 0 & 0 \\
                          0 & -c_{2+} & c_{4+}\mathcal{O}^{(2)}_- & -p & 0 & 0 & -i c_+ & 0 \\
                          0 & -i c_+ & 0 & 0 & -p & -c_{3+}\mathcal{O}^{(2)}_+ & -c_{2+} & 0 \\
                          0 & 0 & 0 & 0 & -c_{4-}\mathcal{O}^{(1)}_- & p & 0 & -c_{1-} \\
                          0 & 0 & 0 & -i c_- & c_{1-} & 0 & p & c_{3-}\mathcal{O}^{(2)}_+ \\
                          0 & 0 & 0 & 0 & 0 & c_{2+} & c_{4+}\mathcal{O}^{(1)}_- & -p
                          \end{array}
                          \right)$}}
\end{align}}

\noindent where the constants $c_{\pm}$ and $c_{1\pm}\dots c_{4\pm}$ are
\begin{align}
  c_{1\pm} &= i \frac{c_{\pm}}{2 T^2}(3T^2 - d^{1/2}T - b_3),~~~c_{2\pm} = i \frac{c_{\pm}}{2 T^2}(3T^2 + d^{1/2}T - b_3) \nonumber\\
  c_{3\pm} &= c_{\pm}(1 - i \frac{b}{T}),~~~c_{4\pm} = c_{\pm}(1 + i \frac{b}{T})\nonumber\\
  c_{\pm} &= -\frac{ 2 b_2 T }{h_0^{1/4} d^{1/2}}(T \pm d^{1/2}) \nonumber\\
  T &= b \tan{\frac{\psi_0}{2}},~~~~d = 1 + b^2\tan^2{\frac{\psi_0}{2}} \nonumber\\
  b_2 &= \sqrt{\frac{\pi T_0}{2b^3M}},~~~b_3 = b^{-1/2} + 3b^{3/2} - 3b^2
\end{align}

\noindent and the operators $\mathcal{O}^{(i)}_{\pm}$ are
\begin{align}
  \mathcal{O}^{(1)}_{\pm} &= \partial_{\theta} \pm i \csc{\theta} \partial_{\phi}\\
  \mathcal{O}^{(2)}_{\pm} &= \cot{\theta} + \mathcal{O}^{(1)}_{\pm}
\end{align}

From inspection of the matrices, Eqs.~(\ref{eq:Hf1}) and~(\ref{eq:Hf2}), and their actions on the spinors $\Phi_{ilm}$ in Eqs.~(\ref{eq:feigenproblems}), we identify the components, $\Phi^A_{ilm}(\theta,\phi),A=1\dots8,i=1,2$, of the spinors $\Phi_{ilm}(\theta,\phi)$ with three distinct functions $Y^+_{lm}(\theta,\phi)$, $Y_{lm}(\theta,\phi)$, and $Y^-_{lm}(\theta,\phi)$
\begin{align}
   \Phi_1^1 &= \Phi_1^3 = \Phi_1^6 = \Phi_1^8 = \Phi_2^2 = \Phi_2^4 = \Phi_2^5 = \Phi_2^7 = Y_{lm}(\theta,\phi) \nonumber\\
   \Phi_1^2 &= \Phi_1^4 = \Phi_2^6 = \Phi_2^8 = Y^-_{lm}(\theta,\phi),~~~\Phi_1^5 = \Phi_1^7 = \Phi_2^1 = \Phi_2^3 = Y^+_{lm}(\theta,\phi)
\end{align}

\noindent which must satisfy four coupled differential equations
\begin{align}
   \label{eq:coupled1}
   \mathcal{O}^{(1)}_{-}Y_{lm}(\theta,\phi) = \lambda_1 Y^-_{lm}(\theta,\phi) \\
   \label{eq:coupled2}
   \mathcal{O}^{(2)}_{+}Y^-_{lm}(\theta,\phi) = \lambda_2 Y_{lm}(\theta,\phi) \\
   \label{eq:coupled3}
   \mathcal{O}^{(1)}_{+}Y_{lm}(\theta,\phi) = \lambda_3 Y^+_{lm}(\theta,\phi) \\
   \label{eq:coupled4}
   \mathcal{O}^{(2)}_{-}Y^+_{lm}(\theta,\phi) = \lambda_4 Y_{lm}(\theta,\phi)
\end{align}

\noindent Eliminating $Y^-_{lm}(\theta,\phi)$ from Eqs.~(\ref{eq:coupled1}) and~(\ref{eq:coupled2}) results in the spherical harmonic eigenvalue problem:
\begin{align}\label{eq:coupled12}
   \hat{L}^2 Y_{lm} = \lambda_1 \lambda_2 Y_{lm}
\end{align}

\noindent So we see that the $Y_{lm}(\theta,\phi)$ are indeed the spherical harmonics, as their name suggests.  Furthermore, Eq.~(\ref{eq:coupled12}) now demands that
\begin{align}
  \lambda_1 \lambda_2 &= -l(l+1)
\end{align}

\noindent Eliminating $Y^+_{lm}(\theta,\phi)$ from Eqs.~(\ref{eq:coupled3}) and~(\ref{eq:coupled4}) results in a similar identity
\begin{align}
   \lambda_3 \lambda_4 &= -l(l+1)
\end{align}

\noindent Consistent with these two constraints, we make the following choices for the $\lambda_i$:
\begin{align}
  \lambda_1 &= \lambda_3 = 1,~~~\lambda_2 = \lambda_4 = -l(l+1)
\end{align}

\noindent and so we find $Y^+_{lm}(\theta,\phi)$ and $Y^-_{lm}(\theta,\phi)$ to be dependent on the spherical harmonics, $Y_{lm}(\theta,\phi)$, in the following way:
\begin{align}
  Y^+_{lm}(\theta,\phi) = \mathcal{O}^{(1)}_{+}Y_{lm}(\theta,\phi) \nonumber\\
  Y^-_{lm}(\theta,\phi) = \mathcal{O}^{(1)}_{-}Y_{lm}(\theta,\phi)
\end{align}

This newfound knowledge allows us to remove all $\theta$ and $\phi$ dependence from the eigenvalue Eqs.~(\ref{eq:feigenproblems}), leaving us with
\begin{align}
   \omega \tilde{\Theta}_1 = \mathcal{H}^{(f)}_1 \tilde{\Theta}_1 \\
   \omega \tilde{\Theta}_2 = \mathcal{H}^{(f)}_2 \tilde{\Theta}_2
\end{align}

\noindent where the matrices $\mathcal{H}^{(f)}_i$ now take the form
\begin{align}\label{eq:KSHf1final}
   \mathcal{H}^{(f)}_1 &= \mbox{{\scriptsize $\left(\begin{array}{c c c c c c c c}
                          -p & c_{3+}l(l+1) & c_{2+} & 0 & 0 & i c_+ & 0 & 0 \\
                          -c_{4-} & p & 0 & c_{1-} & 0 & 0 & 0 & 0 \\
                          -c_{1-} & 0 & p & -c_{3-}l(l+1) & 0 & 0 & 0 & i c_- \\
                          0 & -c_{2+} & c_{4+} & -p & 0 & 0 & 0 & 0 \\
                          0 & 0 & 0 & 0 & -p & -c_{3+} & -c_{2+} & 0 \\
                          i c_- & 0 & 0 & 0 & c_{4-}l(l+1) & p & 0 & -c_{1-} \\
                          0 & 0 & 0 & 0 & c_{1-} & 0 & p & c_{3-} \\
                          0 & 0 & i c_+ & 0 & 0 & c_{2+} & -c_{4+}l(l+1) & -p
                          \end{array}
                          \right)$}}
\end{align}

\noindent and
\begin{align}\label{eq:KSHf2final}
   \mathcal{H}^{(f)}_2 &= \mbox{{\scriptsize $\left(\begin{array}{c c c c c c c c}
                          -p & -c_{3+} & c_{2+} & 0 & 0 & 0 & 0 & 0 \\
                          c_{4-}l(l+1) & p & 0 & c_{1-} & -i c_- & 0 & 0 & 0 \\
                          -c_{1-} & 0 & p & c_{3-} & 0 & 0 & 0 & 0 \\
                          0 & -c_{2+} & -c_{4+}l(l+1) & -p & 0 & 0 & -i c_+ & 0 \\
                          0 & -i c_+ & 0 & 0 & -p & c_{3+}l(l+1) & -c_{2+} & 0 \\
                          0 & 0 & 0 & 0 & -c_{4-} & p & 0 & -c_{1-} \\
                          0 & 0 & 0 & -i c_- & c_{1-} & 0 & p & -c_{3-}l(l+1) \\
                          0 & 0 & 0 & 0 & 0 & c_{2+} & c_{4+} & -p
                          \end{array}
                          \right)$}}
\end{align}

These two matrices have the same eight, massive eigenvalues
\begin{align}
   \omega &= \left\{\begin{array}{l}
                    \pm\sqrt{c_{10}(p,l) + \sqrt{c_8(p,l)} \pm \sqrt{c_{9+}}(p,l)} \\
                    \pm\sqrt{c_{10}(p,l) - \sqrt{c_8(p,l)} \pm \sqrt{c_{9-}}(p,l)}
                    \end{array}
                    \right.
\end{align}

\noindent where
\begin{align}
  c_5 &= c_{12}^2 - 3c_{11}c_{13} + 12 c_{14} \nonumber\\
  c_6 &= 2 c_{12}^3 - 9 c_{12}(c_{11}c_{13} + 8 c_{14}) + 27(c_{13}^2 + c_{11}^2c_{14}) \nonumber\\
  c_7 &= c_6 + \sqrt{-4 c_5^3 + c_6^2} \nonumber\\
  c_8 &= c_{11}^2 + \frac{2}{3}\left(-4 c_{12} + \frac{2^{4/3} c_5}{c_7^{1/3}} + 2^{2/3} c_7^{1/3}\right) \nonumber\\
  c_{9\pm} &= \frac{2}{3}\left(3 c_{11}^2 - 8c_{12} - \frac{2^{4/3} c_5}{c_7^{1/3}} - 2^{2/3} c_{7}^{1/3} \pm \frac{3(4 c_{11}c_{12} - c_{11}^3- 8 c_{13})}{\sqrt{c_8}} \right)\nonumber\\
  c_{10} &= 2(p^2 - (c_{3+}c_{4-} + c_{3-}c_{4+})l(l+1) - c_{-}c_{+} - 2 c_{1-} c_{2+})
\end{align}

\noindent with
\begin{align}
  c_{11} &= 4c_{1-} c_{2+} + 2 c_{-}c_{+} + (2 c_{3+} c_{4-} + 2 c_{3-} c_{4+}) l(l+1) - 4p^2
\end{align}
\begin{align}
  c_{12} &= 6 c_{1-}^2c_{2+}^2 + c_{2+}^2 c_{-}^2 + 4 c_{1-}c_{2+} c_{-}c_{+} + c_{1-}^2 c_{+}^2 + c_{-}^2 c_{+}^2 +\nonumber\\
  &+ (2 c_{2+}^2 c_{3-} c_{4-} + 4 c_{1-} c_{2+} c_{3+} c_{4-} + 4 c_{1-} c_{2+} c_{3-} c_{4+} + 2 c_{1-}^2 c_{3+} c_{4+} + 2 c_{3+} c_{4-} c_{-} c_{+} + \nonumber\\
  &+  2 c_{3-} c_{4+} c_{-} c_{+})l(l+1) + (c_{3+}^2 c_{4-}^2 + 4 c_{3-} c_{3+} c_{4-} c_{4+} + c_{3-}^2 c_{4+}^2) l^2(l+1)^2 + \nonumber\\
  &(-12 c_{1-} c_{2+} - 6 c_{-} c_{+} + (-6 c_{3+} c_{4-} - 6 c_{3-} c_{4+}) l(l+1)) p^2 + 6 p^4
\end{align}
\begin{align}
c_{13} &= 4 c_{1-}^3 c_{2+}^3 + 2 c_{1-} c_{2+}^3 c_{-}^2 + 2 c_{1-}^2 c_{2+}^2 c_{-} c_{+} + 2 c_{1-}^3 c_{2+}c_{+}^2 + \nonumber\\
 &+2 c_{1-} c_{2+} c_{-}^2 c_{+}^2 + (4 c_{1-} c_{2+}^3 c_{3-} c_{4-} + 2 c_{1-}^2 c_{2+}^2 c_{3+} c_{4-} + \nonumber\\
    &+2 c_{1-}^2 c_{2+}^2 c_{3-} c_{4+} + 4 c_{1-}^3 c_{2+} c_{3+} c_{4+} + 2 c_{1-} c_{2+} c_{3+} c_{4+} c_{-}^2 + 2 c_{1-} c_{2+} c_{3+} c_{4-} c_{-} c_{+} + \nonumber\\
    &+ 2 c_{1+} c_{2+} c_{3-} c_{4+} c_{-} c_{+} + 2 c_{1-} c_{2+} c_{3-} c_{4-} c_{+}^2) l^2 (1 + l)^2 + \nonumber\\
    &+ (2 c_{2+}^2 c_{3-} c_{3+} c_{4-}^2 + 2 c_{2+}^2 c_{3-}^2 c_{4-} c_{4+} + 4 c_{1-} c_{2+} c_{3-} c_{3+} c_{4-} c_{4+} + 2 c_{1-}^2 c_{3+}^2 c_{4-} c_{4+} + \nonumber\\
    &+ 2 c_{1-}^2 c_{3-} c_{3+} c_{4+}^2 + c_{3+}^2 c_{4-}^2 c_{-} c_{+} + c_{3-}^2 c_{4+}^2 c_{-} c_{+}) l^4 (1 + l)^4 + (2 c_{3-} c_{3+}^2 c_{4-}^2 c_{4+} +  \nonumber\\
    & + 2 c_{3+}^2 c_{3+} c_{4-} c_{4+}^2) l^6 (1 + l)^6 + (-12 c_{1-}^2 c_{2+}^2 - 2 c_{2+}^2 c_{-}^2 - 8 c_{1-} c_{2+} c_{-} c_{+} - 2 c_{1-}^2 c_{+}^2 +\nonumber\\
    &- 2 c_{-}^2 c_{+}^2 + (-4 c_{2+}^2 c_{3-} c_{4-} - 8 c_{1-} c_{2+} c_{3+} c_{4-} - 8 c_{1-} c_{2+} c_{3-} c_{4+} +\nonumber\\
&- 4 c_{1-}^2 c_{3+} c_{4+} - 4 c_{3+} c_{4-} c_{-} c_{+}
 - 4 c_{3-} c_{4+} c_{-} c_{+}) l^2 (1 + l)^2 + (-2 c_{3+}^2 c_{4-}^2 +\nonumber\\
&- 8 c_{3-} c_{3+} c_{4-} c_{4+} - 2 c_{3-}^2 c_{4+}^2) l^4 (1 + l)^4) p^2 + (12 c_{1-} c_{2+} + \nonumber\\
    &+ 6 c_{-} c_{+} + (6 c_{3+} c_{4-} + 6 c_{3-} c_{4+}) l^2 (1 + l)^2) p^4 - 4 p^6
\end{align}
\begin{align}
c_{14} &= c_{1-}^4 c_{2+}^4 + c_{1-}^2 c_{2+}^4 c_{-}^2 + c_{1-}^4 c_{2+}^2 c_{+}^2 + c_{1-}^2 c_{2+}^2 c_{-}^2 c_{+}^2 + (2 c_{1-}^2 c_{2+}^4 c_{3+} c_{4+} + \nonumber\\
    &+2 c_{1-}^4 c_{2+}^2 c_{3+} c_{4+} + 2 c_{1-}^2 c_{2+}^2 c_{3+} c_{4+} c_{-}^2 + 2 c_{1-}^2 c_{2+}^2 c_{3-} c_{4-} c_{+}^2) l^2 (1 + l)^2 + (c_{2+}^4 c_{3-}^2 c_{4-}^2 + \nonumber\\
    &+4 c_{1-}^2 c_{2+}^2 c_{3-} c_{3+} c_{4-} c_{4+} + c_{1-}^4 c_{3+}^2 c_{4+}^2 + c_{1-}^2 c_{3+}^2 c_{4+}^2 c_{-}^2 + c_{2+}^2 c_{3-}^2 c_{4-}^2 c_{+}^2) l^4 (1 + l)^4 + \nonumber\\
    &+(2 c_{2+}^2 c_{3-}^2 c_{3+} c_{4-}^2 c_{4+} + 2 c_{1-}^2 c_{3-} c_{3+}^2 c_{4-} c_{4+}^2) l^6 (1 + l)^6 +
 c_{3-}^2 c_{3+}^2 c_{4-}^2 c_{4+}^2 l^8 (1 + l)^8 + \nonumber\\
    &+(-4 c_{1-}^3 c_{2+}^3 - 2 c_{1-} c_{2+}^3 c_{-}^2 - 2 c_{1-}^2 c_{2+}^2 c_{-} c_{+} - 2 c_{1-}^3 c_{2+} c_{+}^2 - 2 c_{1-} c_{2+} c_{-}^2 c_{+}^2 + \nonumber\\
    &+ (-4 c_{1-} c_{2+}^3 c_{3-} c_{4-} - 2 c_{1-}^2 c_{2+}^2 c_{3+} c_{4-} - 2 c_{1-}^2 c_{2+}^2 c_{3-} c_{4+} - 4 c_{1-}^3 c_{2+} c_{3+} c_{4+} +\nonumber\\
    &- 2 c_{1-} c_{2+} c_{3+} c_{4+} c_{-}^2 - c_{1-} c_{2+} c_{3+} c_{4+} c_{-} c_{+} - 2 c_{1-} c_{2+} c_{3-} c_{4+} c_{-} c_{+}+\nonumber\\
& - 2 c_{1-} c_{2+} c_{3-} c_{4-} c_{+}^2) l^2 (1 + l)^2 + \nonumber\\
    &+(-2 c_{2+}^2 c_{3-} c_{3+} c_{4-}^2 - 2 c_{2+}^2 c_{3-}^2 c_{4-} c_{4+} - 4 c_{1-} c_{2+} c_{3-} c_{3+} c_{4-} c_{4+} - 2 c_{1-}^2 c_{3+}^2 c_{4-} c_{4+} +\nonumber\\
    &- 2 c_{1-}^2 c_{3-} c_{3+} c_{4+}^2 - c_{3+}^2 c_{4-}^2 c_{-}c_{+} - c_{3-}^2 c_{4+}^2 c_{-} c_{+}) l^4 (1 + l)^4 + (-2 c_{3-} c_{3+}^2 c_{4-}^2 c_{4+} +\nonumber\\
    &- 2 c_{3-}^2 c_{3+} c_{4-} c_{4+}^2) l^6 (1 + l)^6) p^2 + (6 c_{1-}^2 c_{2+}^2 + c_{2+}^2 c_{-}^2 + 4 c_{1-} c_{2+} c_{-} c_{+} + c_{1-}^2 c_{+}^2 + c_{-}^2 c_{+}^2 + \nonumber\\
    &+ (2 c_{2+}^2 c_{3-} c_{4-} + 4 c_{1-} c_{2+} c_{3+} c_{4-} + 4 c_{1-} c_{2+} c_{3-} c_{4+} + 2 c_{1-}^2 c_{3+} c_{4+} + 2 c_{3+} c_{4-} c_{-} c_{+} + \nonumber\\
    &+ 2 c_{3-} c_{4+} c_{-} c_{+}) l^2 (1 + l)^2 + (c_{3+}^2 c_{4-}^2 + 4 c_{3-} c_{3+} c_{4-} c_{4+} + c_{3-}^2 c_{4+}^2) l^4 (1 + l)^4) p^4 + \nonumber\\
    &+(-4 c_{1-} c_{2+} - 2 c_{-} c_{+} + (-2 c_{3+} c_{4-} - 2 c_{3-} c_{4+}) l^2 (1 + l)^2) p^6 + p^8
\end{align}
\section{D4-brane Fermionic Fluctuations in CGLP Background}
The frame-fields for the CGLP background can be written in a 10 dimensional representation as
\begin{align}
   e^{\overline{0}}_{~0} &= e^{\overline{1}}_{~1} = e^{\overline{2}}_{~2} = H^{-1/4},~~~e^{\overline{9}}_{~9} = l f H^{1/4},   \nonumber\\
   e^{\overline{3}}_{~3} &= \csc\theta e^{\overline{4}}_{~4} = \csc\chi\csc\theta e^{\overline{5}}_{~5} = \csc\psi\csc\chi\csc\theta e^{\underline{6}}_{~6} = lbH^{1/4}, \nonumber\\
   e^{\overline{7}}_{~\mu} &= lH^{1/4}~a\frac{\partial{\mu^i}}{\partial\tilde{\theta}} A^j_{\alpha} \epsilon^{ijk}\mu^k,~~~e^{\underline{8}}_{~\mu} = lH^{1/4}a \csc\tilde{\theta}\frac{\partial{\mu^i}}{\partial\tilde{\phi}} A^j_{\alpha} \epsilon^{ijk}\mu^k,~~~\alpha = 4,5,6
\end{align}

\noindent with parametrization of the unit $S^2$, $(\mu^i)^2 = 1$, given by
\begin{align}
   \mu^1 &= \sin\tilde{\theta}~\cos\tilde{\phi},~~~\mu^2 = \sin\tilde{\theta}~\sin\tilde{\phi},~~~\mu^3 = \cos\tilde{\theta}
\end{align}

\noindent In the above, the 10 independent bosonic coordinates are numbered $0\dots9$ as
\begin{align}
   X^{\mu} = (t,x,x^2,\psi,\chi,\theta,\phi,\tilde{\theta},\tilde{\phi},\tau)
\end{align}

We calculate the only non-vanishing components of $\Omega_a^{~\overline{\mu\nu}}$ to be
\begin{align}
  &\Omega_{\chi}^{~\overline{98}} = \Omega_{\chi}^{~\overline{43}} = \cos\psi_0, &&\Omega_{\theta}^{~\overline{53}} =\Omega_{\theta}^{~\overline{79}} = \cos\psi_0~\sin\chi, \nonumber\\
  &\Omega_{\theta}^{~\overline{54}} =\Omega_{\theta}^{~\overline{87}} = \cos\chi, &&\Omega_{\phi}^{~\overline{63}} = \Omega_{\phi}^{~\overline{87}} = \cos\psi_0~\sin\chi~\sin\theta, \nonumber\\
  &\Omega_{\phi}^{~\overline{64}} = \Omega_{\phi}^{~\overline{97}} = \cos\chi~\sin\theta,
  &&\Omega_{\phi}^{~\overline{65}} = \Omega_{\phi}^{~\overline{98}} = \cos\theta,
\end{align}

\noindent With this we calculate the term in the action, Eq. (\ref{eq:fermionicaction}), which contains the spin connection to be
\begin{align}
   \frac{1}{4}\left(\mathcal{M}^{-1}\right)^{ab}\Gamma_a \Omega_b^{~\overline{\mu\nu}}\Gamma_{\overline{\mu\nu}} &= M_c + \cot\chi~M_1 + \csc\chi~\cot\theta~M_2 \nonumber\\
   M_c &= \frac{1}{2}\sqrt{\frac{R}{6}}\cos\psi_0(3\Gamma_{\overline{3}} + \Gamma_{\overline{498}}+\Gamma_{\overline{579}}+\Gamma_{\overline{687}})\nonumber\\
   M_1 &= \frac{1}{2}\sqrt{\frac{R}{6}}(2\Gamma_{\overline{4}} + \Gamma_{\overline{587}} + \Gamma_{\overline{697}}) \nonumber\\
   M_2 &= \frac{1}{2}\sqrt{\frac{R}{6}}(\Gamma_{\overline{5}} + \Gamma_{\overline{698}})
\end{align}

\noindent This is the only term in the action that has $\theta,\chi$ dependence, modulo the measure.  Many of the formulas in the fermionic action, Eq.~(\ref{eq:fermionicaction}), simplify to
\begin{align}
   W_a &= -\frac{1}{8}e^{\Phi_0}\frac{1}{4!}F_{\mu\nu\alpha\beta}\Gamma^{\mu\nu\alpha\beta}\Gamma_a \\
   \Delta^{(1)} &= \frac{1}{4!}H_{\alpha\mu\nu}\Gamma^{\alpha\mu\nu},~~~\Delta^{(2)} = -\frac{1}{8 \cdot 4!}e^{\Phi_0} F_{\alpha\beta\mu\nu}\Gamma^{\alpha\beta\mu\nu}\\
   \Gamma_{D_4}' &= 1 - \frac{\epsilon^{abcde}\Gamma_{abcde}}{5!\sqrt{ \mathcal{M}}}\Gamma^{11}(1 - \frac{1}{2}\Gamma^{11}\Gamma^{ab}\mathcal{F}_{ab}),
\end{align}

\noindent all of which are, again, $\chi,\theta$ independent.

As in the CGLP bosonic case, App.~\ref{app:CGLPfluctuations}, we investigate $S^3$ independent solutions for the fluctuations about the classical solution
\begin{align}
   \Theta = 0 + \delta\Theta(t,x)
\end{align}

\noindent leaving us with an action for fermionic fluctuations, Eq.~(\ref{eq:fermionicaction}), of the form
\begin{align}
   \delta S^{f}_{4eff} \propto \int dt dx\int d\chi d\theta d\phi &\sin^2\chi~\sin\theta \delta\overline{\Theta}\Gamma_{D_4}'((\mathcal{M}^{-1})^{mn}\Gamma_m\partial_n + M_f + \nonumber\\
         & + \cot\chi~M_1 + \csc\chi~\cot\theta~M_2) \delta\Theta,~~~ m,n = t,x,
\end{align}

\noindent where
\begin{align}\label{eq:CGLPMf}
  M_f &= M_c + \left(M^{-1}\right)^{ab}\left(\frac{1}{8}\Gamma_a H_{b\mu\nu}\Gamma^{\mu\nu} + \Gamma_b W_a\right) - \Delta^{(1)} - \Delta^{(2)}.
\end{align}

Integrating out the $S^3$ as in the CGLP bosonic case, it is easy to see that the terms proportional to $M_1$ and $M_2$ integrate to zero, leaving us with
\begin{align}\label{eq:fermioneffectiveaction}
  \delta S^{f}_{eff} \propto \int dt dx \delta\overline{\Theta}\Gamma_{D_4}'((\mathcal{M}^{-1})^{mn}\Gamma_m\partial_n + M_f) \delta\Theta,~~~ m,n = t,x. 
\end{align}

We solve the Euler equation from this action by Fourier transform
\begin{align}\label{eq:DiracFT}
  \Gamma_{D_4}'(i(\mathcal{M}^{-1})^{mn}\Gamma_m p_n + M_f) \delta\Theta=0,~~~ m,n = t,x~~~
  p_t& = -\omega,~~~p_x = p.
\end{align}

\noindent and now pick a representation for the $32 \times 32$ gamma matrices
\begin{align}\label{eq:CGLPflatgamma}
  \Gamma^{\overline{0}} &= i \sigma^1 \otimes \sigma^1 \otimes \sigma^1 \otimes \sigma^3 \otimes \sigma^0,~~~\Gamma^{\overline{1}} = \sigma^1 \otimes \sigma^1 \otimes \sigma^2 \otimes \sigma^0 \otimes\sigma^0 \nonumber\\
  \Gamma^{\overline{2}} &= \sigma^1 \otimes \sigma^1 \otimes \sigma^1 \otimes \sigma^1 \otimes \sigma^3,~~~\Gamma^{\overline{3}} = \sigma^1 \otimes \sigma^1 \otimes \sigma^1 \otimes \sigma^1 \otimes\sigma^2 \nonumber\\
 \Gamma^{\overline{4}} &= \sigma^1 \otimes \sigma^1 \otimes \sigma^3 \otimes \sigma^0 \otimes \sigma^0,~~~\Gamma^{\overline{5}} = \sigma^1 \otimes \sigma^2 \otimes \sigma^0 \otimes \sigma^0 \otimes \sigma^0 \nonumber\\
  \Gamma^{\overline{6}} &= -\sigma^1 \otimes \sigma^1 \otimes \sigma^1 \otimes \sigma^1 \otimes \sigma^1,~~~\Gamma^{\overline{7}} = \sigma^1 \otimes \sigma^3 \otimes \sigma^0 \otimes \sigma^0 \otimes \sigma^0 \nonumber\\
  \Gamma^{\overline{8}} &= \sigma^2 \otimes \sigma^0 \otimes \sigma^0 \otimes \sigma^0 \otimes \sigma^0,~~~\Gamma^{\overline{9}} = \sigma^1 \otimes \sigma^1 \otimes \sigma^1 \otimes \sigma^2 \otimes \sigma^0.
\end{align}

\noindent which leaves us with a diagonal $\Gamma^{11}$, and so through the constraint, Eq.~(\ref{eq:Thetaconstraint}), we are able to set the lower 16 components of $\delta\Theta$ to zero.  At the same time, this reduces the 32 Eqs.~(\ref{eq:DiracFT}) to 16 independent equations.  These equations can be reorganized into the following form
\begin{align}
  \omega \delta\Theta &= H_f \delta\Theta
\end{align}

\noindent where the Matrix, $H_f$, has the block diagonal form
\begin{align}\label{eq:CGLPHf}
   H_f = \left(\begin{array}{lll}
                      H_1 & 0 & 0 \\
                      0 & H_2 & 0 \\
                      0 & 0 & H_3
               \end{array}
         \right),
\end{align}

\noindent and $H_1$ and $H_2$ are $4 \times 4$ matrices
\begin{align}\label{eq:Hi}
   H_1 &= \left(\begin{array}{llll}
                    p & 0 & -c_i & c_a \\
                    0 & p & c_b & -c_i \\
                    -c_j & c_c & -p & 0 \\
                    c_d & -c_j & 0 & -p
                \end{array}
          \right),
   H_2 = \left(\begin{array}{llll}
                    p & 0 & c_i & c_e \\
                    0 & p & c_f & c_i \\
                    c_j & c_g & -p & 0 \\
                    c_h & c_j & 0 & -p
                \end{array}
          \right),
\end{align}
\noindent and $H_3$ is an $8 \times 8$ matrix

\begin{align}
   H_3 &= \left(\begin{array}{llllllll}
                  -p & 0 & c_j & -c_c & 0 & 0 & -c_k & 0 \\
                  0 & -p & -c_d & c_j & 0 & 0 & 0 & -c_k \\
                  c_i & -c_a & p & 0 & -c_n & 0 & 0 & 0 \\
                  -c_b & c_i & 0 & p & 0 & -c_n & 0 & 0 \\
                  0 & 0 & -c_k & 0 & -p & 0 & -c_j & -c_g \\
                  0 & 0 & 0 & -c_k & 0 & -p & -c_h & -c_j \\
                  -c_n & 0 & 0 & 0 -c_i & -c_e & p & 0 \\
                  0 & -c_n & 0 & 0 & -c_f & -c_i & 0 & p
                \end{array}
          \right)
\end{align}

\noindent where the $c's$ are constants.

The eigenvalues of $H_f$ are
\begin{align}\label{eq:wf1}
  \omega &= \pm\sqrt{p^2 + \alpha_{1} \pm \alpha_{2}} \\
  \label{eq:wf2}
  \omega &= \pm\sqrt{p^2 + \alpha_{3} \pm \alpha_{4}} \\
  \label{eq:wf3}
   \omega &= \left\{\begin{array}{l}
                    \pm\sqrt{\alpha_{7}(p) + \alpha_{5}(p) \pm \alpha^{+}_{6}(p)} \\
                    \pm\sqrt{\alpha_{7}(p) - \alpha_{5}(p) \pm \alpha^{-}_{6}(p)}
                    \end{array}
                    \right.,
\end{align}
where $\alpha_1, \alpha_2, \alpha_3,$ and $\alpha_4$ are constants combinations of the $c's$ in Eq.~(\ref{eq:Hi}),  and $\alpha_{5}, \alpha_6^{\pm},$ and $\alpha_7$ are functions of p:

\begin{align}
    \alpha_{7}(p) &= \alpha_{7}^{(0)}+\alpha_{7}^{(2)}p^2,\nonumber\\
    \alpha_{5}^2 &= \sum_{n = 0,2,4}\alpha_{5}^{(n)}p^n + 4\beta_{1}(p),\nonumber\\
    (\alpha_{6}^\pm)^2 &= 2\alpha_{5}^{2}(p) - 3\beta_{1}(p) \pm \beta_{5}(p),\nonumber\\
    \beta_{1}(p) &= \frac{1}{12}\left(\frac{\beta_{3}(p)}{\beta_{2}(p)}+\beta_{2}(p)\right),\nonumber\\
    2\beta_{2}^3(p) &= \beta_{4}(p) +\sqrt{\beta_{4}^3(p) - 4\beta_{3}^3(p)},\nonumber\\
    \beta_{3}(p) &= \sum_{n = 0,2,,8}\beta_{3}^{(n)}p^n,\nonumber\\
    \beta_{4}(p) &= \sum_{n = 0,2,,12}\beta_{4}^{(n)}p^n,\nonumber\\
    \beta_{5}(p) &= \alpha_{5}^{-1}\sum_{n = 0,2,,6}\beta_{5}^{(n)}p^n.
\end{align}

\noindent Here, the $\alpha_{i}^{(n)}$ and $\beta_{i}^{(n)}$ are constant combinations of the $c's$ from Eq.~(\ref{eq:Hi}).

From the regularization procedure used in section~\ref{MassiveModes}, we see that the first two sets of fermionic eigenmodes, Eqs.~(\ref{eq:wf1}) and~(\ref{eq:wf2}), will contribute a constant to the fermionic energy.  The remaining eigenmodes, Eq.~(\ref{eq:wf3}), prove to be very difficult to regulate from their very complicated $p$-dependence.  We assume that they will not contribute to the L\"uscher term, as they are massive, and the propagators will go as $e^{-mL}$, $m$ being the mass of the eigenmode and $L$ the large quark separation. 

\bibliographystyle{utphys}
\bibliography{Bibliography}

\providecommand{\href}[2]{#2}\begingroup\raggedright\begin{thebibliography}{10}

\bibitem{Luscher:1980ac}
M.~Luscher, ``{Symmetry Breaking Aspects of the Roughening Transition in Gauge
  Theories},''
\href{http://dx.doi.org/10.1016/0550-3213(81)90423-5}{{\em Nucl. Phys.} {\bf
  B180} (1981)  317}.

\bibitem{Luscher:1980fr}
M.~Luscher, K.~Symanzik, and P.~Weisz, ``{Anomalies of the Free Loop Wave
  Equation in the WKB Approximation},''
\href{http://dx.doi.org/10.1016/0550-3213(80)90009-7}{{\em Nucl. Phys.} {\bf
  B173} (1980)  365}.

\bibitem{Shifman:2005eb}
M.~Shifman, ``{k strings from various perspectives: QCD, lattices, string
  theory and toy models},'' {\em Acta Phys. Polon.} {\bf B36} (2005)
  3805--3836,
\href{http://arxiv.org/abs/hep-ph/0510098}{{\tt arXiv:hep-ph/0510098}}.

\bibitem{Maldacena:1997re}
J.~M. Maldacena, ``{The large N limit of superconformal field theories and
  supergravity},'' {\em Adv. Theor. Math. Phys.} {\bf 2} (1998)  231--252,
\href{http://arxiv.org/abs/hep-th/9711200}{{\tt arXiv:hep-th/9711200}}.

\bibitem{Klebanov:2000hb}
I.~R. Klebanov and M.~J. Strassler, ``{Supergravity and a confining gauge
  theory: Duality cascades and chiSB-resolution of naked singularities},'' {\em
  JHEP} {\bf 08} (2000)  052,
\href{http://arxiv.org/abs/hep-th/0007191}{{\tt arXiv:hep-th/0007191}}.

\bibitem{Cvetic:2001ma}
M.~Cvetic, G.~W. Gibbons, H.~Lu, and C.~N. Pope, ``{Supersymmetric non-singular
  fractional D2-branes and NS-NS 2-branes},''
  \href{http://dx.doi.org/10.1016/S0550-3213(01)00236-X}{{\em Nucl. Phys.} {\bf
  B606} (2001)  18--44},
\href{http://arxiv.org/abs/hep-th/0101096}{{\tt arXiv:hep-th/0101096}}.

\bibitem{Herzog:2001fq}
C.~P. Herzog and I.~R. Klebanov, ``On string tensions in supersymmetric su(m)
  gauge theory,'' \href{http://dx.doi.org/10.1016/S0370-2693(02)01155-3}{{\em
  Phys. Lett.} {\bf B526} (2002)  388--392},
\href{http://arxiv.org/abs/hep-th/0111078}{{\tt arXiv:hep-th/0111078}}.

\bibitem{Herzog:2002ss}
C.~P. Herzog, ``{String tensions and three dimensional confining gauge
  theories},'' \href{http://dx.doi.org/10.1103/PhysRevD.66.065009}{{\em Phys.
  Rev.} {\bf D66} (2002)  065009},
\href{http://arxiv.org/abs/hep-th/0205064}{{\tt arXiv:hep-th/0205064}}.

\bibitem{Ridgway:2007vh}
J.~M. Ridgway, ``{Confining k-string tensions with D-branes in super Yang-
  Mills theories},''
  \href{http://dx.doi.org/10.1016/j.physletb.2007.02.054}{{\em Phys. Lett.}
  {\bf B648} (2007)  76--83},
\href{http://arxiv.org/abs/hep-th/0701079}{{\tt arXiv:hep-th/0701079}}.

\bibitem{PandoZayas:2008hw}
L.~A. Pando~Zayas, V.~G.~J. Rodgers, and K.~Stiffler, ``{Luscher Term for
  k-string Potential from Holographic One Loop Corrections},''
\href{http://arxiv.org/abs/0809.4119}{{\tt arXiv:0809.4119 [hep-th]}}.

\bibitem{Doran:2009pp}
C.~A. Doran, L.~A. Pando~Zayas, V.~G.~J. Rodgers, and K.~Stiffler, ``{Tensions
  and Luscher Terms for (2+1)-dimensional k-strings from Holographic Models},''
  \href{http://dx.doi.org/10.1088/1126-6708/2009/11/064}{{\em JHEP} {\bf 11}
  (2009)  064},
\href{http://arxiv.org/abs/0907.1331}{{\tt arXiv:0907.1331 [hep-th]}}.

\bibitem{Stiffler:2009ma}
K.~Stiffler, ``{Mesons From String Theory},''
\href{http://arxiv.org/abs/0909.5681}{{\tt arXiv:0909.5681 [hep-th]}}.

\bibitem{Teper:1998te}
M.~J. Teper, ``{SU(N) gauge theories in 2+1 dimensions},''
  \href{http://dx.doi.org/10.1103/PhysRevD.59.014512}{{\em Phys. Rev.} {\bf
  D59} (1999)  014512},
\href{http://arxiv.org/abs/hep-lat/9804008}{{\tt arXiv:hep-lat/9804008}}.

\bibitem{Bringoltz:2006zg}
B.~Bringoltz and M.~Teper, ``{A precise calculation of the fundamental string
  tension in SU(N) gauge theories in 2+1 dimensions},''
  \href{http://dx.doi.org/10.1016/j.physletb.2006.12.056}{{\em Phys. Lett.}
  {\bf B645} (2007)  383--388},
\href{http://arxiv.org/abs/hep-th/0611286}{{\tt arXiv:hep-th/0611286}}.

\bibitem{Bringoltz:2008nd}
B.~Bringoltz and M.~Teper, ``{Closed k-strings in SU(N) gauge theories : 2+1
  dimensions},'' \href{http://dx.doi.org/10.1016/j.physletb.2008.04.052}{{\em
  Phys. Lett.} {\bf B663} (2008)  429--437},
\href{http://arxiv.org/abs/0802.1490}{{\tt arXiv:0802.1490 [hep-lat]}}.

\bibitem{Karabali:1997wk}
D.~Karabali, C.-j. Kim, and V.~P. Nair, ``{Planar Yang-Mills theory:
  Hamiltonian, regulators and mass gap},''
  \href{http://dx.doi.org/10.1016/S0550-3213(98)00309-5}{{\em Nucl. Phys.} {\bf
  B524} (1998)  661--694},
\href{http://arxiv.org/abs/hep-th/9705087}{{\tt arXiv:hep-th/9705087}}.

\bibitem{Karabali:1998yq}
D.~Karabali, C.-j. Kim, and V.~P. Nair, ``{On the vacuum wave function and
  string tension of Yang- Mills theories in (2+1) dimensions},''
  \href{http://dx.doi.org/10.1016/S0370-2693(98)00751-5}{{\em Phys. Lett.} {\bf
  B434} (1998)  103--109},
\href{http://arxiv.org/abs/hep-th/9804132}{{\tt arXiv:hep-th/9804132}}.

\bibitem{Karabali:2000gy}
D.~Karabali, C.-j. Kim, and V.~P. Nair, ``{Manifest covariance and the
  Hamiltonian approach to mass gap in (2+1)-dimensional Yang-Mills theory},''
  \href{http://dx.doi.org/10.1103/PhysRevD.64.025011}{{\em Phys. Rev.} {\bf
  D64} (2001)  025011},
\href{http://arxiv.org/abs/hep-th/0007188}{{\tt arXiv:hep-th/0007188}}.

\bibitem{Karabali:2007mr}
D.~Karabali and V.~P. Nair, ``{The robustness of the vacuum wave function and
  other matters for Yang-Mills theory},''
  \href{http://dx.doi.org/10.1103/PhysRevD.77.025014}{{\em Phys. Rev.} {\bf
  D77} (2008)  025014},
\href{http://arxiv.org/abs/0705.2898}{{\tt arXiv:0705.2898 [hep-th]}}.

\bibitem{Karabali:2009rg}
D.~Karabali, V.~P. Nair, and A.~Yelnikov, ``{The Hamiltonian Approach to
  Yang-Mills (2+1): An Expansion Scheme and Corrections to String Tension},''
\href{http://arxiv.org/abs/0906.0783}{{\tt arXiv:0906.0783 [hep-th]}}.

\bibitem{Firouzjahi:2006vp}
H.~Firouzjahi, L.~Leblond, and S.~H. Henry~Tye, ``{The (p,q) string tension in
  a warped deformed conifold},'' {\em JHEP} {\bf 05} (2006)  047,
\href{http://arxiv.org/abs/hep-th/0603161}{{\tt arXiv:hep-th/0603161}}.

\bibitem{Feynman:1965}
R.~P. Feynman, {\em The Character of Physical Law}.
\newblock MIT Press, 1965.

\bibitem{ps}
M.~Peskin and D.~Schroeder, {\em An Introduction to Quantum Field Theory}.
\newblock Addison-Wesley Publishing Company, 1995.

\bibitem{Aharony:1999ti}
O.~Aharony, S.~S. Gubser, J.~M. Maldacena, H.~Ooguri, and Y.~Oz, ``{Large N
  field theories, string theory and gravity},''
  \href{http://dx.doi.org/10.1016/S0370-1573(99)00083-6}{{\em Phys. Rept.} {\bf
  323} (2000)  183--386},
\href{http://arxiv.org/abs/hep-th/9905111}{{\tt arXiv:hep-th/9905111}}.

\bibitem{bbs1}
K.~Becker, M.~Becker, and J.~Schwarz, {\em String Theory and M-Theory}.
\newblock Cambridge University Press, 2007.

\bibitem{Szabo:2002ca}
R.~J. Szabo, ``{BUSSTEPP lectures on string theory: An introduction to string
  theory and D-brane dynamics},''
\href{http://arxiv.org/abs/hep-th/0207142}{{\tt arXiv:hep-th/0207142}}.

\bibitem{Polchinski:1998v1}
J.~Polchinski, {\em String Theory Volume I: An Introduction to the Bosonic
  String}.
\newblock Cambridge University Press, 1998.

\bibitem{GreenSchwarzWitten:1987v1}
M.~B. Green, J.~H. Schwarz, and E.~Witten, {\em Superstring Theory}, vol.~1.
\newblock Cambridge University Press, 1987.

\bibitem{Polchinski:1998v2}
J.~Polchinski, {\em String Theory Volume II: Superstring Theory and Beyond}.
\newblock Cambridge University Press, 1998.

\bibitem{Polyakov:1981re}
A.~M. Polyakov, ``{Quantum geometry of fermionic strings},''
\href{http://dx.doi.org/10.1016/0370-2693(81)90744-9}{{\em Phys. Lett.} {\bf
  B103} (1981)  211--213}.

\bibitem{GreenSchwarzWitten:1987v2}
M.~B. Green, J.~H. Schwarz, and E.~Witten, {\em Superstring Theory}, vol.~2.
\newblock Cambridge University Press, 1987.

\bibitem{VanNieuwenhuizen:1981ae}
P.~Van~Nieuwenhuizen, ``{Supergravity},''
\href{http://dx.doi.org/10.1016/0370-1573(81)90157-5}{{\em Phys. Rept.} {\bf
  68} (1981)  189--398}.

\bibitem{Herzog:2000rz}
C.~P. Herzog and I.~R. Klebanov, ``{Gravity duals of fractional branes in
  various dimensions},''
  \href{http://dx.doi.org/10.1103/PhysRevD.63.126005}{{\em Phys. Rev.} {\bf
  D63} (2001)  126005},
\href{http://arxiv.org/abs/hep-th/0101020}{{\tt arXiv:hep-th/0101020}}.

\bibitem{Horowitz:1991cd}
G.~T. Horowitz and A.~Strominger, ``{Black strings and P-branes},''
\href{http://dx.doi.org/10.1016/0550-3213(91)90440-9}{{\em Nucl. Phys.} {\bf
  B360} (1991)  197--209}.

\bibitem{Polchinski:1995mt}
J.~Polchinski, ``{Dirichlet-Branes and Ramond-Ramond Charges},''
  \href{http://dx.doi.org/10.1103/PhysRevLett.75.4724}{{\em Phys. Rev. Lett.}
  {\bf 75} (1995)  4724--4727},
\href{http://arxiv.org/abs/hep-th/9510017}{{\tt arXiv:hep-th/9510017}}.

\bibitem{z1}
B.~Zwiebach, {\em A First Course in String Theory}.
\newblock Cambridge University Press, 2005.

\bibitem{'tHooft:1973jz}
G.~'t~Hooft, ``{A PLANAR DIAGRAM THEORY FOR STRONG INTERACTIONS},''
\href{http://dx.doi.org/10.1016/0550-3213(74)90154-0}{{\em Nucl. Phys.} {\bf
  B72} (1974)  461}.

\bibitem{Witten:1995im}
E.~Witten, ``{Bound states of strings and p-branes},''
  \href{http://dx.doi.org/10.1016/0550-3213(95)00610-9}{{\em Nucl. Phys.} {\bf
  B460} (1996)  335--350},
\href{http://arxiv.org/abs/hep-th/9510135}{{\tt arXiv:hep-th/9510135}}.

\bibitem{Martucci:2005rb}
L.~Martucci, J.~Rosseel, D.~Van~den Bleeken, and A.~Van~Proeyen, ``{Dirac
  actions for D-branes on backgrounds with fluxes},''
  \href{http://dx.doi.org/10.1088/0264-9381/22/13/014}{{\em Class. Quant.
  Grav.} {\bf 22} (2005)  2745--2764},
\href{http://arxiv.org/abs/hep-th/0504041}{{\tt arXiv:hep-th/0504041}}.

\bibitem{Gubser:1998bc}
S.~S. Gubser, I.~R. Klebanov, and A.~M. Polyakov, ``{Gauge theory correlators
  from non-critical string theory},''
  \href{http://dx.doi.org/10.1016/S0370-2693(98)00377-3}{{\em Phys. Lett.} {\bf
  B428} (1998)  105--114},
\href{http://arxiv.org/abs/hep-th/9802109}{{\tt arXiv:hep-th/9802109}}.

\bibitem{Witten:1998qj}
E.~Witten, ``{Anti-de Sitter space and holography},'' {\em Adv. Theor. Math.
  Phys.} {\bf 2} (1998)  253--291,
\href{http://arxiv.org/abs/hep-th/9802150}{{\tt arXiv:hep-th/9802150}}.

\bibitem{vgjrp}
V.~G.~J. Rodgers. Private communication.

\bibitem{Douglas:1995nw}
M.~R. Douglas and S.~H. Shenker, ``{Dynamics of SU(N) supersymmetric gauge
  theory},'' \href{http://dx.doi.org/10.1016/0550-3213(95)00258-T}{{\em Nucl.
  Phys.} {\bf B447} (1995)  271--296},
\href{http://arxiv.org/abs/hep-th/9503163}{{\tt arXiv:hep-th/9503163}}.

\bibitem{Maldacena:2000yy}
J.~M. Maldacena and C.~Nunez, ``{Towards the large N limit of pure N = 1 super
  Yang Mills},'' \href{http://dx.doi.org/10.1103/PhysRevLett.86.588}{{\em Phys.
  Rev. Lett.} {\bf 86} (2001)  588--591},
\href{http://arxiv.org/abs/hep-th/0008001}{{\tt arXiv:hep-th/0008001}}.

\bibitem{Canoura:2005uz}
F.~Canoura, J.~D. Edelstein, L.~A.~P. Zayas, A.~V. Ramallo, and D.~Vaman,
  ``{Supersymmetric branes on AdS(5) x Y**(p,q) and their field theory
  duals},'' {\em JHEP} {\bf 03} (2006)  101,
\href{http://arxiv.org/abs/hep-th/0512087}{{\tt arXiv:hep-th/0512087}}.

\bibitem{Klebanov:2000nc}
I.~R. Klebanov and A.~A. Tseytlin, ``{Gravity duals of supersymmetric SU(N) x
  SU(N+M) gauge theories},''
  \href{http://dx.doi.org/10.1016/S0550-3213(00)00206-6}{{\em Nucl. Phys.} {\bf
  B578} (2000)  123--138},
\href{http://arxiv.org/abs/hep-th/0002159}{{\tt arXiv:hep-th/0002159}}.

\bibitem{PandoZayas:2003yb}
L.~A. Pando~Zayas, J.~Sonnenschein, and D.~Vaman, ``{Regge trajectories
  revisited in the gauge / string correspondence},''
  \href{http://dx.doi.org/10.1016/j.nuclphysb.2003.12.006}{{\em Nucl. Phys.}
  {\bf B682} (2004)  3--44},
\href{http://arxiv.org/abs/hep-th/0311190}{{\tt arXiv:hep-th/0311190}}.

\bibitem{Ambjorn:1984me}
J.~Ambjorn, P.~Olesen, and C.~Peterson, ``{OBSERVATION OF A STRING IN
  THREE-DIMENSIONAL SU(2) LATTICE GAUGE THEORY},''
\href{http://dx.doi.org/10.1016/0370-2693(84)91352-2}{{\em Phys. Lett.} {\bf
  B142} (1984)  410}.

\bibitem{Ambjorn:1984mb}
J.~Ambjorn, P.~Olesen, and C.~Peterson, ``{Stochastic Confinement and
  Dimensional Reduction. 1. Four- Dimensional SU(2) Lattice Gauge Theory},''
\href{http://dx.doi.org/10.1016/0550-3213(84)90475-9}{{\em Nucl. Phys.} {\bf
  B240} (1984)  189}.

\bibitem{Ambjorn:1984yu}
J.~Ambjorn, P.~Olesen, and C.~Peterson, ``{THREE-DIMENSIONAL LATTICE GAUGE
  THEORY AND STRINGS},''
\href{http://dx.doi.org/10.1016/0550-3213(84)90193-7}{{\em Nucl. Phys.} {\bf
  B244} (1984)  262}.

\bibitem{Schulman:1981pi}
L.~S. Schulman, {\em Techniques and Applications of Path Integration}.
\newblock John Wiley \& Sons, 1981.

\bibitem{Cvetic:2000mh}
M.~Cvetic, H.~Lu, and C.~N. Pope, ``{Brane resolution through transgression},''
  \href{http://dx.doi.org/10.1016/S0550-3213(01)00050-5}{{\em Nucl. Phys.} {\bf
  B600} (2001)  103--132},
\href{http://arxiv.org/abs/hep-th/0011023}{{\tt arXiv:hep-th/0011023}}.

\bibitem{Candelas:1989js}
P.~Candelas and X.~de~la Ossa, ``Comments on conifolds,'' {\em Nucl. Phys. B}
  {\bf 342} (1990)  246.

\bibitem{Maldacena:2001pb}
J.~M. Maldacena and H.~S. Nastase, ``{The supergravity dual of a theory with
  dynamical supersymmetry breaking},'' {\em JHEP} {\bf 09} (2001)  024,
\href{http://arxiv.org/abs/hep-th/0105049}{{\tt arXiv:hep-th/0105049}}.

\bibitem{Liu:2010i}
X.~Liu and K.~Stiffler, ``A simultaneous calculation of d=3 and d=4 spacetime
  dimensional k-string tensions and luscher terms using holography (manuscript
  in progress),''.

\bibitem{Gomis:2006im}
J.~Gomis and F.~Passerini, ``{Wilson loops as D3-branes},'' {\em JHEP} {\bf 01}
  (2007)  097,
\href{http://arxiv.org/abs/hep-th/0612022}{{\tt arXiv:hep-th/0612022}}.

\bibitem{Gomis:2006sb}
J.~Gomis and F.~Passerini, ``{Holographic Wilson loops},'' {\em JHEP} {\bf 08}
  (2006)  074,
\href{http://arxiv.org/abs/hep-th/0604007}{{\tt arXiv:hep-th/0604007}}.

\bibitem{Bigazzi:2004ze}
F.~Bigazzi, A.~L. Cotrone, L.~Martucci, and L.~A. Pando~Zayas, ``{Wilson loop,
  Regge trajectory and hadron masses in a Yang- Mills theory from semiclassical
  strings},'' \href{http://dx.doi.org/10.1103/PhysRevD.71.066002}{{\em Phys.
  Rev.} {\bf D71} (2005)  066002},
\href{http://arxiv.org/abs/hep-th/0409205}{{\tt arXiv:hep-th/0409205}}.

\bibitem{Bigazzi:2002gw}
F.~Bigazzi, A.~L. Cotrone, L.~Girardello, and A.~Zaffaroni, ``{pp-wave and
  non-supersymmetric gauge theory},'' {\em JHEP} {\bf 10} (2002)  030,
\href{http://arxiv.org/abs/hep-th/0205296}{{\tt arXiv:hep-th/0205296}}.

\bibitem{Bertoldi:2004rn}
G.~Bertoldi, F.~Bigazzi, A.~L. Cotrone, C.~Nunez, and L.~A. Pando~Zayas, ``{On
  the universality class of certain string theory hadrons},''
  \href{http://dx.doi.org/10.1016/j.nuclphysb.2004.08.044}{{\em Nucl. Phys.}
  {\bf B700} (2004)  89--139},
\href{http://arxiv.org/abs/hep-th/0401031}{{\tt arXiv:hep-th/0401031}}.

\bibitem{Elizalde:1995a}
E.~Elizalde, {\em Ten Physical Applications of Spectral Zeta Functions}.
\newblock Springer, 1995.

\bibitem{kachru}
S.~Kachru, ``Tasi lectures.'' Lectures given at tasi 2007.

\bibitem{Ohta:1999we}
K.~Ohta and T.~Yokono, ``{Deformation of conifold and intersecting branes},''
  {\em JHEP} {\bf 02} (2000)  023,
\href{http://arxiv.org/abs/hep-th/9912266}{{\tt arXiv:hep-th/9912266}}.

\bibitem{Gubser:1998fp}
S.~S. Gubser and I.~R. Klebanov, ``{Baryons and domain walls in an N = 1
  superconformal gauge theory},''
  \href{http://dx.doi.org/10.1103/PhysRevD.58.125025}{{\em Phys. Rev.} {\bf
  D58} (1998)  125025},
\href{http://arxiv.org/abs/hep-th/9808075}{{\tt arXiv:hep-th/9808075}}.

\bibitem{Klebanov:1999rd}
I.~R. Klebanov and N.~A. Nekrasov, ``{Gravity duals of fractional branes and
  logarithmic RG flow},''
  \href{http://dx.doi.org/10.1016/S0550-3213(00)00016-X}{{\em Nucl. Phys.} {\bf
  B574} (2000)  263--274},
\href{http://arxiv.org/abs/hep-th/9911096}{{\tt arXiv:hep-th/9911096}}.

\bibitem{Higuchi:1986wu}
A.~Higuchi, ``{SYMMETRIC TENSOR SPHERICAL HARMONICS ON THE N SPHERE AND THEIR
  APPLICATION TO THE DE SITTER GROUP SO(N,1)},''
\href{http://dx.doi.org/10.1063/1.527513}{{\em J. Math. Phys.} {\bf 28} (1987)
  1553}.

\bibitem{Marolf:2003ye}
D.~Marolf, L.~Martucci, and P.~J. Silva, ``{Fermions, T-duality and effective
  actions for D-branes in bosonic backgrounds},'' {\em JHEP} {\bf 04} (2003)
  051,
\href{http://arxiv.org/abs/hep-th/0303209}{{\tt arXiv:hep-th/0303209}}.

\bibitem{Marolf:2003vf}
D.~Marolf, L.~Martucci, and P.~J. Silva, ``{Actions and fermionic symmetries
  for D-branes in bosonic backgrounds},'' {\em JHEP} {\bf 07} (2003)  019,
\href{http://arxiv.org/abs/hep-th/0306066}{{\tt arXiv:hep-th/0306066}}.

\bibitem{Marolf:2004jb}
D.~Marolf, L.~Martucci, and P.~J. Silva, ``{The explicit form of the effective
  action for F1 and D- branes},'' {\em Class. Quant. Grav.} {\bf 21} (2004)
  S1385--S1390,
\href{http://arxiv.org/abs/hep-th/0404197}{{\tt arXiv:hep-th/0404197}}.

\bibitem{Lawrie:1990}
I.~D. Lawrie, {\em A Unified Grand Tour of Theoretical Physics}.
\newblock IOP Publishing Ltd., 1990.

\bibitem{Kirsch:2006he}
I.~Kirsch, ``{Spectroscopy of fermionic operators in AdS/CFT},'' {\em JHEP}
  {\bf 09} (2006)  052,
\href{http://arxiv.org/abs/hep-th/0607205}{{\tt arXiv:hep-th/0607205}}.

\end{thebibliography}\endgroup

\end{document}